\documentclass[11pt]{article}
\usepackage[a4paper,margin=2.3cm]{geometry}
\usepackage[T1]{fontenc}
\usepackage{amsmath,amsthm,mathtools}
\usepackage{newtxtext,newtxmath}
\usepackage[scaled=0.92]{helvet}
\DeclareMathVersion{sfdigits}
\SetSymbolFont{operators}{sfdigits}{OT1}{phv}{m}{n}
\SetMathAlphabet{\mathbf}{sfdigits}{OT1}{phv}{b}{n}
\usepackage{microtype}
\usepackage{aliascnt}
\usepackage{physics}
\usepackage{graphicx}
\usepackage{booktabs}
\usepackage{array}
\usepackage{longtable}
\usepackage{algorithm}
\usepackage{algpseudocode}
\usepackage{placeins}
\usepackage{tikz}
\usepackage{xcolor}
\usepackage[numbers,super,sort&compress]{natbib}
\usepackage{bibunits}
\usepackage{caption}
\DeclareCaptionLabelSeparator{pipe}{\ $|$\ }
\DeclareCaptionFont{sfmath}{\sffamily\mathversion{sfdigits}}
\usepackage{titlesec}
\usepackage{hyperref}
\usepackage[capitalize,nameinlink]{cleveref}
\RemoveFromHook{label}[firstaid/cleveref] %
\usepackage{lineno}
\hypersetup{hidelinks,pdftitle={Certifying quantum code and circuit distances at scale}}
\allowdisplaybreaks
\titleformat{\section}{\sffamily\bfseries\large}{\thesection}{.7em}{}
\titleformat{\subsection}{\sffamily\bfseries\normalsize}{\thesubsection}{.7em}{}
\titlespacing*{\section}{0pt}{1.3em plus .3em minus .2em}{.4em}
\titlespacing*{\subsection}{0pt}{1em plus .2em minus .2em}{.25em}
\renewcommand{\bibnumfmt}[1]{#1.}

\newif\ifreview \reviewfalse  %
\newcommand{\fullwidthfigure}[1]{\makebox[\linewidth][c]{\includegraphics[width=183mm]{#1}}}
\makeatletter
\renewcommand{\maketitle}{\begingroup\raggedright\sffamily
  {\fontsize{20}{24}\selectfont\bfseries\@title\par}\vspace{10pt}
  {\fontsize{11}{14}\selectfont\@author\par}\vspace{5pt}
  {\fontsize{8.5}{11}\selectfont\affiliations\par}\endgroup\vspace{12pt}}
\renewenvironment{abstract}{\par\begingroup\noindent\sffamily\mathversion{sfdigits}\small\ignorespaces}{\par\endgroup}
\makeatother

\newtheorem{theorem}{Theorem}[section]
\newtheorem{proposition}[theorem]{Proposition}
\newtheorem{lemma}[theorem]{Lemma}

\newtheorem{corollary}[theorem]{Corollary}

\theoremstyle{definition}
\newtheorem{definition}[theorem]{Definition}

\newtheorem{remark}[theorem]{Remark}

\newcommand{\FF}{\mathbb{F}}
\newcommand{\CC}{\mathbb{C}}
\newcommand{\RR}{\mathbb{R}}

\newcommand{\TT}{\mathbb{T}}

\newcommand{\calB}{\mathcal{B}}
\newcommand{\calC}{\mathcal{C}}
\newcommand{\calD}{\mathcal{D}}

\newcommand{\calP}{\mathcal{P}}
\newcommand{\calQ}{\mathcal{Q}}
\newcommand{\calR}{\mathcal{R}}

\newcommand{\calO}{\mathcal{O}}

\newcommand{\calG}{\mathcal{G}}

\newcommand{\calT}{\mathcal{T}}

\newcommand{\sfP}{\mathsf{P}}
\newcommand{\sfS}{\mathsf{S}}
\newcommand{\sfN}{\mathsf{N}}
\newcommand{\sfT}{\mathsf{T}}

\newcommand{\trop}{\mathrm{trop}}

\newcommand{\im}{\operatorname{im}}
\newcommand{\leaf}{\operatorname{leaf}}
\newcommand{\rw}{\operatorname{rw}}

\newcommand{\poly}{\operatorname{poly}}

\newcommand{\supp}{\operatorname{supp}}

\newcommand{\wt}{\operatorname{wt}}

\newcommand{\rowsp}{\operatorname{rowsp}}

\newcommand{\Ret}{\operatorname{Ret}_2}
\newcommand{\disjoint}{\mathbin{\dot{\cup}}}

\renewcommand{\qc}[1]{\ensuremath{[\![#1]\!]}}  %
\renewcommand{\figurename}{Fig.}
\crefname{table}{Supplementary Table}{Supplementary Tables}
\Crefname{table}{Supplementary Table}{Supplementary Tables}
\crefname{figure}{Supplementary Fig.}{Supplementary Figs.}
\Crefname{figure}{Supplementary Fig.}{Supplementary Figs.}
\crefname{section}{Section}{Sections}
\Crefname{section}{Section}{Sections}

\title{Certifying quantum code and circuit distances at scale}
\author{Bin Cheng\textsuperscript{1}, Naixu Guo\textsuperscript{1} and Feng Pan\textsuperscript{1,2,*}}
\newcommand{\affiliations}{%
\textsuperscript{1}Centre for Quantum Technologies, National University of Singapore, Singapore\\
\textsuperscript{2}Science, Mathematics and Technology Cluster, Singapore University of Technology and Design, Singapore\\
\textsuperscript{*}Corresponding author: \href{mailto:feng\_pan@sutd.edu.sg}{feng\_pan@sutd.edu.sg}}
\date{}

\begin{document}
\maketitle
\ifreview\linenumbers\fi

\begin{abstract}
Code and circuit distances characterize how strongly quantum error correction suppresses logical errors and guide code comparisons, schedule choices and resource estimates.
Yet many published distances are only upper bounds from heuristic searches.
Here, we express both distances as tropical partition functions and evaluate them with rank-based dynamic programming and weight-bounded search.
Circuit reductions and relaxations shrink the problems, and proved lower bounds with explicit logical errors of matching weight certify exact values.
On two benchmarks, our method certifies more instances than all other tested exact solvers combined.
We certify distances for codes with up to 31,200 qubits, such as Google's modular hyperbolic codes, and for memory circuits with up to $8\times10^5$ fault mechanisms, such as those of high-rate Cornucopia codes.
Our audit of 26 recent papers identifies overestimated distances and non-tight upper bounds in 10 papers, with a maximum discrepancy of at least eight.
Certified distances also guide circuit design and show which defective chips remain usable.
In particular, we certify circuit distance 16 for IBM's two-gross memory circuits with the standard schedule, not the assumed 18, whereas redesigned schedules reach 18 at one noisy cycle, at the cost of one more timestep per cycle.
At 99.98\% junction yield, keeping gross-code memory blocks with dead checks that are certified at distance 10 or more raises the estimated usable share from 53\% for perfect blocks alone to 94\% with an improved schedule.
\end{abstract}

\medskip
Quantum computers promise to simulate complex materials and solve classically intractable problems such as factoring, yet these computations require logical error rates of $10^{-10}$ or less, far below the physical error rates of any hardware\cite{beverland_2022_requirements,gidney_2025_rsa}.
Quantum error correction bridges this gap by encoding logical information across many physical qubits and repeatedly measuring parity checks to detect errors as they occur\cite{shor_1995,dennis_2002,fowler_2012}.
Experiments have demonstrated surface-code memories below threshold on superconducting qubits\cite{google_2025_below} and logical processors on neutral atoms\cite{bluvstein_2024}.
With the surface code, however, a cryptographically relevant computation needs about a million physical qubits\cite{gidney_2025_rsa}.
Quantum low-density parity-check (qLDPC) codes encode many logical qubits in one block and promise to cut this overhead by an order of magnitude\cite{gottesman_2014,breuckmann_2021,bravyi_2024_bb}.
These codes now appear in hardware roadmaps, including IBM's plan for a processor with 200 logical qubits running $10^8$ gates by 2029 on bivariate bicycle codes\cite{yoder_2025_tour,ibm_roadmap}.

A code of distance $d$ corrects any combination of fewer than $d/2$ errors, and below threshold its logical error rate falls roughly as $p^{d/2}$ with the physical error rate $p$\cite{dennis_2002,fowler_2012}.
Codes are therefore ranked by their distance, and resource estimates use it to set the size of each code block, the number of physical qubits and the error budget of a computation\cite{beverland_2022_requirements,yoder_2025_tour}.
In hardware the relevant quantity is the circuit distance, the smallest number of faults in the syndrome-measurement circuit that flip a logical observable without changing any detector.
A detector is a parity of measurement outcomes that is fixed when no fault occurs\cite{mcewen_2023_relaxing,stim}.
Because the circuit spreads a single fault onto several qubits, the circuit distance can be smaller than the code distance.
This reduction is common in qLDPC codes, whose checks involve many qubits.
IBM's gross code, for example, has code distance 12 but circuit distance 10 with its standard schedule\cite{bravyi_2024_bb}.
Overestimating the circuit distance by two increases the predicted exponent of $p$ in the low-noise logical error rate by one.
At a fixed $p$, the error in the predicted rate also depends on the number of low-weight fault sets.

Computing the minimum distance of a classical linear code is already NP-hard\cite{vardy_1997}.
Computing circuit distance is more challenging in practice because the search problems are much larger, with $10^4$ to $10^5$ fault mechanisms in current circuit designs.
In practice, papers report upper bounds from heuristic searches, which decode random logical constraints\cite{roffe_2020_decoding}, sample random information sets\cite{qdistrnd} or use Stim's search for undetectable errors\cite{stim}.
Such searches provide upper bounds on the distance, but do not certify that the errors found have minimum weight.
The resulting estimates may therefore exceed the true distance.
Exact methods, among them integer programming, MaxSAT on the circuit's error model\cite{beni_2025_tesseract} and connected-cluster enumeration\cite{dumer_2017_verification,distm4ri_code}, certify small instances and some memory circuits, such as that of IBM's gross code, but stop short of most circuits proposed for hardware\cite{webster_2026_distance}.
Resource estimates for planned machines therefore rest on assumed distances.
The architecture study of IBM's processor, for example, takes the circuit distance of its two-gross code to be 18 and notes that the true value is not known\cite{yoder_2025_tour}.

Here, we compute code and circuit distances exactly, at the scale of current proposals.
We express each distance as a tropical partition function\cite{liu_2021_tropical} and evaluate it with three exact algorithms, after certified relaxations have reduced the problem.
Each exact distance is certified by a lower bound established through exhaustive computation and an explicit error of matching weight.
For circuits, we verify this error by replaying it in simulation.
When a search is incomplete, we report the interval between the certified lower bound and the weight of the lightest logical error found.

On two benchmarks, our method certifies more instances than all other exact solvers combined.
The method reaches the scale needed to check published distances and compare circuit designs.
Our audit of 26 recent papers finds that code distances reported as exact are reliable.
In 10 papers, however, at least one reported circuit distance or code-distance estimate exceeds the true distance.
The largest overestimate is at least eight.
Beyond checking published values, our methods support circuit design, architecture assessment, logical measurements and chip certification.
For circuit design, we compute the exact circuit distances of all 8,496 uniform depth-seven schedules for each of four bivariate bicycle codes; the standard schedule is optimal for only one of them.
For architecture assessment, we evaluate memory instructions in IBM's planned processor: with one noisy cycle, the two-gross code has circuit distance 16 with the standard schedule and 18 with optimized schedules.
Explicit fault sets yield decoder-independent lower bounds on logical error rates and low-noise estimates for a minimum-weight decoder.
For logical measurements, we certify the merged-code distances of two published constructions.
One construction preserves the distance in every merge only when the gauge qubits introduced by merging are fixed.
For chip certification, we certify lower bounds on the circuit distances of individual gross-code chips with dead checks and estimate the increase in usable yield.

\section*{Results}

\subsection*{Certified distances from a tropical partition function}

\begin{figure}[tbp]
\centering
\fullwidthfigure{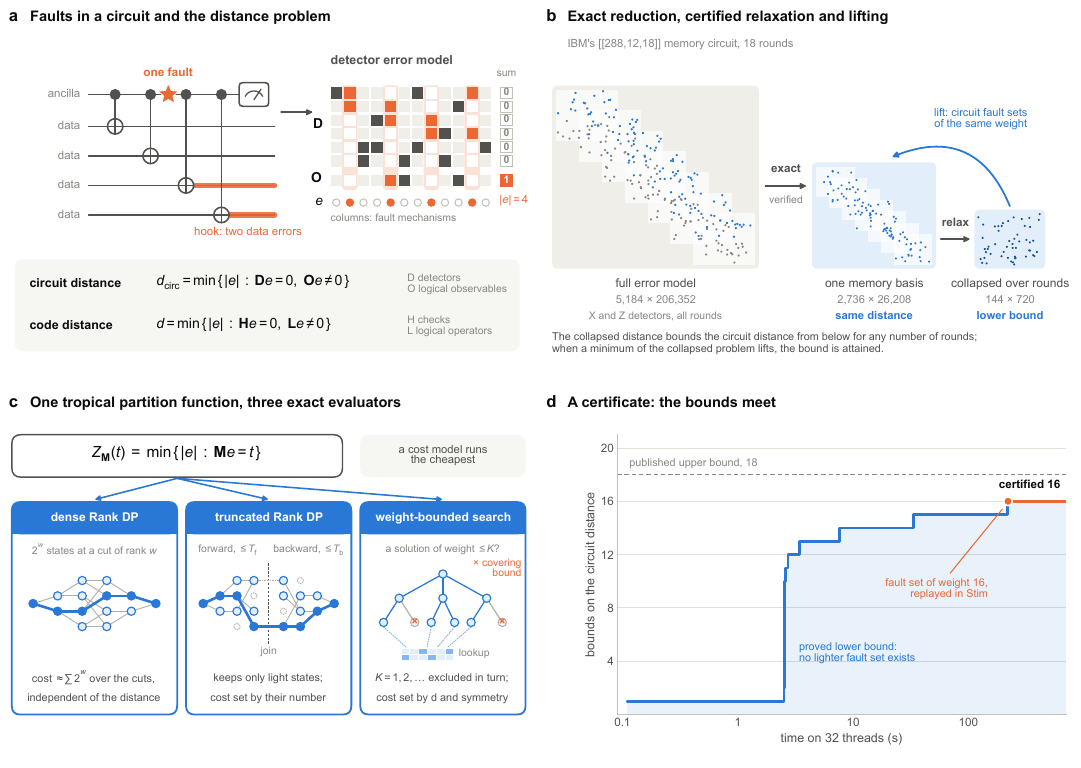}
\caption{\textbf{Certified code and circuit distances from one tropical partition function.} \textbf{a}, In a
syndrome-measurement circuit, one fault on the ancilla of a check spreads to the data qubits of its later gates
(a hook error). The detector error model lists every fault mechanism as a column of the detector matrix
$\vb{D}$ and the observable matrix $\vb{O}$.
A fault set (orange columns) is an undetected logical
error when its columns sum to zero on $\vb{D}$ and not on $\vb{O}$; the circuit distance is the least
weight of such a set. The code distance is the same problem for the check matrix $\vb{H}$ and the logical
operators $\vb{L}$; for a CSS code, it is the smaller of the values for the two Pauli types. \textbf{b}, Reductions of a CSS memory circuit, with the sizes for IBM's
\qc{288,12,18} circuit over 18 rounds: keeping the detectors of one memory basis is exact (checked for every
circuit); summing each check's detectors over the rounds gives a small problem whose distance is a lower bound
for any number of identical rounds; lifting maps its minimum solutions back to circuit fault sets. \textbf{c}, Three exact
algorithms compute the distance, the least value over the logical rows of the tropical partition function
$Z_{\vb{M}}(\vb{t})$, the least weight of a solution of $\vb{M}\vb{e}=\vb{t}$. The two dynamic programs evaluate it for one
logical row at a time: the dense Rank DP sweeps a trellis with $2^w$ states at a cut of rank $w$, and the truncated Rank DP
keeps only light partial solutions, up to weights $T_{\mathrm{f}}$ and $T_{\mathrm{b}}$ in forward and backward sweeps that are joined at every cut. The
weight-bounded search decides for all logical rows at once whether the distance is at most $K$, for $K=1,2,\ldots$ in turn, by canonical branching, pruning with a covering
bound and completion by table lookup. A cost model schedules them. \textbf{d}, The certificate for the
\qc{288,12,18} circuit (one memory basis, 32 threads): the proved lower bound rises until it meets a fault set of
weight 16 that is replayed in Stim, below the published upper bound of 18.}
\label{fig:overview}
\end{figure}

Finding a logical error gives an upper bound on the distance, but certifying a distance requires proving that no lighter logical error exists.
We obtain such proofs by treating code and circuit distances as a unified problem, shown in Fig.~\ref{fig:overview}a.
We write \qc{n,k,d} for a code of $n$ physical qubits, $k$ logical qubits and distance $d$.
For a CSS code, whose checks are each of purely $X$ or purely $Z$ type, with $X$-check matrix $\vb{H}_X$ and $X$-logical operators $\vb{L}_X$, the distance $d_Z$ is the least weight of a vector $\vb{e}$ with $\vb{H}_X\vb{e}=\vb{0}$ and $\vb{L}_X\vb{e}\neq\vb{0}$, and $d_X$ is defined analogously.
For a circuit, a detector error model lists the fault mechanisms by the detectors and logical observables they flip\cite{stim}.
With $\vb{D}$ and $\vb{O}$ for these two binary matrices, the circuit distance is the least weight of a fault vector with $\vb{D}\vb{e}=\vb{0}$ and $\vb{O}\vb{e}\neq\vb{0}$.
For depolarizing noise, counting merged mechanisms counts faulty circuit locations.
For each logical row $\vb{r}^{\sfT}$ of $\vb{O}$, we add the parity constraint $\vb{r}^{\sfT}\vb{e}=1$ to the zero-syndrome equations.
We then find the minimum weight of a solution to the resulting linear system.
The distance is the minimum of these weights over all logical rows.
We express this minimum as a tropical partition function, the zero-temperature limit of the corresponding free energy in the min-plus algebra\cite{liu_2021_tropical}.
This form was previously used for the distances of concatenated codes\cite{sommers_2025_concatenated}.
We evaluate it with three algorithms, shown in Fig.~\ref{fig:overview}c, whose costs depend on different features of a problem:

\begin{itemize}
    \item The dense rank-based dynamic program (Rank~DP) stores the minimum weight of partial assignments with the same boundary state.
    It is the min-plus counterpart of our previous maximum-likelihood decoding algorithm\cite{cheng_pan_2026_rankdp}.
    For a rank decomposition of the reduced Tanner graph with width $\omega$, each join takes $O(4^\omega)$ updates, as in dynamic programs over branch decompositions\cite{cunningham_2007_branchwidth,kashyap_2009_tree}.
    The linear implementation used here takes $O(2^\omega)$ updates per column, as in trellis decoding\cite{forney_1994,jeong_2017_trellis}.
    Minimizing $\omega$ over general or linear decompositions gives rank-width or linear rank-width, respectively.
    These bounds have no explicit dependence on the distance.
    \item The weight-truncated Rank DP stores only entries within prescribed weight budgets in sparse tables.
    It joins forward and backward tables at each cut, combining BEAST's bidirectional search\cite{bocharova_2004_beast} with min-plus state merging.
    Its cost depends on the number of table entries.
    \item The weight-bounded search explores fault sets within increasing weight budgets to find an undetectable logical error.
    It branches on faults that flip a selected nonzero syndrome bit\cite{dumer_2017_verification}.
    Covering bounds and completion tables\cite{stern_1989} prune branches.
    Its cost depends on the distance, branching factor, and effectiveness of pruning and symmetry reduction.
\end{itemize}

We use circuit reductions, shown in Fig.~\ref{fig:overview}b, to obtain a smaller problem and a lower bound on the distance.
Sector reduction first restricts the detectors to one memory basis.
It preserves the distance when every sector column has a fault representative that fires no detector outside the sector.
We verify this condition for every circuit.
Time collapse then sums each check's detectors over the rounds.
The reductions and column merging take polynomial time in the size of the detector error model.
For the repeated rounds considered here, the collapsed problem is independent of the number of rounds after identical columns are merged.
We compute its distance with the algorithms above to obtain a lower bound on the circuit distance\cite{strikis_2026_leftright,liang_2026_amplified,distm4ri_code}.

We try to match this bound by lifting minimum-weight collapsed solutions to circuit fault sets of the same weight.
A successful lift certifies the distance; ruling out all such lifts raises the lower bound by one.
To close any remaining gap, a cost model uses the predicted work of the two dynamic programs to coordinate them with the search.
The certificate pairs a proved lower bound with a logical error of matching weight, verified against the constraints and, for a circuit, replayed in Stim, as for IBM's two-gross circuit in Fig.~\ref{fig:overview}d.
See Methods for details of the evaluators and reductions.

\subsection*{Benchmarking exact distance computation}

\begin{figure}[tbp]
\centering
\fullwidthfigure{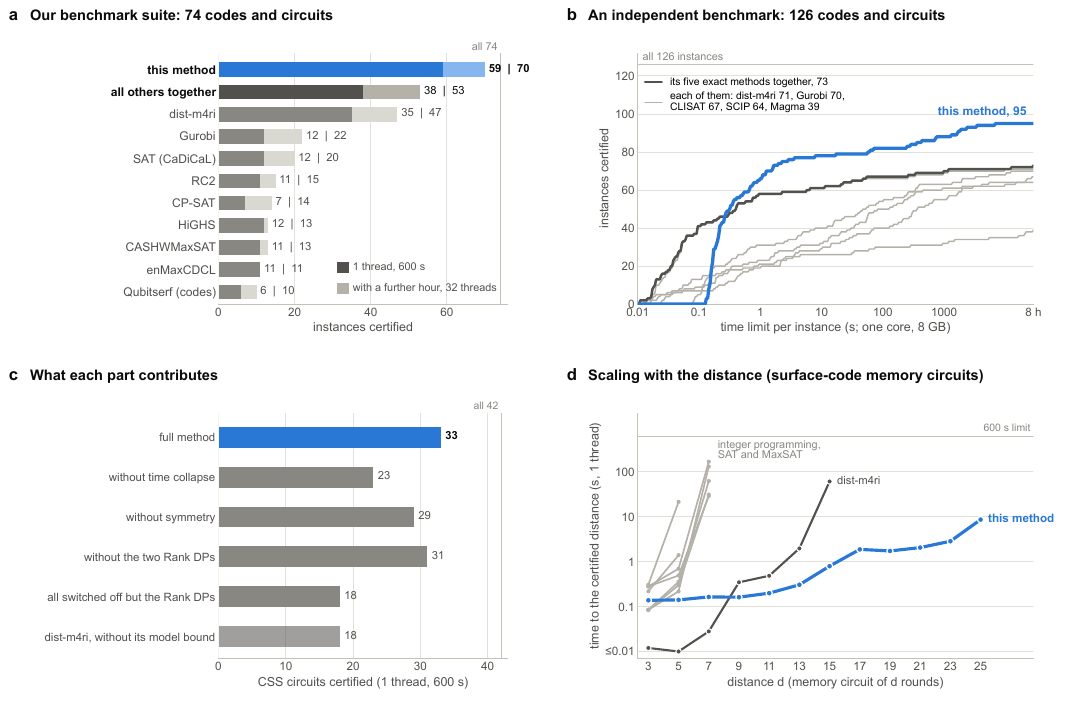}
\caption{\textbf{Two benchmarks of exact distance solvers.} \textbf{a},
Our suite contains 74 instances (47 memory circuits and 27 codes).
Each tool used the same matrices and server, skipping larger instances within a family after two consecutive failures to certify.
Dark bars show instances certified within 600 seconds on one thread; light bars show totals after a further hour with 32 threads when supported.
An instance counts for a tool if any of its configurations certifies it.
\textbf{b}, Independent benchmark of ref.~\citenum{webster_2026_distance} (103 codes and 23 circuits).
The curves show the number of instances certified within each time limit per instance.
Our method used the study's limits (one core, 8~GB, eight hours per instance).
We use the study's published results for five exact methods.
Our timings include process start-up, which dominates below 1 second; the fastest run takes 0.13 seconds.
\textbf{c}, Ablation on the 42 CSS memory circuits in our suite (one memory basis; one thread; 600 seconds).
We compare the full method, variants with components disabled, and dist-m4ri without its model bound.
The variant without time collapse, lifting, round windows, symmetry and the covering bound uses both Rank DPs.
Disabling truncated Rank DP or lifting individually does not reduce the number of certified instances in this tier; Supplementary Fig.~\ref{fig:si-ablation} gives the full ablation.
\textbf{d}, Certification time for Stim's rotated surface-code memory circuits of distance $d$ with $d$ rounds (one memory basis; one thread; 600 seconds).
For dist-m4ri, we show the fastest configuration at each $d$.
For $d\ge17$, we solve the time-collapsed problem using dense Rank DP after giving the search a head start of 1 second.}
\label{fig:benchmark}
\end{figure}

We compared the method with other exact solvers on two complementary benchmarks, summarized in Fig.~\ref{fig:benchmark}.
We assembled the first benchmark to cover current hardware proposals, with 74 instances: 47 memory circuits and 27 codes.
The circuit instances include bivariate bicycle memories with several schedules, circuits for nine recent code proposals, and Stim's surface- and color-code memories.
The code instances include color, concatenated, tile and qLDPC codes.
The codes have up to 2,401 qubits, and the circuits have up to $3.3\times10^5$ fault mechanisms.
On it we ran every exact distance solver we could obtain: connected-cluster enumeration and related searches (dist-m4ri, with and without its model lower bound\cite{distm4ri_code}, and the three methods of Qubitserf\cite{qubitserf_code}), integer programming (HiGHS\cite{huangfu_2018_highs}, CP-SAT\cite{cpsat_code} and Gurobi\cite{gurobi_manual}), SAT\cite{biere_2024_cadical,ignatiev_2018_pysat} and three MaxSAT solvers\cite{ignatiev_2019_rc2,maxsat_2023,enmaxcdcl_code}.
Each tool received the same matrices and ran on the same server.
The first tier allowed 600 seconds per instance on one thread.
The second tier allowed a further hour, with 32 threads for multithreaded solvers and one thread otherwise.
A harness verified each reported error and compared each bound with those of all other runs.
The second is the benchmark of distance-finding methods of Webster, Jacob and Higgott\cite{webster_2026_distance}, selected independently of us.
It has 126 CSS instances, of which 103 are codes (quantum Tanner, hyperbolic surface and color, lifted-product and bivariate bicycle codes) and 23 are memory circuits.
We ran it from its files under its own limits (one core, 8~GB and eight hours per instance) and compared with its published results for five exact methods, obtained on other hardware.

On our suite, the method certifies 59 instances in the first tier and 70 across both tiers.
All other methods combined certify 38 and 53, respectively, as Fig.~\ref{fig:benchmark}a shows.
Across both tiers, our method certifies 45 of the 47 circuit instances, compared with 34 for the other methods combined.
Eighteen instances are certified only by our method, including qLDPC circuits with $2\times10^5$ fault mechanisms such as IBM's \qc{288,12,18} memory.
One instance is certified only by another method: a many-hypercube code with 256 logical qubits.
After failing to certify two consecutive instances in a family, a solver skips larger instances in that family.
Consequently, eight of these 18 instances were not attempted by any other solver.
Ref.~\citenum{beni_2025_tesseract} certified the two smallest bivariate bicycle circuits using MaxSAT on Stim's export of the full circuit.
This approach certifies none of our qLDPC circuits within one hour.
On the independent benchmark, shown in Fig.~\ref{fig:benchmark}b, the method certifies 95 instances, compared with 73 for the study's five exact methods combined.
Our runs with the current dist-m4ri version raise the latter total to 80.
All their certified distances agree with ours.
With 32 threads and a one-hour limit, the method certifies 97 instances, including the superdense color-code circuit whose distance of 11 was previously uncertified.

The advantage comes from the reductions and from having more than one evaluator.
With the reductions, symmetry and covering bound switched off, our method still certifies 18 of the 42 CSS circuits in the first tier, as many as dist-m4ri without its model bound.
Of the 33 that the full method certifies, switching off the time collapse loses 10, symmetry 4 and the two dynamic programs 2, as the ablation in Fig.~\ref{fig:benchmark}c shows.
On Stim's surface-code memory, shown in Fig.~\ref{fig:benchmark}d, the time to the certified distance stays below 10 seconds up to distance 25 on one thread, and within 600 seconds on one thread no other general-purpose exact solver we ran goes beyond distance 15.
These problems are graph-like, so a dedicated shortest-path solver is also exact on them.
See Supplementary Section~\ref{sec:benchmark} for details of the benchmark.

\subsection*{Reassessing published code and circuit distances}

\begin{figure}[tbp]
\centering
\fullwidthfigure{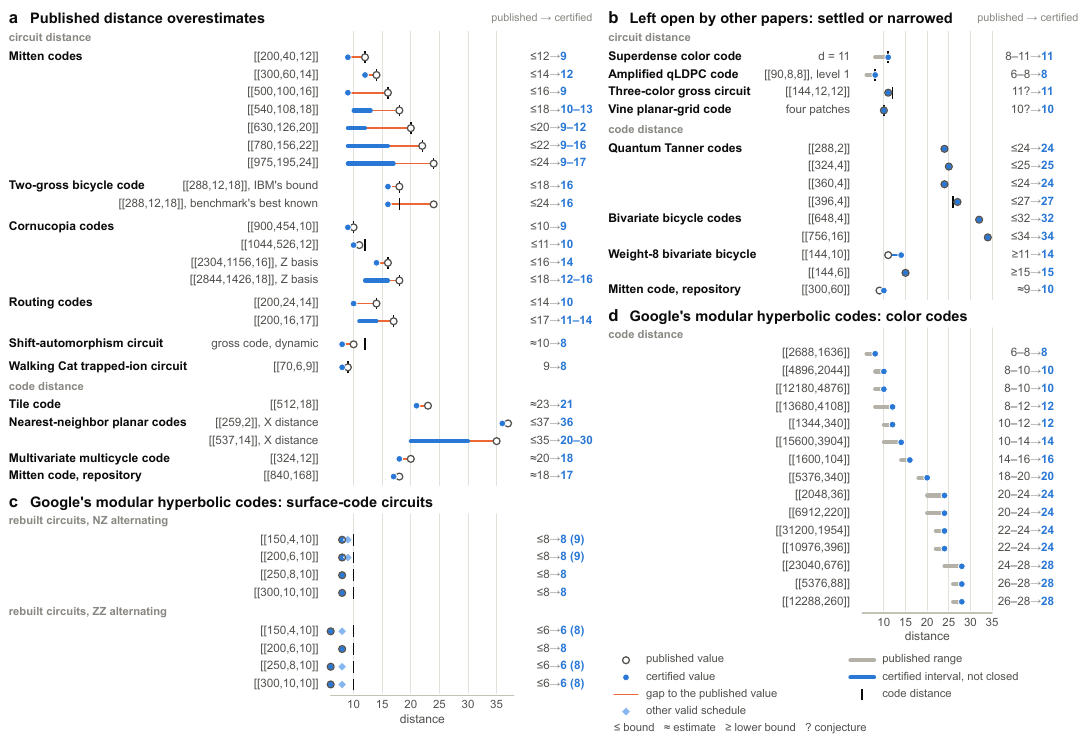}
\caption{\textbf{Published code and circuit distances against certified values.}
\textbf{a}, The 22 overestimated published values.
Published values (open circles) are compared with certified values (blue circles) or intervals from proved lower bounds to the weights of the lightest errors found (blue bars).
Orange lines mark the gaps; ticks mark code distances.
Both values appear on the right ($\le$, upper bound; $\approx$, estimate or expectation).
The repository entry is from the mitten repository.
The \qc{288,12,18} circuit appears with IBM's bound and the best known value from the benchmark of distance methods.
The \qc{2304,1156,16} and \qc{2844,1426,18} values are for the $Z$ basis at 8 cycles.
Walking Cat and shift-automorphism values hold for all numberings of the check terms; values that depend on the numbering and circuits without published distances are not counted and appear only in Supplementary Table~\ref{tab:audit}.
\textbf{b}, Previously unresolved distances, reported as conjectures, ranges (gray bars), lower bounds or best known values, now exact (blue points) or more tightly bounded (blue bars).
Tanner and bivariate bicycle results come from longer runs outside the benchmark, described in Supplementary Section~\ref{subsec:extended}; published values are the best known value from the benchmark of distance methods.
For Tanner codes, published and certified values are $Z$ distances; the tick marks the code distance 26 of \qc{396,4}, its $X$ distance.
\textbf{c}, Reconstructed surface-code circuits of ref.~\citenum{higgott_2026_hyperbolic}, matching the published bounds.
Diamonds and parenthesized values indicate alternative valid schedules.
NZ and ZZ alternating are the two lifted schedules with alternating rounds.
\textbf{d}, Fifteen modular hyperbolic color codes from Google's study\cite{higgott_2026_hyperbolic}, with previously unresolved distances certified at the upper endpoints of the published intervals; Supplementary Fig.~\ref{fig:si-hyperbolic} gives the constructions.
Sources: mitten\cite{bhardwaj_2026_mitten}, two-gross\cite{bravyi_2024_bb}, the benchmark of distance methods\cite{webster_2026_distance} (superdense, Tanner and bivariate bicycle rows), Cornucopia\cite{lu_2026_cornucopia}, routing\cite{zhang_2026_routing},
shift automorphism\cite{kim_2026_shift}, Walking Cat\cite{tripier_2026_walkingcat}, three-color\cite{strikis_2026_leftright},
vine\cite{nixon_2026_vine}, amplified\cite{liang_2026_amplified}, tile\cite{steffan_2025_tile},
nearest-neighbor planar\cite{gu_2026_planar}, multivariate multicycle\cite{mian_2026_multicycle}, weight-8 bivariate bicycle\cite{lu_2026_weight8} and modular
hyperbolic\cite{higgott_2026_hyperbolic}.}
\label{fig:audit}
\end{figure}

We audited code and circuit distances reported in 26 papers from 2023 to 2026, restricting the audit to instances we could reconstruct.
We reconstructed each instance from the authors' circuit or data files when available, and from the paper's description otherwise.
Each reconstruction was checked against values reported as exact.
Figure~\ref{fig:audit}a summarizes the overestimated distance values.
Figure~\ref{fig:audit}b shows previously unresolved distances, and Fig.~\ref{fig:audit}c,d show the results for Google's modular hyperbolic codes.
Supplementary Section~\ref{sec:audit} gives the audit protocol, and Supplementary Table~\ref{tab:audit} documents the 69 audited values and their provenance.
We confirm all code distances reported as exact, together with hundreds of additional values.
In 10 papers, however, at least one published value overestimates the distance.
The affected values comprise 17 circuit distances from 7 papers and 5 code distances from 4 papers.
None of these five code distances was computed exactly in the original papers.
The \qc{288,12,18} circuit is counted twice because two papers report different values.
Of these 22 reported values, one is stated as the distance, 4 are estimates or expectations, and 17 are upper bounds or best known values.
Seven of the 17 values are bounds obtained with an estimator for schedules described as preserving the code distance.
The upper bounds remain valid, but they overestimate the distance if interpreted as exact values.

The largest discrepancies occur for schedules presented as preserving the code distance.
The high-rate mitten schedules were selected using a randomized estimator\cite{bhardwaj_2026_mitten}.
The \qc{200,40,12}, \qc{300,60,14} and \qc{500,100,16} circuits have actual distances 9, 12 and 9, respectively.
IBM's \qc{288,12,18} circuit\cite{bravyi_2024_bb} has distance 16 in both the $X$- and $Z$-basis memory experiments.
This holds for the public Stim files\cite{tesseract_code}, circuits generated by Bravyi et al.'s code, and the architecture's memory instruction discussed below.
Bravyi et al.\ report an upper bound of 18 and note that it is unlikely to be tight.
The architecture study assumes a distance of 18\cite{yoder_2025_tour}, while the benchmark of distance methods lists 24 as its best known value\cite{webster_2026_distance}.
The Cornucopia circuits\cite{lu_2026_cornucopia} at $n=900$ and $n=1{,}044$ have distances one below their published upper bounds.
At eight cycles in the $Z$ basis, the \qc{2304,1156,16} circuit has distance 14, while the largest Cornucopia circuit, \qc{2844,1426,18}, has a distance interval of 12 to 16.
For the largest circuit, the published extrapolation gives a logical error rate of $2.6\times10^{-16}$ per qubit and cycle\cite{lu_2026_cornucopia}.
This estimate is used in a modular architecture\cite{lu_2026_modular}.
For memories with eight or more cycles, our distance bound implies a leading order of at most $p^8$, rather than the $p^9$ assumed in the extrapolation.
The \qc{200,24,14} routing code with weight-11 checks\cite{zhang_2026_routing} has circuit distance 10, four below its published upper bound.
Smaller discrepancies occur for a time-dynamic schedule of the gross code\cite{kim_2026_shift} and a trapped-ion circuit\cite{tripier_2026_walkingcat}, under every numbering of their check terms.
We exclude three further values from these papers because they depend on unspecified numberings of the check terms.

The distance losses occur in circuits with checks acting on many qubits and newly proposed schedules.
Their minimum fault sets combine hook errors from several checks.
The structured \qc{330,66,12} code in the mitten repository illustrates this mechanism.
For its supplied schedule, we certify a previously unreported circuit distance of 7.
Supplementary Fig.~\ref{fig:si-hook} shows a minimum fault set: seven ancilla faults, each after the second or third of its check's nine gates.
Together, these faults produce an undetected data error of weight 37.
Randomized searches use one decoding trial or sampled information set at a time and rarely find these combinations.
Without a lower bound, they cannot certify that no lighter fault set exists.
These audit conclusions hold under every fault model used in the source papers.
We first validated the reconstructed circuits by reproducing the papers' exact values.

The audit also determines previously unresolved distances, shown in Fig.~\ref{fig:audit}b,d.
We confirm two conjectured circuit distances: 11 for a three-color schedule of the gross code\cite{strikis_2026_leftright} and 10 for four planar vine patches\cite{nixon_2026_vine}.
We also certify distance 11 for the superdense color-code circuit.
For modular hyperbolic codes with up to 31,200 qubits, we certify 15 previously unresolved distances, each equal to the upper bound of its published interval\cite{higgott_2026_hyperbolic}.
Longer runs outside the benchmark, described in Supplementary Section~\ref{subsec:extended}, certify quantum Tanner code distances of 24 for \qc{288,2} and \qc{360,4}, 25 for \qc{324,4}, and 26 for \qc{396,4}.
These runs also certify distances of 32 for the bivariate bicycle code \qc{648,4} and 34 for \qc{756,16}.
The benchmark's value of 27 for \qc{396,4} is its $Z$ distance.
Its $X$ distance is 26, so its code distance is 26.
The largest certified code instance is the modular hyperbolic code \qc{31200,1954}, with distance 24.
The largest certified circuit instance is the Cornucopia \qc{2304,1156,16} memory, with 3,456 qubits and $8\times10^5$ fault mechanisms.
Its circuit distance is 14 in the $Z$ basis over eight cycles.
A revised distance changes the leading order of the logical error rate whenever it changes $\lceil d/2\rceil$.
The effect at a fixed noise level also depends on the number of low-weight fault sets, as illustrated below for IBM's codes.

\subsection*{Certified design of syndrome-extraction circuits}

\begin{figure}[tbp]
\centering
\fullwidthfigure{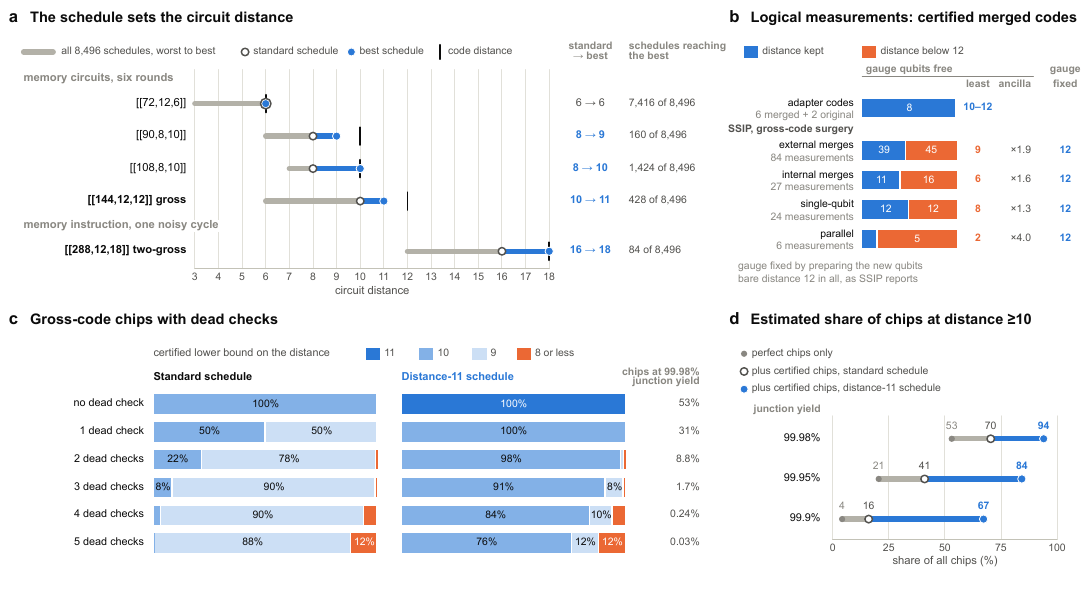}
\caption{\textbf{The schedule sets the circuit distance of IBM's bicycle codes, and certificates extend to logical measurements and to every chip.} \textbf{a}, Certified circuit distances over all 8,496 uniform depth-seven one-ancilla schedules of five bivariate bicycle codes\cite{bravyi_2024_bb}.
Bars show the distance range, with values above the standard schedule in blue.
Open circles, blue circles and ticks mark the standard schedule, best schedule and code distance, respectively.
The right columns give the standard and best distances and the number of optimal schedules.
For the four smaller codes, distances are the minimum over both bases of six-round memory circuits.
For two-gross, we use the memory instruction of ref.~\citenum{yoder_2025_tour} with one noisy cycle; its distances are upper bounds for more cycles.
The minimum, maximum and number of optimal schedules are exact; distances of schedules with a hook family are only bounded above by 16.
\textbf{b}, Merged codes for logical measurements: adapter codes from ref.~\citenum{swaroop_2024_adapters}, whose published distances are certified, and gross-code merges produced by SSIP\cite{cowtan_2024_ssip} at the published merge depths.
For SSIP, bars count measurements with dressed distance 12 (blue) or below 12 (orange), with the gauge qubits introduced by merging left unfixed, as in SSIP's definition.
Adjacent columns give the minimum certified distance (the range for adapter codes) and the ratio of ancilla counts at depths preserving distance 12 to the published counts.
Preparing the new qubits fixes the eigenvalues of the $Z$-type gauge operators for $X$ measurements, and vice versa.
Every SSIP merged code then has distance 12 (right column).
The bare distances evaluated by SSIP's routine are also 12.
\textbf{c}, Gross-code chips with dead check qubits or couplers: the share of chips at each certified lower bound on the circuit distance, three noisy cycles, for the standard
schedule and for a distance-11 schedule of eight timesteps; on the right, how common each number of dead checks is at a junction yield of 99.98\%. Chips with 1 or 2 dead checks are all cases, up to the code's translations; chips with 3 or more are 200 random chips, the same on both schedules.
A chip's bound is the smaller of two exact problems with dead checks of one type, each with its fault set replayed in Stim. \textbf{d}, Estimated share of all chips certified at distance 10 or more at three junction yields: perfect chips only (gray circle), and with the certified chips added, on the standard schedule (open circle) and on the distance-11 schedule (blue). Shares combine the fabrication statistics with the sampled chips, interpolated at 7 and 9 dead checks; the sampling error is about one percentage point at most.}
\label{fig:design}
\end{figure}

Schedules are usually chosen among a few candidates by estimated distances.
With exact distances one can evaluate every candidate in a design space.
For the weight-six bivariate bicycle codes\cite{bravyi_2024_bb}, we enumerated all uniform depth-seven syndrome-extraction schedules with one ancilla per check, where uniform means that all checks of a type use the same gate order.
There are 8,496 such schedules satisfying two constraints: $X$- and $Z$-check gates never act on the same block in a layer, and each overlapping pair of checks is crossed consistently.
Of these, 936 belong to the same family as the standard schedule.
We computed the exact circuit distance of each of the 8,496 schedules for the \qc{72,12,6}, \qc{90,8,10}, \qc{108,8,10} and \qc{144,12,12} codes.
This gives 33,984 certified circuit distances, shown in Fig.~\ref{fig:design}a; the two-gross code is considered below.
The standard schedule is optimal for the \qc{72,12,6} code only.
For \qc{90,8,10}, the best schedules lie outside the standard family and reach 9 instead of 8.
For \qc{108,8,10}, they reach the full code distance of 10 instead of 8.
For the gross code \qc{144,12,12}, 428 schedules reach 11 instead of 10.
Because all schedules were computed, these optima are certified for this family.
No uniform depth-seven one-ancilla schedule of the gross code reaches 12, which complements a bound of 10 for uniformly tiled non-interleaved circuits\cite{strikis_2026_leftright}.

We also designed syndrome-extraction circuits for codes whose original publications did not include circuit designs; Supplementary Fig.~\ref{fig:si-design}a,b shows the results.
For two weight-eight bivariate bicycle codes\cite{lu_2026_weight8}, 12 and 482 of the 3,456 mirror-symmetric depth-eight gate orders reach the full distances 8 and 10 in the $Z$ basis.
We designed parallel depth-eight circuits for 32 self-dual bivariate bicycle codes\cite{liang_2025_selfdual} with $n\le132$.
These circuits reach the code distance for 28 of the 32 codes.
The best sequential circuits reach the code distance for 10 of these codes, as certified by an exhaustive search over all 5,040 gate orders.
For the GALA code \qc{132,30,12}\cite{yang_2026_gala}, a parallel 12-layer circuit has distance 10.
The published layer order gives distance 8 when implemented as sequential $X$- and $Z$-check rounds.

\subsection*{The error-correction layer of IBM's planned Starling processor}

IBM's bicycle architecture, the basis of its planned Starling processor\cite{ibm_roadmap}, stores logical
qubits in modules of the gross code \qc{144,12,12}, with the two-gross code \qc{288,12,18} as the option for
larger computations\cite{yoder_2025_tour}. Its resource estimates add up a logical error rate for each instruction.
The IBM team simulated each instruction at higher noise, fitted the fraction of fault configurations of each size
that fail, and used the fit to extrapolate the rates to $p=10^{-3}$ and $10^{-4}$, taking the circuit distance to be
10 for the gross code and 18 for the two-gross code. They note that the two-gross code's circuit distance might be
lower, and they ask how accurate the extrapolation is\cite{yoder_2025_tour}.

We rebuilt the architecture's memory instruction using the paper's setting and circuit noise.
The instruction starts from an error-free code state, runs noisy syndrome cycles of eight timesteps each with the standard schedule\cite{bravyi_2024_bb}, and ends with a noise-free final round. With one noisy cycle, the certified circuit
distance is 10 in both bases for the gross code, as assumed, and 16 in both bases for the two-gross code. A set of
16 faults within one cycle can therefore flip logical qubits of the two-gross code without firing any detector.
An IBM study of rare failures\cite{beverland_2025_failfast} likewise assumes distance 18 for the two-gross memory circuit of ref.~\citenum{bravyi_2024_bb}.
Under this assumption, it counts minimum-weight failures involving nine faults using logical operators of weight 18.
We matched our reconstructed circuits to those generated by the IBM team's public code.
We certify distance 16 for the two-gross circuit, so some configurations of eight faults already cause logical failure.

Other schedules reach the code distance of 18 at one noisy cycle, with the same couplers, as Fig.~\ref{fig:design}a shows.
For 7,056 of the 8,496 schedules a family of hook errors lies within one round, which limits them to 16.
Their one-round problems fall into six kinds, each holding a fault set of weight 16 or less that we replayed in every corresponding circuit.
Among the remaining 1,440 schedules, our search certifies that 600 have circuit distance 15, 144 have distance 16, 612 have distance 17 and 84 have distance 18.
The worst schedules have distance 12.
All 1,872 schedules that fit the paper's eight timesteps per cycle have such a family, so none exceeds the standard schedule's 16, and every schedule above 16 takes nine timesteps.
The symmetries of the code reduce the 16,992 problems of all schedules in two memory bases to 708.

A value at one cycle is an upper bound for more cycles.
For the four smaller codes, two cycles reproduce the six-round circuit distances in all 67,968 problems.
One cycle reproduces these distances in all but 616 cases.
For each of these four codes, schedules attaining the maximum circuit distance at one cycle have the same circuit distance at six rounds, provided that no family of hook errors lies within a single round.
For one of the best two-gross schedules, we excluded all undetectable logical fault sets of weight at most 17 with two cycles and at most 16 with three cycles.
Undetectable logical fault sets of weight 18 give an upper bound of 18 in both cases.
Thus, the circuit distance is 18 with two cycles and either 17 or 18 with three cycles.
In both cases, a minimum-weight decoder can correct any configuration of at most eight faults.

Because minimum fault sets are explicit, the same enumeration also predicts logical error rates at noise levels that sampling cannot reach.
The two halves of a minimum fault set fire the same detectors, so every decoder fails on one of them: with the standard schedule, every decoder fails on some configurations of eight faults, whereas the extrapolation model of ref.~\citenum{yoder_2025_tour} sets the failing fraction to zero for all fault counts below nine.
Supplementary Section~\ref{subsec:ibm-bound} turns such fault sets into lower bounds on the logical error rate that hold for every decoder and compares them with the rates that the paper extrapolates.

For the gross code, all 428 schedules of distance 11 keep that distance in the instruction, with the same couplers, and 100 of them fit eight timesteps per cycle in the paper's timing.
The gain in logical error rate depends on all low-weight failures, not only the minimum ones.
Undetectable fault sets of up to 12 faults in the paper's setting of 10 cycles give an estimate of the low-noise expansion of a minimum-weight decoder's logical error rate to order $p^6$.
A distance-11 schedule removes the failures of five faults, but halves of fault sets of weight 12 fail at the next order with every schedule.
At $p=10^{-4}$ an eight-timestep schedule selected by its minimum fault sets therefore lowers the estimated rate only three- to sixfold, and more at lower noise, as Supplementary Fig.~\ref{fig:si-design}e shows.
See Methods and Supplementary Section~\ref{sec:ibm} for details of these computations.

\subsection*{Certifying logical measurements}

Computing with qLDPC codes needs logical measurements, which join code blocks to an ancilla system for several rounds.
The merged code must keep the distance, which published constructions check with integer programming or randomized searches.
The adapter codes joining a bivariate bicycle code and a lifted-product code\cite{swaroop_2024_adapters} have 115 to 355 qubits.
We certify their published distances of 10 and 12, with each computation taking at most 0.2 seconds.
The surgery tool SSIP\cite{cowtan_2024_ssip} merges gross-code blocks, and its benchmark scripts raise the merge depth until the randomized search QDistRnd reports distance 12.
SSIP treats additional logical qubits introduced by a merge as gauge qubits.
The SSIP paper\cite{cowtan_2024_ssip} uses the dressed distance: the minimum physical weight of a logical Pauli operator acting nontrivially on the protected logical qubits.
This definition allows arbitrary action on the gauge qubits.
SSIP's distance routine instead estimates the bare distance, obtained by additionally requiring these operators to act as the identity on the gauge subsystem.
We reran SSIP's own benchmarks with exact distances, as described in Supplementary Section~\ref{sec:surgery}.
The bare distances are 12 in all 141 measurements, so the searches were tight for what they computed.
Whether 12 is the distance depends on the gauge qubits.
Preparing the new qubits fixes the eigenvalues of the $Z$-type gauge operators for $X$ measurements, and vice versa.
Ancilla readout and the next round of checks verify these eigenvalues.
With the gauge fixed, every merged code has distance 12, since dressed distances below 12 occur only in the measured Pauli sector.
With the gauge qubits left unfixed, as in SSIP's definition, the dressed distance is below 12 in 78 measurements.
It can be as low as 2 when several logical qubits are measured simultaneously, as Fig.~\ref{fig:design}b shows.
Achieving dressed distance 12 requires deeper merges, increasing the total ancilla count from 5,664 to 12,834.
Inside such a tool, an exact solver tells which distance a construction has under which protocol and can choose the smallest ancilla system that keeps it.

\subsection*{Certifying chips with dead checks}

Fabricated chips have defects, and each defective chip is a new circuit.
The architecture study estimates that about two in five gross-code modules, with their logical processing units, come out perfect at a junction yield of 99.98\%, and it notes without numbers that encoding around defects typically lowers the distance\cite{yoder_2025_tour}.
We certified such chips one at a time.
A dead check qubit or a dead coupler removes its check.
The check's gates, preparation, measurement and detectors drop out in every round, its data qubits idle in those slots, and the logical operators stay those of the perfect code.
A chip with a dead data qubit counts as rejected.
We ran the instruction in the paper's timing with three noisy cycles, because the loss appears only from two cycles on.
Ten cycles give the same value on all 78 problems tested, with the standard schedule.
A run took seconds to minutes on one thread.

With the standard schedule, one dead $Z$ check lowers the distance from 10 to 9 and one dead $X$ check leaves it at 10, as Fig.~\ref{fig:design}c shows.
The distance-11 schedule of eight timesteps keeps distance 10 under either dead check, and 76\% of the 200 sampled chips with five dead checks are certified at 10.
Most schedules of distance 11 behave this way, with 400 of the 428 keeping 10 under either kind of dead check.
Most chips that fall below 9 have also lost a stabilizer, so a rank test finds them quickly.
Among 1,228 sampled problems with four or five dead checks of one type, however, five keep every stabilizer and still fall to 8, and only the certificate finds those.

Keeping defective chips raises the usable yield under this model, as summarized in Fig.~\ref{fig:design}d.
For the memory block alone, with the upper ends of IBM's junction counts, two per qubit and three per coupler, 53\% of chips are perfect at 99.98\% junction yield.
Keeping chips certified at distance 10 or more raises the estimated share to 70\% with the standard schedule and to 94\% with the distance-11 schedule.
At 99.9\% junction yield the three shares are 4\%, 16\% and 67\%.
A dead check has a modest rate penalty in the cases studied.
To order $p^5$ the logical error rate of a minimum-weight decoder at $p=10^{-4}$ on the standard schedule is 1.09 times the perfect chip's for one dead $Z$ check and 1.03 times for one dead $X$ check.
On the distance-11 schedule the same term, which the perfect chip does not have, is 1.6\% and 0.4\% of the perfect standard chip's.
For the tested chips with up to five dead checks, BP-OSD simulations at $p=3\times10^{-3}$ find logical error rates at most 1.6 times the perfect chip's on either schedule.
See Supplementary Section~\ref{sec:chips} for details of the chip study.

\FloatBarrier %
\section*{Discussion}

Certified exact distances turn the distance of a code or circuit from an estimate into a verified quantity, at the scale of the circuits now proposed for hardware.
Because the distance sets the leading order of the logical error rate, this removes a source of error from code comparisons, resource estimates and the choice between architectures.
For example, IBM's two-gross memory instruction was assumed to have distance 18 but is certified at 16.
Moreover, a certificate states for which protocol a distance holds: SSIP's merged codes keep distance 12 when their gauge qubits are fixed, but not as the subsystem codes its paper defines.
The exact value may still be out of reach; in that case, the time-collapsed relaxation gives a cheap lower bound, so that an interval, rather than an upper bound alone, describes what is known.

Reaching this scale took more than one algorithm, because no single exact algorithm is best on all problems.
Their costs depend on different features, and each settles instances that the others cannot.
The dense dynamic program's cost depends on the cut rank, and it computes distances in seconds to minutes for color codes of distance 19 to 31 and in seconds for the concatenated Steane code \qc{2401,1,81}.
Only the truncated program, whose cost depends on the number of light partial solutions, settles the superdense circuit of distance 11.
The search, whose cost depends on the distance and the symmetry, settles qLDPC circuits such as IBM's \qc{288,12,18} memory.
Because all three evaluate the same tropical partition function, they can be scheduled by predicting the costs of the dynamic programs.
All three build on coding theory, from trellis decoding to sequential and cluster search\cite{forney_1994,bocharova_2004_beast,dumer_2017_verification}.
The new step is to specialize them to detector error models and combine them, which decides most distances of current proposals.

However, problems that are both wide and of large distance remain expensive.
For IBM's \qc{756,16,34} circuit, for example, we certify only a lower bound of 15, and many other instances, including the Cornucopia \qc{2304,1156,16} circuit and a many-hypercube code with 256 logical qubits\cite{goto_2024_hypercube}, still require substantial computing resources.
Three directions would lower these costs.
First, the search splits into independent subtrees, which can be distributed over many machines, as for the Cornucopia circuit, and the table updates of the dense dynamic program are data-parallel and suited to GPUs.
Second, tree-shaped rank decompositions are never wider than the linear column orders used here, and their dynamic program, which we develop in the Supplementary Information, could reach wider codes at a higher cost per join.
Third, the search could adapt its bounding rule to each problem: its covering bound uses one simple feasible dual solution, whereas weights optimized for the problem, or bounds learned from pruned branches, would prune more.
On the many-hypercube code, SAT, which learns constraints from failed branches, certifies the distance in about 20 minutes, against 5.3 hours for our search.

Even at the present cost, circuit searches, exhaustive or driven by machine learning\cite{cruzbenito_2026_llm}, can now use certified distances as their objective, and the same search enumerates the fault sets a few faults above the distance that also set low-noise performance.
The logical-measurement instructions of IBM's architecture, whose circuit distances its study could only bound\cite{yoder_2025_tour}, are a natural next target.
Beyond quantum codes, the minimum distance and the light codewords set the low-noise error rate of classical LDPC codes, including the quasi-cyclic codes of communication standards, whose cyclic symmetry our search already exploits.
The same problem underlies code-based cryptography: the security of the McEliece cryptosystem\cite{mceliece_1978} and of HQC, the quasi-cyclic scheme selected by NIST\cite{nist_ir_8545}, rests on the hardness of finding light vectors with a given syndrome, and the strongest attacks are information-set decoders, from which our completion tables descend\cite{stern_1989}.
Although cryptographic sizes are out of reach by design, certified distances of smaller structured instances could test the assumptions behind these security estimates.

\section*{Methods}

\subsection*{Distance problems and certificates}

For binary matrices $\vb{M}\in\FF_2^{m\times \ell}$ and $\vb{R}\in\FF_2^{q\times \ell}$ let
\begin{equation}
    \Delta(\vb{M},\vb{R})=\min\{|\vb{e}| : \vb{e}\in\FF_2^\ell,\ \vb{M}\vb{e}=\vb{0},\
    \vb{R}\vb{e}\neq\vb{0}\}.
    \label{eq:delta}
\end{equation}
Consider a CSS code with check matrices $\vb{H}_X,\vb{H}_Z$ and logical operators $\vb{L}_X,\vb{L}_Z$, chosen so that $\ker\vb{H}_X\cap\ker\vb{L}_X$ is the row space of $\vb{H}_Z$ and $\ker\vb{H}_Z\cap\ker\vb{L}_Z$ that of $\vb{H}_X$.
The distances are $d_Z=\Delta(\vb{H}_X,\vb{L}_X)$ and $d_X=\Delta(\vb{H}_Z,\vb{L}_Z)$, and the code distance is the smaller of the two.
For a circuit, the detector error model gives $\vb{D}$
(detectors by fault mechanisms) and $\vb{O}$ (logical observables by fault mechanisms), obtained by
propagating each elementary fault through the Clifford circuit\cite{stim}, and the circuit distance is
$d_{\mathrm{circ}}=\Delta(\vb{D},\vb{O})$. When the allowed faults at each location form a linear space,
as for depolarizing channels, Pauli flips and measurement flips, and every nonzero fault is listed as a mechanism
with equal mechanisms merged, the minimum number of mechanisms equals the minimum number of faulty locations,
as shown in Supplementary Section~\ref{sec:distance}. Columns with no effect are removed and equal columns merged before any evaluation,
which preserves $\Delta$.

To unify the computational problems, we consider the tropical partition function over the min-plus semiring
\begin{equation}
    Z_{\vb{M}}(\vb{t})=\min\{|\vb{e}| : \vb{M}\vb{e}=\vb{t}\}
    =\lim_{\beta\to\infty}\Bigl(-\tfrac{1}{\beta}\log\sum_{\vb{M}\vb{e}=\vb{t}}e^{-\beta|\vb{e}|}\Bigr),
    \label{eq:Z}
\end{equation}
which is the zero-temperature limit of the corresponding free energy\cite{liu_2021_tropical}. 
Since $\vb{R} \vb{e} \neq \vb{0}$ requires at least one logical observable to flip, we impose this condition one logical row at a time.
With $\vb{M}^{(a)}$ the matrix
$\vb{M}$ extended by row $a$ of $\vb{R}$ and $\vb{t}^{(a)}=(\vb{0},1)$, we have
\begin{equation}
\Delta(\vb{M},\vb{R})=\min_{1\le a \le q}Z_{\vb{M}^{(a)}}(\vb{t}^{(a)}).
\label{eq:reduction}
\end{equation}
The evaluators below compute $Z$ or decide whether $Z\le K$. A certificate for $\Delta=d$ consists of a proof
that no vector of weight below $d$ satisfies the constraints, produced by an evaluator whose completeness is
proved in Supplementary Sections~\ref{sec:tropical-rank-dp} to~\ref{sec:weight-search}, and an explicit $\vb{e}$ of weight $d$ that satisfies them.
Every reported error is checked against the matrices and, for a circuit, inserted into the noiseless Stim circuit as deterministic Pauli errors: no detector fires and an observable flips.
The kernels are tested against brute force on small instances, and in the benchmark a harness checks every record against the ground truth and the bounds of every other tool; Supplementary Section~\ref{subsec:software-verification} describes these tests.

\subsection*{Dense Rank DP}

Fix an order of the columns of $\vb{M}$. 
At the cut after column $j$, define the boundary space $I_j=\im\vb{M}_{\le j} \cap \im\vb{M}_{>j}$~\cite{jeong_2017_trellis}.
Its dimension is
\begin{equation}
    w_j=\rank\vb{M}_{\le j}+\rank\vb{M}_{>j}-\rank\vb{M},
\label{eq:cutrank}
\end{equation}
which equals the cut rank of the corresponding cut in a fundamental graph of $\vb{M}$\cite{oum_2005_rankwidth,jeong_2017_trellis}. 
The largest cut rank over $j$ is the width of this linear order.
Assume that $\vb{M}\vb{e}=\vb{t}$ is consistent; otherwise, $Z_{\vb{M}}(\vb{t})=+\infty$.
A partial assignment $\vb{e}_{\le j}$ to the first $j$ coordinates has state $\vb{s}=\vb{M}_{\le j}\vb{e}_{\le j}$.
The extendible states form $\Sigma_j:=\im\vb{M}_{\le j}\cap(\vb{t}+\im\vb{M}_{>j})$, an affine translate of $I_j$ with $2^{w_j}$ elements.
We maintain a DP table $T_j$, whose entry $T_j(\vb{s})$ stores the minimum weight among the partial assignments with state $\vb{s}$.
We initialize $T_0(\vb{0})=0$ and assign value $+\infty$ to entries outside each state space.
Let $\vb{m}_j$ be column $j$ of $\vb{M}$.
The choices $e_j=0$ and $e_j=1$ give the min-plus update
\begin{equation}
    T_j(\vb{s})=\min\bigl\{T_{j-1}(\vb{s}),\
    T_{j-1}(\vb{s}+\vb{m}_j)+1\bigr\},
    \qquad\vb{s}\in\Sigma_j,
    \label{eq:dp}
\end{equation}
where syndrome addition is over $\FF_2$.
Partial assignments with the same state admit the same completions, so storing the minimum weight for each state preserves the optimum.
The final table entry $T_\ell(\vb{t})$ therefore equals $Z_{\vb{M}}(\vb{t})$, and traceback recovers a minimizing assignment $\vb{e}$ as a witness.
The table $T_j$ has at most $2^{w_j}$ entries and the DP sweep evaluates $Z_{\vb{M}}(\vb{t})$ using $O(\sum_{j=0}^{\ell}2^{w_j})$ state updates.
This bound has no explicit dependence on the distance.
The DP described here uses a linear order and corresponds to the minimal syndrome trellis of refs.~\citenum{forney_1994,ollivier_2006_trellis}.
Supplementary Section~\ref{subsec:dense-rank-dp} gives the state invariant, correctness proof and complexity of this linear sweep.
We also consider general branch decompositions, as in ref.~\citenum{cheng_pan_2026_rankdp}, and combine child tables by min-plus convolution; Supplementary Section~\ref{subsec:rank-dp} develops this extension.
For a detector error model ordered by time in which every fault fires detectors of a bounded number of consecutive rounds, as when detectors compare successive measurements of a check, the width of the linear order is bounded independently of the number of rounds.
We choose the order with the smallest width among the given order, its reverse and, for problems of up to 2,000 columns other than circuit sectors, greedy orders, and use the corresponding table sizes to determine the number of state updates before running the DP.

\subsection*{Weight-truncated Rank DP}

We combine the bidirectional weight-bounded search principle of BEAST~\cite{bocharova_2004_beast} with min-plus state merging on the syndrome trellis.
The weight-truncated method uses the same states and update~\eqref{eq:dp} as dense Rank DP, but stores only entries within prescribed weight budgets in sparse tables.
For a fixed threshold $K$, it decides whether $Z_{\vb{M}}(\vb{t})\le K$.
We split $K$ into integer budgets $a$ and $K-a$ for the forward and backward sweeps.
At cut $j$, $\widehat{T}_j(\vb{s})$ is the least stored prefix weight at state $\vb{s}$, and $\widehat{B}_j(\vb{s})$ is the least stored weight of a suffix compatible with that state; absent entries have value $+\infty$.
At every cut, we join matching states by testing
\begin{equation}
    \min_{\vb{s}}\bigl[\widehat{T}_j(\vb{s})+\widehat B_j(\vb{s})\bigr]\le K.
    \label{eq:join}
\end{equation}
Each accepted pair gives a complete solution, and every solution of weight at most $K$ fits both budgets at some cut.
Thus, a completed run either finds a solution of weight at most $K$ or certifies that $Z_{\vb{M}}(\vb{t})>K$, and when $Z_{\vb{M}}(\vb{t})\le K$ the least accepted sum equals $Z_{\vb{M}}(\vb{t})$.
To obtain the distance, we test the logical-row queries at successive integer thresholds $K$, starting from a certified lower bound.
If all queries fail, the distance exceeds $K$; the first successful threshold gives the distance, and traceback recovers a minimum-weight witness.
The implementation also supports pruning with admissible lower bounds on the weight still needed, which the reported runs leave off.
We also vary the split with the cut according to estimated table sizes and avoid redundant joins.
These refinements preserve the decision guarantee; correctness arguments and implementation details are given in Supplementary Section~\ref{sec:sparse-dp}.

\subsection*{Weight-bounded search}

The search decides, for $K=L,L+1,\ldots$ from a certified lower bound $L$, whether a solution of weight at most $K$ exists. 
A node holds a selected set $A$ of columns and a set $U$ of available ones. 
The search starts from one-column sets (anchors), since the empty set has zero syndrome and would be closed at once. 
If the residual syndrome $\mathbf{s}(A)=\mathbf{M}\mathbf{1}_A$ is nonzero, the search chooses a row $r$ with $s_r(A)=1$.
Let $j_1<\cdots<j_t$ list the columns $j\in U$ satisfying $M_{rj}=1$.
Child $p$ replaces $A$ by $A\cup\{j_p\}$ and $U$ by $U\setminus\{j_1,\ldots,j_p\}$.
Every completion contains at least one column in this list.
It belongs to exactly one child, indexed by the first listed column that it contains.
This assignment avoids duplicate enumeration, as in connected-cluster enumeration\cite{dumer_2017_verification}.
Search decoders also add faults one at a time\cite{beni_2025_tesseract,ott_2025_decisiontree}.
A node with zero residual syndrome is a solution if $\mathbf{R}\mathbf{1}_A\neq0$ and is closed otherwise. 
Every nonempty proper subset of a minimum-weight solution has nonzero syndrome.
Thus, stopping at zero syndrome preserves the completeness of the search.
If the search at threshold $K$ finishes without finding a solution, then the distance exceeds $K$, as proved in Supplementary Section~\ref{sec:weight-search}.
Symmetries of the problem (lattice translations of bivariate
bicycle codes, cyclic shifts of quasi-cyclic codes and automorphisms of the time-collapsed problem) restrict the
first column to one representative per orbit. 
A node with remaining budget $b=K-|A|$ is pruned if $h(A,U)>b$.
The covering bound is
\begin{equation}
    h(A,U)=\sum_{r:\,s_r(A)=1}y_r,\qquad y_r\ge0,\quad \sum_{r:\,s_r(A)=1,\,M_{rj}=1}y_r\le1\ \ \text{for all } j\in U.
    \label{eq:cover}
\end{equation}
Here, the weights $(y_r)$ form a feasible solution of the dual LP for covering the unsatisfied rows with available columns.
Their objective value $h(A,U)$ is a lower bound on the number of additional faults needed for a completion.
Thus, $h(A,U)$ is an admissible heuristic in the sense of IDA*\cite{korf_1985}.
The last levels are completed by a lookup in a table of the
syndromes of all column sets up to a fixed weight, which keeps the lightest record for each syndrome, or for each syndrome and logical label; this is a meet in the
middle\cite{stern_1989} and a pattern database\cite{culberson_1998_patterns}. The table only filters: every
positive lookup is verified by reconstructing a completion from the columns outside the selected set.

\subsection*{Reductions, relaxation and lifting}

For any binary matrix $\mathbf{T}$, every solution of $\mathbf{M}\mathbf{e}=\vb{0}$, $\mathbf{R}\mathbf{e}\neq\vb{0}$
satisfies $\mathbf{T}\mathbf{M}\mathbf{e}=\vb{0}$, so $\Delta(\mathbf{T}\mathbf{M},\mathbf{R})\le\Delta(\mathbf{M},\mathbf{R})$.
The sector problem of a CSS memory circuit keeps the detectors of the measured basis, and the time-collapsed
problem adds the detectors of each check over the rounds; both are row projections of this kind. The sector
problem is exact when each of its columns is the image of a full column that fires no detector outside the sector, so that
every sector solution is itself a circuit fault set of the same weight; we check this condition for every
circuit. After equal columns are merged, the collapsed problem has one row per check and a number of columns that
does not grow with the number of rounds, and for repeated identical rounds that carry data errors through unchanged
its distance is a lower bound for any number of rounds, as proved in Supplementary Section~\ref{subsec:search-reductions}. For
single-ancilla CSS circuits it is the extended-code bound of Strikis, Browne and
Beverland\cite{strikis_2026_leftright}; related bounds appear in ref.~\citenum{liang_2026_amplified}, and dist-m4ri
computes a code-sized model of the same kind\cite{distm4ri_code}.

A minimum-weight collapsed solution is lifted by searching, over the rounds, for circuit fault sets of the same
weight whose collapse it is. If one exists, the collapsed lower bound is matched and the distance is exact. If no
collapsed solution of weight at most $W$ lifts to a circuit fault set of weight at most $W$, the circuit distance
exceeds $W$, because every such fault set collapses to one of them; the collapsed solutions are listed by the
recording variant of the search, up to symmetry, for $W$ up to three above a collapsed distance of at least four. Lifting is for a fixed number of rounds.
Restricting the columns to a window of rounds gives upper bounds in the same way, since every solution of the
restricted problem is a circuit fault set.

\subsection*{Choosing an evaluator}

For each problem of up to 30,000 columns, the dense Rank DP's peak table size and work are computed from its plan
before anything is run, the work estimated from the plan of one observable when there are several. On the main
problem (a code, or the sector of a circuit), a DP predicted to finish within 10~s, after a head start of 1~s for
the search, gives the distance, and its traceback (for at most $2^{28}$ table entries) or a search at that single
threshold gives a witness; a DP that fits in memory but is slower races the search, which runs first for the DP's
predicted time. On the time-collapsed problem the DP runs only when it is predicted within 10~s, and its value is
the threshold at which the collapsed search starts. For circuits, after the collapsed bound, lifting and windows,
the remaining gap is given to the truncated DP at threshold $d_{\mathrm{up}}-1$, where $d_{\mathrm{up}}$ is the best upper bound, when its
tables fit (at most 56 active rows): its time is predicted from probe sweeps, the search runs first for that time,
shortened where needed to leave the DP 1.5 times its prediction before the deadline, and the DP then returns the
distance or certifies that it is at least $d_{\mathrm{up}}$. We call this configuration
\texttt{dp=2}; the benchmark, the instruction distances, the two-gross scan and the chips use it with default
settings, and the audit runs use the search with its reductions.

\subsection*{Memory instructions of the bicycle architecture}

We built the memory instruction of ref.~\citenum{yoder_2025_tour} in its own setting. A noise-free data
preparation and a noise-free syndrome round give the error-free code state and the reference outcomes; then
follow $C$ noisy cycles of a depth-seven schedule, in which the check qubits are measured and reset in one
timestep so that $C$ cycles take $8C+1$ timesteps, and finally a noise-free stabilizer round and a noise-free data
measurement. Every CNOT is followed by one of the 15 non-identity two-qubit Paulis with probability $p/15$, every
idle qubit in a timestep by $X$, $Y$ or $Z$ with probability $p/3$, and preparations and measurements fail with
probability $p$\cite{yoder_2025_tour}. We certify distances for $C=1$ unless stated otherwise. The rates of ref.~\citenum{yoder_2025_tour} are defined differently, as the
failure probability of $C=d^*$ cycles divided by $C$, with preparations and measurements in separate timesteps and with
faults linearized to independent events of probability $p/15$; to compare with them we built that circuit as well,
as described in Supplementary Section~\ref{subsec:ibm-fit}.
Distances are certified as in the rest of this work, with the lattice translations of each code as symmetries.

\subsection*{All schedules of the two-gross code}

A group automorphism of $\mathbb{Z}_{l}\times\mathbb{Z}_{m}$ that maps the terms of the code's polynomials $A$ and $B$ to translates of themselves, of each other, or of their negatives relabels a schedule and
carries its circuit to the circuit of the relabeled schedule, with $X$ and $Z$ exchanged for the negatives. Reversing the layers keeps the basis. These maps reduce the 16,992 problems of a code to orbits, 708 for the two-gross code;
all generators reproduce every exact value of the four smaller codes, and a direct comparison of the detector error models of six schedules confirms them on two-gross. A one-round window is the problem on the columns whose detectors all lie in one round. It holds the single-qubit columns and possibly one family of hook columns, so windows repeat: over the 708 problems seven occur, one with single-qubit columns only and six with a family of hook columns. A fault set found once in a window therefore holds in every problem with that window, and we replayed it in each problem's own circuit. Only the 120 problems of the 78 classes of schedules without a hook family remain, and for those we ran the search up to weight 17, stopping at the first fault set. A weight-18 logical operator at one time step bounds every schedule by 18. Supplementary Section~\ref{subsec:ibm-scan} gives the details of the scan.

\subsection*{Low-weight fault sets and logical error rates}

The two halves of an undetectable logical fault set fire the same detectors and differ by a logical operator, so every decoder fails on one of them;
summed over disjoint halves, this gives a lower bound on the logical error rate for every decoder, derived in Supplementary Section~\ref{subsec:ibm-bound}.

For the gross code we also enumerated the undetectable logical fault sets of up to 12 faults in the 10-cycle setting,
with the search in enumeration mode but without its dominance exclusions, which keep every set of minimum weight but may
drop heavier ones; these sets are local in time, so they follow from enumerations of 3 and 4 cycles by translation, which
a direct enumeration with 10 cycles confirms for the standard schedule in the $Z$ basis; sets spanning more cycles are
treated in Supplementary Section~\ref{subsec:ibm-leading}. A minimum-weight decoder fails on a configuration contained in such a set when it is heavier than its
complement in the set, and on the less likely of the two when they are equally heavy, unless a configuration of its own
logical class with the same detectors is lighter, or as heavy and likelier than every one of the other class. The sets of weight up to 12
therefore give the failing configurations of up to six faults, and the sum of their probabilities estimates the logical
error rate of such a decoder to order $p^{6}$, with the approximations listed in Supplementary Section~\ref{subsec:ibm-leading}.

\subsection*{Chips with dead checks}

A dead check is removed from every round of the instruction, so with memory basis $Z$ a dead $X$ check only removes faults: a fault on a data qubit in its slot has the same effect on the $Z$ detectors whether the slot holds a gate or nothing. Hence $d_{\mathrm{circ}}^{(Z)}(\text{dead }X,\text{dead }Z)\ge d_{\mathrm{circ}}^{(Z)}(\text{none},\text{dead }Z)$, and the same with $X$ and $Z$ exchanged. We therefore run only the problems with dead checks of the memory basis's own type, one thread per run, and take the minimum of the two bases as the chip's value, a certified lower bound. Whole-chip runs on 29 chips agree in 56 of 58 runs; in the other two a dead check of the other type also removes the hook faults of the lightest set, so the whole-chip value is 10 instead of 9, and the chip values, the minimum over the two bases, agree on all 29. We count a dead qubit or coupler from the upper ends of IBM's junction counts, with all 864 data-check couplers of the memory block, and the shares at each junction yield come from the binomial law for the number of dead checks and the sampled distance tables, as detailed in Supplementary Section~\ref{sec:chips}.

\section*{Data and code availability}

The software is available at \url{https://github.com/Fanerst/tropidist} under the MIT
license\cite{circuit_distance_code}. It is implemented in Python with C kernels parallelized with OpenMP, takes a
Stim circuit or a pair of matrices and returns a certified interval with its error. The repository also holds the
benchmark instances, which its scripts export with their metadata, every result record, the errors behind all
certified values, the audit tables and the scripts that regenerate every figure and table of this paper. Circuits
and data files of the audited papers were taken from the sources cited with each paper.

\section*{Acknowledgments}
We thank Dongyang Feng for bringing up the idea of this work, and the authors of the audited papers for publishing
their circuits, code and data, which made most of this audit possible.
F.P. acknowledges the support from the Ministry of Education Singapore under Grant No.~SKI~2021\_07\_03, A*STAR
under its MTC Young Individual Research Grant No.~H25-MRG3466, and the National Research Foundation, Singapore
through the National Quantum Office, hosted in A*STAR, under its Advanced Quantum Algorithms and Solutions Funding
Initiative (S25Q9DA001 and S25Q9DA002).
B.C. acknowledges the support by the CQT Young Researcher Career Development Grant (26-YRCDG-BC).

\section*{Author contributions}
F.P. conceived the project. B.C. and F.P. developed the theory and the algorithms. F.P. conducted numerical experiments. All authors contributed to the research and to the writing of the manuscript.

\section*{Competing interests}
The authors declare no competing interests.

\bibliographystyle{naturemag_ncs}
\bibliography{refs,ncs}

\begin{thebibliography}{10}
\expandafter\ifx\csname url\endcsname\relax
  \def\url#1{\texttt{#1}}\fi
\expandafter\ifx\csname urlprefix\endcsname\relax\def\urlprefix{URL }\fi
\providecommand{\bibinfo}[2]{#2}
\providecommand{\eprint}[2][]{\url{#2}}

\bibitem{gottesman_1997_stabilizer}
\bibinfo{author}{Gottesman, D.}
\newblock \emph{\bibinfo{title}{{Stabilizer codes and quantum error
  correction}}}.
\newblock Ph.D. thesis, \bibinfo{school}{California Institute of Technology}
  (\bibinfo{year}{1997}).
\newblock
  \bibinfo{note}{\href{https://arxiv.org/abs/quant-ph/9705052}{arXiv:quant-ph/9705052}}.

\bibitem{oum_2005_rankwidth}
\bibinfo{author}{Oum, S.}
\newblock \bibinfo{title}{{Rank-width and vertex-minors}}.
\newblock \emph{\bibinfo{journal}{J. Comb. Theory Ser. B}}
  \textbf{\bibinfo{volume}{95}}, \bibinfo{pages}{79--100}
  (\bibinfo{year}{2005}).

\bibitem{oum_seymour_2006_approximating}
\bibinfo{author}{Oum, S.} \& \bibinfo{author}{Seymour, P.}
\newblock \bibinfo{title}{{Approximating clique-width and branch-width}}.
\newblock \emph{\bibinfo{journal}{J. Comb. Theory Ser. B}}
  \textbf{\bibinfo{volume}{96}}, \bibinfo{pages}{514--528}
  (\bibinfo{year}{2006}).

\bibitem{jeong_2017_trellis}
\bibinfo{author}{Jeong, J.}, \bibinfo{author}{Kim, E.~J.} \&
  \bibinfo{author}{Oum, S.}
\newblock \bibinfo{title}{{The ``art of trellis decoding'' is fixed-parameter
  tractable}}.
\newblock \emph{\bibinfo{journal}{IEEE Trans. Inf. Theory}}
  \textbf{\bibinfo{volume}{63}}, \bibinfo{pages}{7178--7205}
  (\bibinfo{year}{2017}).

\bibitem{mcewen_2023_relaxing}
\bibinfo{author}{McEwen, M.}, \bibinfo{author}{Bacon, D.} \&
  \bibinfo{author}{Gidney, C.}
\newblock \bibinfo{title}{{Relaxing hardware requirements for surface code
  circuits using time-dynamics}}.
\newblock \emph{\bibinfo{journal}{Quantum}} \textbf{\bibinfo{volume}{7}},
  \bibinfo{pages}{1172} (\bibinfo{year}{2023}).

\bibitem{stim}
\bibinfo{author}{Gidney, C.}
\newblock \bibinfo{title}{{{Stim}: a fast stabilizer circuit simulator}}.
\newblock \emph{\bibinfo{journal}{Quantum}} \textbf{\bibinfo{volume}{5}},
  \bibinfo{pages}{497} (\bibinfo{year}{2021}).

\bibitem{cheng_pan_2026_rankdp}
\bibinfo{author}{Cheng, B.} \& \bibinfo{author}{Pan, F.}
\newblock \bibinfo{title}{{Exact maximum likelihood decoding beyond treewidth
  via rank-decomposition dynamic programming}}.
\newblock Preprint at \url{https://arxiv.org/abs/2609.39556}
  (\bibinfo{year}{2026}).

\bibitem{bocharova_2004_beast}
\bibinfo{author}{Bocharova, I.~E.}, \bibinfo{author}{Handlery, M.},
  \bibinfo{author}{Johannesson, R.} \& \bibinfo{author}{Kudryashov, B.~D.}
\newblock \bibinfo{title}{A {BEAST} for prowling in trees}.
\newblock \emph{\bibinfo{journal}{IEEE Trans. Inf. Theory}}
  \textbf{\bibinfo{volume}{50}}, \bibinfo{pages}{1295--1302}
  (\bibinfo{year}{2004}).

\bibitem{webster_2026_distance}
\bibinfo{author}{Webster, M.}, \bibinfo{author}{Jacob, A.} \&
  \bibinfo{author}{Higgott, O.}
\newblock \bibinfo{title}{{Distance-finding algorithms for quantum codes and
  circuits}}.
\newblock Preprint at \url{https://arxiv.org/abs/2603.22532}
  (\bibinfo{year}{2026}).

\bibitem{beni_2025_tesseract}
\bibinfo{author}{Beni, L.~A.}, \bibinfo{author}{Higgott, O.} \&
  \bibinfo{author}{Shutty, N.}
\newblock \bibinfo{title}{{Tesseract: a search-based decoder for quantum error
  correction}}.
\newblock Preprint at \url{https://arxiv.org/abs/2503.10988}
  (\bibinfo{year}{2025}).

\bibitem{bravyi_2024_bb}
\bibinfo{author}{Bravyi, S.} \emph{et~al.}
\newblock \bibinfo{title}{{High-threshold and low-overhead fault-tolerant
  quantum memory}}.
\newblock \emph{\bibinfo{journal}{Nature}} \textbf{\bibinfo{volume}{627}},
  \bibinfo{pages}{778--782} (\bibinfo{year}{2024}).

\bibitem{tesseract_code}
\bibinfo{title}{tesseract-decoder} (\bibinfo{year}{2025}).
\newblock \bibinfo{note}{Commit 4fd104d;
  \href{https://github.com/quantumlib/tesseract-decoder}{github.com/quantumlib/tesseract-decoder}}.

\bibitem{lu_2026_cornucopia}
\bibinfo{author}{Lu, Z.}, \bibinfo{author}{Li, W.} \& \bibinfo{author}{Deng,
  D.-L.}
\newblock \bibinfo{title}{{Quantum error correction at ultra-low overhead}}.
\newblock Preprint at \url{https://arxiv.org/abs/2608.02773}
  (\bibinfo{year}{2026}).

\bibitem{bhardwaj_2026_mitten}
\bibinfo{author}{Bhardwaj, A.} \emph{et~al.}
\newblock \bibinfo{title}{{High-rate qLDPC processors}}.
\newblock Preprint at \url{https://arxiv.org/abs/2607.28795}
  (\bibinfo{year}{2026}).

\bibitem{zhang_2026_routing}
\bibinfo{author}{Zhang, J.} \emph{et~al.}
\newblock \bibinfo{title}{{Routing codes: high-rate quantum LDPC codes with
  short, parallel non-local connectivity}}.
\newblock Preprint at \url{https://arxiv.org/abs/2606.25330}
  (\bibinfo{year}{2026}).

\bibitem{kim_2026_shift}
\bibinfo{author}{Kim, Y.}, \bibinfo{author}{Gicev, S.},
  \bibinfo{author}{Sevior, M.} \& \bibinfo{author}{Usman, M.}
\newblock \bibinfo{title}{{Time-dynamic circuits for fault-tolerant shift
  automorphisms in quantum LDPC codes}}.
\newblock Preprint at \url{https://arxiv.org/abs/2601.09911}
  (\bibinfo{year}{2026}).

\bibitem{ismail_2026_star}
\bibinfo{author}{Ismail, R.} \emph{et~al.}
\newblock \bibinfo{title}{{Fast and parallel high-rate STAR architecture for
  megaquop quantum simulation}}.
\newblock Preprint at \url{https://arxiv.org/abs/2606.25011}
  (\bibinfo{year}{2026}).

\bibitem{tripier_2026_walkingcat}
\bibinfo{author}{Tripier, F.} \emph{et~al.}
\newblock \bibinfo{title}{{Fault-tolerant quantum computing with trapped ions:
  the Walking Cat architecture}}.
\newblock Preprint at \url{https://arxiv.org/abs/2604.19481}
  (\bibinfo{year}{2026}).

\bibitem{strikis_2026_leftright}
\bibinfo{author}{Strikis, A.}, \bibinfo{author}{Browne, D.~E.} \&
  \bibinfo{author}{Beverland, M.~E.}
\newblock \bibinfo{title}{{High-performance syndrome extraction circuits for
  quantum codes}}.
\newblock Preprint at \url{https://arxiv.org/abs/2603.05481}
  (\bibinfo{year}{2026}).

\bibitem{liang_2025_selfdual}
\bibinfo{author}{Liang, Z.} \& \bibinfo{author}{Chen, Y.-A.}
\newblock \bibinfo{title}{{Self-dual bivariate bicycle codes with transversal
  Clifford gates}}.
\newblock \emph{\bibinfo{journal}{npj Quantum Inf.}}  (\bibinfo{year}{2026}).

\bibitem{lu_2026_weight8}
\bibinfo{author}{Lu, L.}, \bibinfo{author}{Yang, R.} \& \bibinfo{author}{Guo,
  G.}
\newblock \bibinfo{title}{{Quantum bivariate bicycle codes with weight-8 checks
  surpassing the BB benchmark}}.
\newblock Preprint at \url{https://arxiv.org/abs/2609.06572}
  (\bibinfo{year}{2026}).

\bibitem{okada_2026_pair}
\bibinfo{author}{Okada, K.} \& \bibinfo{author}{Kasai, K.}
\newblock \bibinfo{title}{{Pair-partition constructions for CPM-based quantum
  LDPC codes}}.
\newblock Preprint at \url{https://arxiv.org/abs/2607.14091}
  (\bibinfo{year}{2026}).

\bibitem{lee_2026_ultrahigh}
\bibinfo{author}{Lee, J.~Y.}, \bibinfo{author}{Okada, K.},
  \bibinfo{author}{Maskara, N.}, \bibinfo{author}{Kasai, K.} \&
  \bibinfo{author}{Zhou, H.}
\newblock \bibinfo{title}{{Design principles for ultra-high-rate quantum
  codes}}.
\newblock Preprint at \url{https://arxiv.org/abs/2609.30069}
  (\bibinfo{year}{2026}).

\bibitem{hong_2026_rate}
\bibinfo{author}{Hong, Y.}
\newblock \bibinfo{title}{{Quantum LDPC codes with design rate 1/5 and good
  performance below 1000 physical qubits}}.
\newblock Preprint at \url{https://arxiv.org/abs/2607.27644}
  (\bibinfo{year}{2026}).

\bibitem{mian_2026_multicycle}
\bibinfo{author}{Mian, F.~A.}, \bibinfo{author}{Gwilliam, O.} \&
  \bibinfo{author}{Krastanov, S.}
\newblock \bibinfo{title}{{Multivariate multicycle codes for complete
  single-shot decoding}}.
\newblock Preprint at \url{https://arxiv.org/abs/2601.18879}
  (\bibinfo{year}{2026}).

\bibitem{cruzbenito_2026_llm}
\bibinfo{author}{Cruz-Benito, J.}, \bibinfo{author}{Cross, A.~W.},
  \bibinfo{author}{Kremer, D.} \& \bibinfo{author}{Faro, I.}
\newblock \bibinfo{title}{{Evolutionary discovery of bivariate bicycle codes
  with LLM-guided search}}.
\newblock Preprint at \url{https://arxiv.org/abs/2606.02418}
  (\bibinfo{year}{2026}).

\bibitem{yang_2026_gala}
\bibinfo{author}{Yang, W.}, \bibinfo{author}{Duckering, C.} \&
  \bibinfo{author}{Dua, A.}
\newblock \bibinfo{title}{{Designer codes from GALA: compact, self-dual, and
  rate-1/2 QEC on reconfigurable atom arrays}}.
\newblock Preprint at \url{https://arxiv.org/abs/2608.07431}
  (\bibinfo{year}{2026}).

\bibitem{goto_2024_hypercube}
\bibinfo{author}{Goto, H.}
\newblock \bibinfo{title}{High-performance fault-tolerant quantum computing
  with many-hypercube codes}.
\newblock \emph{\bibinfo{journal}{Science Advances}}
  \textbf{\bibinfo{volume}{10}}, \bibinfo{pages}{eadp6388}
  (\bibinfo{year}{2024}).

\bibitem{gu_2026_planar}
\bibinfo{author}{Gu, B.} \emph{et~al.}
\newblock \bibinfo{title}{Nearest-neighbour gates are all you need: high-rate
  quantum low-density parity-check codes on a planar grid}.
\newblock Preprint at \url{https://arxiv.org/abs/2606.19482}
  (\bibinfo{year}{2026}).

\bibitem{dumer_2017_verification}
\bibinfo{author}{Dumer, I.}, \bibinfo{author}{Kovalev, A.~A.} \&
  \bibinfo{author}{Pryadko, L.~P.}
\newblock \bibinfo{title}{Distance verification for classical and quantum
  {LDPC} codes}.
\newblock \emph{\bibinfo{journal}{IEEE Trans. Inf. Theory}}
  \textbf{\bibinfo{volume}{63}}, \bibinfo{pages}{4675--4686}
  (\bibinfo{year}{2017}).

\bibitem{distm4ri_code}
\bibinfo{title}{{dist-m4ri}} (\bibinfo{year}{2026}).
\newblock \bibinfo{note}{Version 0.14.7 (commit bed621e);
  \href{https://github.com/QEC-pages/dist-m4ri}{github.com/QEC-pages/dist-m4ri}}.

\bibitem{qubitserf_code}
\bibinfo{title}{{Qubitserf}} (\bibinfo{year}{2026}).
\newblock \bibinfo{note}{Version 0.1.0 (commit ab96b9c);
  \href{https://github.com/Quantinuum/Qubitserf}{github.com/Quantinuum/Qubitserf}}.

\bibitem{huangfu_2018_highs}
\bibinfo{author}{Huangfu, Q.} \& \bibinfo{author}{Hall, J. A.~J.}
\newblock \bibinfo{title}{{Parallelizing the dual revised simplex method}}.
\newblock \emph{\bibinfo{journal}{Math. Program. Comput.}}
  \textbf{\bibinfo{volume}{10}}, \bibinfo{pages}{119--142}
  (\bibinfo{year}{2018}).

\bibitem{cpsat_code}
\bibinfo{author}{Perron, L.} \& \bibinfo{author}{Didier, F.}
\newblock \bibinfo{title}{{CP-SAT}} (\bibinfo{year}{2026}).
\newblock \bibinfo{note}{Google OR-Tools, version 9.15;
  \href{https://developers.google.com/optimization/cp/cp_solver}{developers.google.com/optimization}}.

\bibitem{gurobi_manual}
\bibinfo{author}{{Gurobi Optimization, LLC}}.
\newblock \bibinfo{title}{{Gurobi Optimizer Reference Manual}}
  (\bibinfo{year}{2026}).
\newblock \bibinfo{note}{Version 13.0.3;
  \href{https://www.gurobi.com}{www.gurobi.com}}.

\bibitem{biere_2024_cadical}
\bibinfo{author}{Biere, A.} \emph{et~al.}
\newblock \bibinfo{title}{{CaDiCaL} 2.0}.
\newblock In \emph{\bibinfo{booktitle}{Computer Aided Verification (CAV
  2024)}}, vol. \bibinfo{volume}{14681} of \emph{\bibinfo{series}{Lecture Notes
  in Computer Science}}, \bibinfo{pages}{133--152}
  (\bibinfo{publisher}{Springer}, \bibinfo{year}{2024}).

\bibitem{ignatiev_2018_pysat}
\bibinfo{author}{Ignatiev, A.}, \bibinfo{author}{Morgado, A.} \&
  \bibinfo{author}{Marques-Silva, J.}
\newblock \bibinfo{title}{{PySAT}: A {Python} toolkit for prototyping with
  {SAT} oracles}.
\newblock In \emph{\bibinfo{booktitle}{Theory and Applications of
  Satisfiability Testing (SAT 2018)}}, vol. \bibinfo{volume}{10929} of
  \emph{\bibinfo{series}{Lecture Notes in Computer Science}},
  \bibinfo{pages}{428--437} (\bibinfo{publisher}{Springer},
  \bibinfo{year}{2018}).

\bibitem{ignatiev_2019_rc2}
\bibinfo{author}{Ignatiev, A.}, \bibinfo{author}{Morgado, A.} \&
  \bibinfo{author}{Marques-Silva, J.}
\newblock \bibinfo{title}{{RC2}: an efficient {MaxSAT} solver}.
\newblock \emph{\bibinfo{journal}{J. Satisf. Boolean Model. Comput.}}
  \textbf{\bibinfo{volume}{11}}, \bibinfo{pages}{53--64}
  (\bibinfo{year}{2019}).

\bibitem{maxsat_2023}
\bibinfo{title}{{MaxSAT Evaluation 2023}} (\bibinfo{year}{2023}).
\newblock \bibinfo{note}{Solver CASHWMaxSAT-CorePlus;
  \href{https://maxsat-evaluations.github.io/2023/}{maxsat-evaluations.github.io/2023}}.

\bibitem{zhang_2026_enmaxcdcl}
\bibinfo{author}{Zhang, J.}, \bibinfo{author}{Li, C.-M.},
  \bibinfo{author}{Cherif, S.} \& \bibinfo{author}{Li, S.}
\newblock \bibinfo{title}{Enhanced lower bound computation in branch-and-bound
  for {MaxSAT}}.
\newblock In \emph{\bibinfo{booktitle}{32nd International Conference on
  Principles and Practice of Constraint Programming (CP 2026)}}, vol.
  \bibinfo{volume}{379} of \emph{\bibinfo{series}{Leibniz International
  Proceedings in Informatics (LIPIcs)}}, \bibinfo{pages}{60:1--60:18}
  (\bibinfo{year}{2026}).

\bibitem{enmaxcdcl_code}
\bibinfo{title}{{enMaxCDCL}} (\bibinfo{year}{2026}).
\newblock
  \bibinfo{note}{\href{https://github.com/zjl9959/enMaxCDCL_CP26}{github.com/zjl9959/enMaxCDCL\_CP26}}.

\bibitem{roffe_2020_decoding}
\bibinfo{author}{Roffe, J.}, \bibinfo{author}{White, D.~R.},
  \bibinfo{author}{Burton, S.} \& \bibinfo{author}{Campbell, E.~T.}
\newblock \bibinfo{title}{{Decoding across the quantum LDPC code landscape}}.
\newblock \emph{\bibinfo{journal}{Phys. Rev. Res.}}
  \textbf{\bibinfo{volume}{2}}, \bibinfo{pages}{043423} (\bibinfo{year}{2020}).

\bibitem{wang_2026_bbsurface}
\bibinfo{author}{Wang, R.} \& \bibinfo{author}{Pryadko, L.~P.}
\newblock \bibinfo{title}{Algebra of bivariate-bicycle surface codes}.
\newblock Preprint at \url{https://arxiv.org/abs/2606.08771}
  (\bibinfo{year}{2026}).

\bibitem{nixon_2026_vine}
\bibinfo{author}{Nixon, G.~M.}, \bibinfo{author}{McLauchlan, C.~K.} \&
  \bibinfo{author}{van Rest, C. C.~L.}
\newblock \bibinfo{title}{Vine codes: low-overhead quantum {LDPC} codes on a
  planar square grid}.
\newblock Preprint at \url{https://arxiv.org/abs/2606.20263}
  (\bibinfo{year}{2026}).

\bibitem{liang_2026_amplified}
\bibinfo{author}{Liang, Z.}, \bibinfo{author}{Gu, B.}, \bibinfo{author}{Chen,
  Y.-A.}, \bibinfo{author}{Eisert, J.} \& \bibinfo{author}{Wang, Z.}
\newblock \bibinfo{title}{Ultra-high-distance quantum memories from amplified
  {qLDPC} codes}.
\newblock Preprint at \url{https://arxiv.org/abs/2609.37231}
  (\bibinfo{year}{2026}).

\bibitem{steffan_2025_tile}
\bibinfo{author}{Steffan, V.}, \bibinfo{author}{Choe, S.~H.},
  \bibinfo{author}{Breuckmann, N.~P.}, \bibinfo{author}{Pereira, F. R.~F.} \&
  \bibinfo{author}{Eberhardt, J.~N.}
\newblock \bibinfo{title}{Tile codes: high-efficiency quantum codes on a
  lattice with boundary}.
\newblock \emph{\bibinfo{journal}{Phys. Rev. Lett.}}
  \textbf{\bibinfo{volume}{135}}, \bibinfo{pages}{170601}
  (\bibinfo{year}{2025}).

\bibitem{higgott_2026_hyperbolic}
\bibinfo{author}{Higgott, O.} \emph{et~al.}
\newblock \bibinfo{title}{{Low-overhead quantum error correction with
  boundary-connected planar modules}}.
\newblock Preprint at \url{https://arxiv.org/abs/2610.03682}
  (\bibinfo{year}{2026}).

\bibitem{gu_2026_qgpu}
\bibinfo{author}{Gu, B.} \emph{et~al.}
\newblock \bibinfo{title}{{QGPU}: parallel logic in quantum {LDPC} codes}.
\newblock Preprint at \url{https://arxiv.org/abs/2603.05398}
  (\bibinfo{year}{2026}).

\bibitem{bae_2026_tile}
\bibinfo{author}{Bae, Y.} \& \bibinfo{author}{Park, C.-Y.}
\newblock \bibinfo{title}{Strictly local tile-code architectures on
  two-dimensional planar lattices}.
\newblock Preprint at \url{https://arxiv.org/abs/2607.05897}
  (\bibinfo{year}{2026}).

\bibitem{choe_2026_barbell}
\bibinfo{author}{Choe, S.~H.} \emph{et~al.}
\newblock \bibinfo{title}{Barbell codes: {qLDPC} codes for superconducting
  quantum hardware}.
\newblock Preprint at \url{https://arxiv.org/abs/2606.06062}
  (\bibinfo{year}{2026}).

\bibitem{xu_2026_spacegroups}
\bibinfo{author}{Xu, C.-Y.}, \bibinfo{author}{Liu, Z.-C.} \&
  \bibinfo{author}{Xu, Y.}
\newblock \bibinfo{title}{Topological codes from space groups: a route beyond
  translation invariance}.
\newblock Preprint at \url{https://arxiv.org/abs/2606.20548}
  (\bibinfo{year}{2026}).

\bibitem{lu_2026_modular}
\bibinfo{author}{Lu, Z.} \emph{et~al.}
\newblock \bibinfo{title}{A modular quantum computer based on {Cornucopia}
  codes}.
\newblock Preprint at \url{https://arxiv.org/abs/2610.19073}
  (\bibinfo{year}{2026}).

\bibitem{yoder_2025_tour}
\bibinfo{author}{Yoder, T.~J.} \emph{et~al.}
\newblock \bibinfo{title}{Tour de gross: a modular quantum computer based on
  bivariate bicycle codes}.
\newblock Preprint at \url{https://arxiv.org/abs/2506.03094}
  (\bibinfo{year}{2025}).

\bibitem{mueller_2025_relay}
\bibinfo{author}{M{\"u}ller, T.} \emph{et~al.}
\newblock \bibinfo{title}{Improved belief propagation is sufficient for
  real-time decoding of quantum memory}.
\newblock Preprint at \url{https://arxiv.org/abs/2506.01779}
  (\bibinfo{year}{2025}).

\bibitem{beverland_2025_failfast}
\bibinfo{author}{Beverland, M.~E.}, \bibinfo{author}{Carroll, M.},
  \bibinfo{author}{Cross, A.~W.} \& \bibinfo{author}{Yoder, T.~J.}
\newblock \bibinfo{title}{Fail fast: techniques to probe rare events in quantum
  error correction}.
\newblock Preprint at \url{https://arxiv.org/abs/2511.15177}
  (\bibinfo{year}{2025}).

\bibitem{bravyi_bbcode}
\bibinfo{author}{Bravyi, S.}
\newblock \bibinfo{title}{{BivariateBicycleCodes}} (\bibinfo{year}{2024}).
\newblock \bibinfo{note}{Commit fa77e33;
  \href{https://github.com/sbravyi/BivariateBicycleCodes}{github.com/sbravyi/BivariateBicycleCodes}}.

\bibitem{swaroop_2024_adapters}
\bibinfo{author}{Swaroop, E.}, \bibinfo{author}{Jochym-O'Connor, T.} \&
  \bibinfo{author}{Yoder, T.~J.}
\newblock \bibinfo{title}{Universal adapters between quantum low-density parity
  check codes}.
\newblock \emph{\bibinfo{journal}{PRX Quantum}} \textbf{\bibinfo{volume}{7}},
  \bibinfo{pages}{010324} (\bibinfo{year}{2026}).

\bibitem{cowtan_2024_ssip}
\bibinfo{author}{Cowtan, A.}
\newblock \bibinfo{title}{{SSIP}: automated surgery with quantum {LDPC} codes}.
\newblock Preprint at \url{https://arxiv.org/abs/2407.09423}
  (\bibinfo{year}{2024}).

\bibitem{ssip_code}
\bibinfo{title}{{SSIP}} (\bibinfo{year}{2024}).
\newblock \bibinfo{note}{Commit 8e4d771;
  \href{https://github.com/CQCL/SSIP}{github.com/CQCL/SSIP}}.

\bibitem{qdistrnd}
\bibinfo{author}{Pryadko, L.~P.}, \bibinfo{author}{Shabashov, V.~A.} \&
  \bibinfo{author}{Kozin, V.~K.}
\newblock \bibinfo{title}{{QDistRnd}: A {GAP} package for computing the
  distance of quantum error-correcting codes}.
\newblock \emph{\bibinfo{journal}{J. Open Source Softw.}}
  \textbf{\bibinfo{volume}{7}}, \bibinfo{pages}{4120} (\bibinfo{year}{2022}).

\bibitem{williamson_2024_gauging}
\bibinfo{author}{Williamson, D.~J.} \& \bibinfo{author}{Yoder, T.~J.}
\newblock \bibinfo{title}{Low-overhead fault-tolerant quantum computation by
  gauging logical operators}.
\newblock \emph{\bibinfo{journal}{Nature Physics}}
  \textbf{\bibinfo{volume}{22}}, \bibinfo{pages}{598--603}
  (\bibinfo{year}{2026}).

\bibitem{cross_2024_surgery}
\bibinfo{author}{Cross, A.~W.}, \bibinfo{author}{He, Z.},
  \bibinfo{author}{Rall, P.~J.} \& \bibinfo{author}{Yoder, T.~J.}
\newblock \bibinfo{title}{Improved {QLDPC} surgery: logical measurements and
  bridging codes}.
\newblock Preprint at \url{https://arxiv.org/abs/2407.18393}
  (\bibinfo{year}{2024}).

\bibitem{circuit_distance_code}
\bibinfo{author}{Pan, F.}
\newblock \bibinfo{title}{{tropidist: certified code and circuit distances}}
  (\bibinfo{year}{2026}).
\newblock
  \bibinfo{note}{\href{https://github.com/Fanerst/tropidist}{github.com/Fanerst/tropidist}}.

\end{thebibliography}


\begin{thebibliography}{10}
\expandafter\ifx\csname url\endcsname\relax
  \def\url#1{\texttt{#1}}\fi
\expandafter\ifx\csname urlprefix\endcsname\relax\def\urlprefix{URL }\fi
\providecommand{\bibinfo}[2]{#2}
\providecommand{\eprint}[2][]{\url{#2}}

\bibitem{beverland_2022_requirements}
\bibinfo{author}{Beverland, M.~E.} \emph{et~al.}
\newblock \bibinfo{title}{Assessing requirements to scale to practical quantum
  advantage}.
\newblock Preprint at \url{https://arxiv.org/abs/2211.07629}
  (\bibinfo{year}{2022}).

\bibitem{gidney_2025_rsa}
\bibinfo{author}{Gidney, C.}
\newblock \bibinfo{title}{How to factor 2048 bit {RSA} integers with less than
  a million noisy qubits}.
\newblock Preprint at \url{https://arxiv.org/abs/2505.15917}
  (\bibinfo{year}{2025}).

\bibitem{shor_1995}
\bibinfo{author}{Shor, P.~W.}
\newblock \bibinfo{title}{Scheme for reducing decoherence in quantum computer
  memory}.
\newblock \emph{\bibinfo{journal}{Phys. Rev. A}} \textbf{\bibinfo{volume}{52}},
  \bibinfo{pages}{R2493--R2496} (\bibinfo{year}{1995}).

\bibitem{dennis_2002}
\bibinfo{author}{Dennis, E.}, \bibinfo{author}{Kitaev, A.},
  \bibinfo{author}{Landahl, A.} \& \bibinfo{author}{Preskill, J.}
\newblock \bibinfo{title}{Topological quantum memory}.
\newblock \emph{\bibinfo{journal}{J. Math. Phys.}}
  \textbf{\bibinfo{volume}{43}}, \bibinfo{pages}{4452--4505}
  (\bibinfo{year}{2002}).

\bibitem{fowler_2012}
\bibinfo{author}{Fowler, A.~G.}, \bibinfo{author}{Mariantoni, M.},
  \bibinfo{author}{Martinis, J.~M.} \& \bibinfo{author}{Cleland, A.~N.}
\newblock \bibinfo{title}{Surface codes: towards practical large-scale quantum
  computation}.
\newblock \emph{\bibinfo{journal}{Phys. Rev. A}} \textbf{\bibinfo{volume}{86}},
  \bibinfo{pages}{032324} (\bibinfo{year}{2012}).

\bibitem{google_2025_below}
\bibinfo{author}{{Google Quantum AI and Collaborators}}.
\newblock \bibinfo{title}{Quantum error correction below the surface code
  threshold}.
\newblock \emph{\bibinfo{journal}{Nature}} \textbf{\bibinfo{volume}{638}},
  \bibinfo{pages}{920--926} (\bibinfo{year}{2025}).

\bibitem{bluvstein_2024}
\bibinfo{author}{Bluvstein, D.} \emph{et~al.}
\newblock \bibinfo{title}{Logical quantum processor based on reconfigurable
  atom arrays}.
\newblock \emph{\bibinfo{journal}{Nature}} \textbf{\bibinfo{volume}{626}},
  \bibinfo{pages}{58--65} (\bibinfo{year}{2024}).

\bibitem{gottesman_2014}
\bibinfo{author}{Gottesman, D.}
\newblock \bibinfo{title}{Fault-tolerant quantum computation with constant
  overhead}.
\newblock \emph{\bibinfo{journal}{Quantum Inf. Comput.}}
  \textbf{\bibinfo{volume}{14}}, \bibinfo{pages}{1338--1372}
  (\bibinfo{year}{2014}).

\bibitem{breuckmann_2021}
\bibinfo{author}{Breuckmann, N.~P.} \& \bibinfo{author}{Eberhardt, J.~N.}
\newblock \bibinfo{title}{Quantum low-density parity-check codes}.
\newblock \emph{\bibinfo{journal}{PRX Quantum}} \textbf{\bibinfo{volume}{2}},
  \bibinfo{pages}{040101} (\bibinfo{year}{2021}).

\bibitem{bravyi_2024_bb}
\bibinfo{author}{Bravyi, S.} \emph{et~al.}
\newblock \bibinfo{title}{{High-threshold and low-overhead fault-tolerant
  quantum memory}}.
\newblock \emph{\bibinfo{journal}{Nature}} \textbf{\bibinfo{volume}{627}},
  \bibinfo{pages}{778--782} (\bibinfo{year}{2024}).

\bibitem{yoder_2025_tour}
\bibinfo{author}{Yoder, T.~J.} \emph{et~al.}
\newblock \bibinfo{title}{Tour de gross: a modular quantum computer based on
  bivariate bicycle codes}.
\newblock Preprint at \url{https://arxiv.org/abs/2506.03094}
  (\bibinfo{year}{2025}).

\bibitem{ibm_roadmap}
\bibinfo{author}{{IBM Quantum}}.
\newblock \bibinfo{title}{{IBM Quantum} roadmap} (\bibinfo{year}{2025}).
\newblock \bibinfo{note}{\url{https://www.ibm.com/quantum-computing/roadmap},
  accessed October 2026}.

\bibitem{mcewen_2023_relaxing}
\bibinfo{author}{McEwen, M.}, \bibinfo{author}{Bacon, D.} \&
  \bibinfo{author}{Gidney, C.}
\newblock \bibinfo{title}{{Relaxing hardware requirements for surface code
  circuits using time-dynamics}}.
\newblock \emph{\bibinfo{journal}{Quantum}} \textbf{\bibinfo{volume}{7}},
  \bibinfo{pages}{1172} (\bibinfo{year}{2023}).

\bibitem{stim}
\bibinfo{author}{Gidney, C.}
\newblock \bibinfo{title}{{{Stim}: a fast stabilizer circuit simulator}}.
\newblock \emph{\bibinfo{journal}{Quantum}} \textbf{\bibinfo{volume}{5}},
  \bibinfo{pages}{497} (\bibinfo{year}{2021}).

\bibitem{vardy_1997}
\bibinfo{author}{Vardy, A.}
\newblock \bibinfo{title}{The intractability of computing the minimum distance
  of a code}.
\newblock \emph{\bibinfo{journal}{IEEE Trans. Inf. Theory}}
  \textbf{\bibinfo{volume}{43}}, \bibinfo{pages}{1757--1766}
  (\bibinfo{year}{1997}).

\bibitem{roffe_2020_decoding}
\bibinfo{author}{Roffe, J.}, \bibinfo{author}{White, D.~R.},
  \bibinfo{author}{Burton, S.} \& \bibinfo{author}{Campbell, E.~T.}
\newblock \bibinfo{title}{{Decoding across the quantum LDPC code landscape}}.
\newblock \emph{\bibinfo{journal}{Phys. Rev. Res.}}
  \textbf{\bibinfo{volume}{2}}, \bibinfo{pages}{043423} (\bibinfo{year}{2020}).

\bibitem{qdistrnd}
\bibinfo{author}{Pryadko, L.~P.}, \bibinfo{author}{Shabashov, V.~A.} \&
  \bibinfo{author}{Kozin, V.~K.}
\newblock \bibinfo{title}{{QDistRnd}: A {GAP} package for computing the
  distance of quantum error-correcting codes}.
\newblock \emph{\bibinfo{journal}{J. Open Source Softw.}}
  \textbf{\bibinfo{volume}{7}}, \bibinfo{pages}{4120} (\bibinfo{year}{2022}).

\bibitem{beni_2025_tesseract}
\bibinfo{author}{Beni, L.~A.}, \bibinfo{author}{Higgott, O.} \&
  \bibinfo{author}{Shutty, N.}
\newblock \bibinfo{title}{{Tesseract: a search-based decoder for quantum error
  correction}}.
\newblock Preprint at \url{https://arxiv.org/abs/2503.10988}
  (\bibinfo{year}{2025}).

\bibitem{dumer_2017_verification}
\bibinfo{author}{Dumer, I.}, \bibinfo{author}{Kovalev, A.~A.} \&
  \bibinfo{author}{Pryadko, L.~P.}
\newblock \bibinfo{title}{Distance verification for classical and quantum
  {LDPC} codes}.
\newblock \emph{\bibinfo{journal}{IEEE Trans. Inf. Theory}}
  \textbf{\bibinfo{volume}{63}}, \bibinfo{pages}{4675--4686}
  (\bibinfo{year}{2017}).

\bibitem{distm4ri_code}
\bibinfo{title}{{dist-m4ri}} (\bibinfo{year}{2026}).
\newblock \bibinfo{note}{Version 0.14.7 (commit bed621e);
  \href{https://github.com/QEC-pages/dist-m4ri}{github.com/QEC-pages/dist-m4ri}}.

\bibitem{webster_2026_distance}
\bibinfo{author}{Webster, M.}, \bibinfo{author}{Jacob, A.} \&
  \bibinfo{author}{Higgott, O.}
\newblock \bibinfo{title}{{Distance-finding algorithms for quantum codes and
  circuits}}.
\newblock Preprint at \url{https://arxiv.org/abs/2603.22532}
  (\bibinfo{year}{2026}).

\bibitem{liu_2021_tropical}
\bibinfo{author}{Liu, J.-G.}, \bibinfo{author}{Wang, L.} \&
  \bibinfo{author}{Zhang, P.}
\newblock \bibinfo{title}{Tropical tensor network for ground states of spin
  glasses}.
\newblock \emph{\bibinfo{journal}{Phys. Rev. Lett.}}
  \textbf{\bibinfo{volume}{126}}, \bibinfo{pages}{090506}
  (\bibinfo{year}{2021}).

\bibitem{sommers_2025_concatenated}
\bibinfo{author}{Sommers, G.~M.}, \bibinfo{author}{Huse, D.~A.} \&
  \bibinfo{author}{Gullans, M.~J.}
\newblock \bibinfo{title}{Dynamically generated concatenated codes and their
  phase diagrams}.
\newblock \emph{\bibinfo{journal}{Phys. Rev. Res.}}
  \textbf{\bibinfo{volume}{7}}, \bibinfo{pages}{023086} (\bibinfo{year}{2025}).

\bibitem{cheng_pan_2026_rankdp}
\bibinfo{author}{Cheng, B.} \& \bibinfo{author}{Pan, F.}
\newblock \bibinfo{title}{{Exact maximum likelihood decoding beyond treewidth
  via rank-decomposition dynamic programming}}.
\newblock Preprint at \url{https://arxiv.org/abs/2609.39556}
  (\bibinfo{year}{2026}).

\bibitem{cunningham_2007_branchwidth}
\bibinfo{author}{Cunningham, W.~H.} \& \bibinfo{author}{Geelen, J.}
\newblock \bibinfo{title}{On integer programming and the branch-width of the
  constraint matrix}.
\newblock In \emph{\bibinfo{booktitle}{Integer Programming and Combinatorial
  Optimization ({IPCO} 2007)}}, vol. \bibinfo{volume}{4513} of
  \emph{\bibinfo{series}{Lecture Notes in Computer Science}},
  \bibinfo{pages}{158--166} (\bibinfo{publisher}{Springer},
  \bibinfo{year}{2007}).

\bibitem{kashyap_2009_tree}
\bibinfo{author}{Kashyap, N.}
\newblock \bibinfo{title}{On minimal tree realizations of linear codes}.
\newblock \emph{\bibinfo{journal}{IEEE Trans. Inf. Theory}}
  \textbf{\bibinfo{volume}{55}}, \bibinfo{pages}{3501--3519}
  (\bibinfo{year}{2009}).

\bibitem{forney_1994}
\bibinfo{author}{Forney, G.~D.}
\newblock \bibinfo{title}{Dimension/length profiles and trellis complexity of
  linear block codes}.
\newblock \emph{\bibinfo{journal}{IEEE Trans. Inf. Theory}}
  \textbf{\bibinfo{volume}{40}}, \bibinfo{pages}{1741--1752}
  (\bibinfo{year}{1994}).

\bibitem{jeong_2017_trellis}
\bibinfo{author}{Jeong, J.}, \bibinfo{author}{Kim, E.~J.} \&
  \bibinfo{author}{Oum, S.}
\newblock \bibinfo{title}{{The ``art of trellis decoding'' is fixed-parameter
  tractable}}.
\newblock \emph{\bibinfo{journal}{IEEE Trans. Inf. Theory}}
  \textbf{\bibinfo{volume}{63}}, \bibinfo{pages}{7178--7205}
  (\bibinfo{year}{2017}).

\bibitem{bocharova_2004_beast}
\bibinfo{author}{Bocharova, I.~E.}, \bibinfo{author}{Handlery, M.},
  \bibinfo{author}{Johannesson, R.} \& \bibinfo{author}{Kudryashov, B.~D.}
\newblock \bibinfo{title}{A {BEAST} for prowling in trees}.
\newblock \emph{\bibinfo{journal}{IEEE Trans. Inf. Theory}}
  \textbf{\bibinfo{volume}{50}}, \bibinfo{pages}{1295--1302}
  (\bibinfo{year}{2004}).

\bibitem{stern_1989}
\bibinfo{author}{Stern, J.}
\newblock \bibinfo{title}{A method for finding codewords of small weight}.
\newblock In \emph{\bibinfo{booktitle}{Coding Theory and Applications}}, vol.
  \bibinfo{volume}{388} of \emph{\bibinfo{series}{Lecture Notes in Computer
  Science}}, \bibinfo{pages}{106--113} (\bibinfo{publisher}{Springer},
  \bibinfo{year}{1989}).

\bibitem{strikis_2026_leftright}
\bibinfo{author}{Strikis, A.}, \bibinfo{author}{Browne, D.~E.} \&
  \bibinfo{author}{Beverland, M.~E.}
\newblock \bibinfo{title}{{High-performance syndrome extraction circuits for
  quantum codes}}.
\newblock Preprint at \url{https://arxiv.org/abs/2603.05481}
  (\bibinfo{year}{2026}).

\bibitem{liang_2026_amplified}
\bibinfo{author}{Liang, Z.}, \bibinfo{author}{Gu, B.}, \bibinfo{author}{Chen,
  Y.-A.}, \bibinfo{author}{Eisert, J.} \& \bibinfo{author}{Wang, Z.}
\newblock \bibinfo{title}{Ultra-high-distance quantum memories from amplified
  {qLDPC} codes}.
\newblock Preprint at \url{https://arxiv.org/abs/2609.37231}
  (\bibinfo{year}{2026}).

\bibitem{qubitserf_code}
\bibinfo{title}{{Qubitserf}} (\bibinfo{year}{2026}).
\newblock \bibinfo{note}{Version 0.1.0 (commit ab96b9c);
  \href{https://github.com/Quantinuum/Qubitserf}{github.com/Quantinuum/Qubitserf}}.

\bibitem{huangfu_2018_highs}
\bibinfo{author}{Huangfu, Q.} \& \bibinfo{author}{Hall, J. A.~J.}
\newblock \bibinfo{title}{{Parallelizing the dual revised simplex method}}.
\newblock \emph{\bibinfo{journal}{Math. Program. Comput.}}
  \textbf{\bibinfo{volume}{10}}, \bibinfo{pages}{119--142}
  (\bibinfo{year}{2018}).

\bibitem{cpsat_code}
\bibinfo{author}{Perron, L.} \& \bibinfo{author}{Didier, F.}
\newblock \bibinfo{title}{{CP-SAT}} (\bibinfo{year}{2026}).
\newblock \bibinfo{note}{Google OR-Tools, version 9.15;
  \href{https://developers.google.com/optimization/cp/cp_solver}{developers.google.com/optimization}}.

\bibitem{gurobi_manual}
\bibinfo{author}{{Gurobi Optimization, LLC}}.
\newblock \bibinfo{title}{{Gurobi Optimizer Reference Manual}}
  (\bibinfo{year}{2026}).
\newblock \bibinfo{note}{Version 13.0.3;
  \href{https://www.gurobi.com}{www.gurobi.com}}.

\bibitem{biere_2024_cadical}
\bibinfo{author}{Biere, A.} \emph{et~al.}
\newblock \bibinfo{title}{{CaDiCaL} 2.0}.
\newblock In \emph{\bibinfo{booktitle}{Computer Aided Verification (CAV
  2024)}}, vol. \bibinfo{volume}{14681} of \emph{\bibinfo{series}{Lecture Notes
  in Computer Science}}, \bibinfo{pages}{133--152}
  (\bibinfo{publisher}{Springer}, \bibinfo{year}{2024}).

\bibitem{ignatiev_2018_pysat}
\bibinfo{author}{Ignatiev, A.}, \bibinfo{author}{Morgado, A.} \&
  \bibinfo{author}{Marques-Silva, J.}
\newblock \bibinfo{title}{{PySAT}: A {Python} toolkit for prototyping with
  {SAT} oracles}.
\newblock In \emph{\bibinfo{booktitle}{Theory and Applications of
  Satisfiability Testing (SAT 2018)}}, vol. \bibinfo{volume}{10929} of
  \emph{\bibinfo{series}{Lecture Notes in Computer Science}},
  \bibinfo{pages}{428--437} (\bibinfo{publisher}{Springer},
  \bibinfo{year}{2018}).

\bibitem{ignatiev_2019_rc2}
\bibinfo{author}{Ignatiev, A.}, \bibinfo{author}{Morgado, A.} \&
  \bibinfo{author}{Marques-Silva, J.}
\newblock \bibinfo{title}{{RC2}: an efficient {MaxSAT} solver}.
\newblock \emph{\bibinfo{journal}{J. Satisf. Boolean Model. Comput.}}
  \textbf{\bibinfo{volume}{11}}, \bibinfo{pages}{53--64}
  (\bibinfo{year}{2019}).

\bibitem{maxsat_2023}
\bibinfo{title}{{MaxSAT Evaluation 2023}} (\bibinfo{year}{2023}).
\newblock \bibinfo{note}{Solver CASHWMaxSAT-CorePlus;
  \href{https://maxsat-evaluations.github.io/2023/}{maxsat-evaluations.github.io/2023}}.

\bibitem{enmaxcdcl_code}
\bibinfo{title}{{enMaxCDCL}} (\bibinfo{year}{2026}).
\newblock
  \bibinfo{note}{\href{https://github.com/zjl9959/enMaxCDCL_CP26}{github.com/zjl9959/enMaxCDCL\_CP26}}.

\bibitem{higgott_2026_hyperbolic}
\bibinfo{author}{Higgott, O.} \emph{et~al.}
\newblock \bibinfo{title}{{Low-overhead quantum error correction with
  boundary-connected planar modules}}.
\newblock Preprint at \url{https://arxiv.org/abs/2610.03682}
  (\bibinfo{year}{2026}).

\bibitem{bhardwaj_2026_mitten}
\bibinfo{author}{Bhardwaj, A.} \emph{et~al.}
\newblock \bibinfo{title}{{High-rate qLDPC processors}}.
\newblock Preprint at \url{https://arxiv.org/abs/2607.28795}
  (\bibinfo{year}{2026}).

\bibitem{lu_2026_cornucopia}
\bibinfo{author}{Lu, Z.}, \bibinfo{author}{Li, W.} \& \bibinfo{author}{Deng,
  D.-L.}
\newblock \bibinfo{title}{{Quantum error correction at ultra-low overhead}}.
\newblock Preprint at \url{https://arxiv.org/abs/2608.02773}
  (\bibinfo{year}{2026}).

\bibitem{zhang_2026_routing}
\bibinfo{author}{Zhang, J.} \emph{et~al.}
\newblock \bibinfo{title}{{Routing codes: high-rate quantum LDPC codes with
  short, parallel non-local connectivity}}.
\newblock Preprint at \url{https://arxiv.org/abs/2606.25330}
  (\bibinfo{year}{2026}).

\bibitem{kim_2026_shift}
\bibinfo{author}{Kim, Y.}, \bibinfo{author}{Gicev, S.},
  \bibinfo{author}{Sevior, M.} \& \bibinfo{author}{Usman, M.}
\newblock \bibinfo{title}{{Time-dynamic circuits for fault-tolerant shift
  automorphisms in quantum LDPC codes}}.
\newblock Preprint at \url{https://arxiv.org/abs/2601.09911}
  (\bibinfo{year}{2026}).

\bibitem{tripier_2026_walkingcat}
\bibinfo{author}{Tripier, F.} \emph{et~al.}
\newblock \bibinfo{title}{{Fault-tolerant quantum computing with trapped ions:
  the Walking Cat architecture}}.
\newblock Preprint at \url{https://arxiv.org/abs/2604.19481}
  (\bibinfo{year}{2026}).

\bibitem{nixon_2026_vine}
\bibinfo{author}{Nixon, G.~M.}, \bibinfo{author}{McLauchlan, C.~K.} \&
  \bibinfo{author}{van Rest, C. C.~L.}
\newblock \bibinfo{title}{Vine codes: low-overhead quantum {LDPC} codes on a
  planar square grid}.
\newblock Preprint at \url{https://arxiv.org/abs/2606.20263}
  (\bibinfo{year}{2026}).

\bibitem{steffan_2025_tile}
\bibinfo{author}{Steffan, V.}, \bibinfo{author}{Choe, S.~H.},
  \bibinfo{author}{Breuckmann, N.~P.}, \bibinfo{author}{Pereira, F. R.~F.} \&
  \bibinfo{author}{Eberhardt, J.~N.}
\newblock \bibinfo{title}{Tile codes: high-efficiency quantum codes on a
  lattice with boundary}.
\newblock \emph{\bibinfo{journal}{Phys. Rev. Lett.}}
  \textbf{\bibinfo{volume}{135}}, \bibinfo{pages}{170601}
  (\bibinfo{year}{2025}).

\bibitem{gu_2026_planar}
\bibinfo{author}{Gu, B.} \emph{et~al.}
\newblock \bibinfo{title}{Nearest-neighbour gates are all you need: high-rate
  quantum low-density parity-check codes on a planar grid}.
\newblock Preprint at \url{https://arxiv.org/abs/2606.19482}
  (\bibinfo{year}{2026}).

\bibitem{mian_2026_multicycle}
\bibinfo{author}{Mian, F.~A.}, \bibinfo{author}{Gwilliam, O.} \&
  \bibinfo{author}{Krastanov, S.}
\newblock \bibinfo{title}{{Multivariate multicycle codes for complete
  single-shot decoding}}.
\newblock Preprint at \url{https://arxiv.org/abs/2601.18879}
  (\bibinfo{year}{2026}).

\bibitem{lu_2026_weight8}
\bibinfo{author}{Lu, L.}, \bibinfo{author}{Yang, R.} \& \bibinfo{author}{Guo,
  G.}
\newblock \bibinfo{title}{{Quantum bivariate bicycle codes with weight-8 checks
  surpassing the BB benchmark}}.
\newblock Preprint at \url{https://arxiv.org/abs/2609.06572}
  (\bibinfo{year}{2026}).

\bibitem{tesseract_code}
\bibinfo{title}{tesseract-decoder} (\bibinfo{year}{2025}).
\newblock \bibinfo{note}{Commit 4fd104d;
  \href{https://github.com/quantumlib/tesseract-decoder}{github.com/quantumlib/tesseract-decoder}}.

\bibitem{lu_2026_modular}
\bibinfo{author}{Lu, Z.} \emph{et~al.}
\newblock \bibinfo{title}{A modular quantum computer based on {Cornucopia}
  codes}.
\newblock Preprint at \url{https://arxiv.org/abs/2610.19073}
  (\bibinfo{year}{2026}).

\bibitem{swaroop_2024_adapters}
\bibinfo{author}{Swaroop, E.}, \bibinfo{author}{Jochym-O'Connor, T.} \&
  \bibinfo{author}{Yoder, T.~J.}
\newblock \bibinfo{title}{Universal adapters between quantum low-density parity
  check codes}.
\newblock \emph{\bibinfo{journal}{PRX Quantum}} \textbf{\bibinfo{volume}{7}},
  \bibinfo{pages}{010324} (\bibinfo{year}{2026}).

\bibitem{cowtan_2024_ssip}
\bibinfo{author}{Cowtan, A.}
\newblock \bibinfo{title}{{SSIP}: automated surgery with quantum {LDPC} codes}.
\newblock Preprint at \url{https://arxiv.org/abs/2407.09423}
  (\bibinfo{year}{2024}).

\bibitem{liang_2025_selfdual}
\bibinfo{author}{Liang, Z.} \& \bibinfo{author}{Chen, Y.-A.}
\newblock \bibinfo{title}{{Self-dual bivariate bicycle codes with transversal
  Clifford gates}}.
\newblock \emph{\bibinfo{journal}{npj Quantum Inf.}}  (\bibinfo{year}{2026}).

\bibitem{yang_2026_gala}
\bibinfo{author}{Yang, W.}, \bibinfo{author}{Duckering, C.} \&
  \bibinfo{author}{Dua, A.}
\newblock \bibinfo{title}{{Designer codes from GALA: compact, self-dual, and
  rate-1/2 QEC on reconfigurable atom arrays}}.
\newblock Preprint at \url{https://arxiv.org/abs/2608.07431}
  (\bibinfo{year}{2026}).

\bibitem{beverland_2025_failfast}
\bibinfo{author}{Beverland, M.~E.}, \bibinfo{author}{Carroll, M.},
  \bibinfo{author}{Cross, A.~W.} \& \bibinfo{author}{Yoder, T.~J.}
\newblock \bibinfo{title}{Fail fast: techniques to probe rare events in quantum
  error correction}.
\newblock Preprint at \url{https://arxiv.org/abs/2511.15177}
  (\bibinfo{year}{2025}).

\bibitem{goto_2024_hypercube}
\bibinfo{author}{Goto, H.}
\newblock \bibinfo{title}{High-performance fault-tolerant quantum computing
  with many-hypercube codes}.
\newblock \emph{\bibinfo{journal}{Science Advances}}
  \textbf{\bibinfo{volume}{10}}, \bibinfo{pages}{eadp6388}
  (\bibinfo{year}{2024}).

\bibitem{cruzbenito_2026_llm}
\bibinfo{author}{Cruz-Benito, J.}, \bibinfo{author}{Cross, A.~W.},
  \bibinfo{author}{Kremer, D.} \& \bibinfo{author}{Faro, I.}
\newblock \bibinfo{title}{{Evolutionary discovery of bivariate bicycle codes
  with LLM-guided search}}.
\newblock Preprint at \url{https://arxiv.org/abs/2606.02418}
  (\bibinfo{year}{2026}).

\bibitem{mceliece_1978}
\bibinfo{author}{McEliece, R.~J.}
\newblock \bibinfo{title}{A public-key cryptosystem based on algebraic coding
  theory} (\bibinfo{year}{1978}).
\newblock \bibinfo{note}{DSN Progress Report 42--44, 114--116, Jet Propulsion
  Laboratory}.

\bibitem{nist_ir_8545}
\bibinfo{author}{Alagic, G.} \emph{et~al.}
\newblock \bibinfo{title}{Status report on the fourth round of the {NIST}
  post-quantum cryptography standardization process} (\bibinfo{year}{2025}).
\newblock \bibinfo{note}{NIST IR 8545, National Institute of Standards and
  Technology}.

\bibitem{oum_2005_rankwidth}
\bibinfo{author}{Oum, S.}
\newblock \bibinfo{title}{{Rank-width and vertex-minors}}.
\newblock \emph{\bibinfo{journal}{J. Comb. Theory Ser. B}}
  \textbf{\bibinfo{volume}{95}}, \bibinfo{pages}{79--100}
  (\bibinfo{year}{2005}).

\bibitem{ollivier_2006_trellis}
\bibinfo{author}{Ollivier, H.} \& \bibinfo{author}{Tillich, J.-P.}
\newblock \bibinfo{title}{Trellises for stabilizer codes: definition and uses}.
\newblock \emph{\bibinfo{journal}{Phys. Rev. A}} \textbf{\bibinfo{volume}{74}},
  \bibinfo{pages}{032304} (\bibinfo{year}{2006}).

\bibitem{ott_2025_decisiontree}
\bibinfo{author}{Ott, K.~R.}, \bibinfo{author}{Het{\'e}nyi, B.} \&
  \bibinfo{author}{Beverland, M.~E.}
\newblock \bibinfo{title}{Decision-tree decoders for general quantum {LDPC}
  codes}.
\newblock Preprint at \url{https://arxiv.org/abs/2502.16408}
  (\bibinfo{year}{2025}).

\bibitem{korf_1985}
\bibinfo{author}{Korf, R.~E.}
\newblock \bibinfo{title}{Depth-first iterative-deepening: an optimal
  admissible tree search}.
\newblock \emph{\bibinfo{journal}{Artif. Intell.}}
  \textbf{\bibinfo{volume}{27}}, \bibinfo{pages}{97--109}
  (\bibinfo{year}{1985}).

\bibitem{culberson_1998_patterns}
\bibinfo{author}{Culberson, J.~C.} \& \bibinfo{author}{Schaeffer, J.}
\newblock \bibinfo{title}{Pattern databases}.
\newblock \emph{\bibinfo{journal}{Comput. Intell.}}
  \textbf{\bibinfo{volume}{14}}, \bibinfo{pages}{318--334}
  (\bibinfo{year}{1998}).

\bibitem{circuit_distance_code}
\bibinfo{author}{Pan, F.}
\newblock \bibinfo{title}{{tropidist: certified code and circuit distances}}
  (\bibinfo{year}{2026}).
\newblock
  \bibinfo{note}{\href{https://github.com/Fanerst/tropidist}{github.com/Fanerst/tropidist}}.

\end{thebibliography}

\clearpage
\setcounter{section}{0}
\setcounter{figure}{0}
\setcounter{table}{0}
\setcounter{equation}{0}
\renewcommand{\figurename}{Supplementary Fig.}
\renewcommand{\tablename}{Supplementary Table}
\renewcommand{\theequation}{S\arabic{equation}}
\renewcommand{\theHequation}{S\arabic{equation}}
\renewcommand{\theHfigure}{S\arabic{figure}}
\renewcommand{\theHtable}{S\arabic{table}}
\begin{bibunit}[naturemag_ncs]
\makeatletter  %
\let\SI@natlinkstart\hyper@natlinkstart\def\hyper@natlinkstart#1{\SI@natlinkstart{S#1}}
\let\SI@natanchorstart\hyper@natanchorstart\def\hyper@natanchorstart#1{\SI@natanchorstart{S#1}}
\makeatother
\renewcommand{\citenumfont}[1]{S#1}
\renewcommand{\bibnumfmt}[1]{S#1.}
\renewcommand{\refname}{Supplementary References}

{\raggedright\sffamily
{\fontsize{20}{24}\selectfont\bfseries Supplementary Information\par}\vspace{6pt}
{\large Certifying quantum code and circuit distances at scale\par}\vspace{10pt}}
\setcounter{tocdepth}{1}
\tableofcontents

\section*{Overview}

Sections~\ref{sec:preliminaries} to~\ref{sec:weight-search} give the theory behind the method of the main text, with proofs: the distance problems and their tropical formulation (\cref{sec:preliminaries,sec:distance}), the dense Rank DP (\cref{sec:tropical-rank-dp}), the weight-truncated Rank DP (\cref{sec:sparse-dp}), and the weight-bounded search with its relaxations, lifting and certificates (\cref{sec:weight-search}).
Sections~\ref{sec:benchmark} to~\ref{sec:software} describe the benchmark (\cref{sec:benchmark}), the audit of published distances (\cref{sec:audit}), the design studies (\cref{sec:design}), the memory instruction of IBM's bicycle architecture (\cref{sec:ibm}), codes of logical measurements (\cref{sec:surgery}), chips with dead checks (\cref{sec:chips}) and the software (\cref{sec:software}).
Throughout, a distance is certified when a lower bound proved by a complete evaluator meets the weight of an
explicit error that has been checked against the matrices.

\section{Preliminaries}
\label{sec:preliminaries}

For a positive integer $\ell$, write $[\ell]:=\{1,\ldots,\ell\}$.
For a binary vector $\vb{e}\in\FF_2^\ell$, its Hamming weight $|\vb{e}|$ is the number of its nonzero entries.
For a matrix $\vb{M}$, its image $\im\vb{M}$ is the space spanned by its columns.
Vectors are column vectors unless stated otherwise.

\subsection{Stabilizer formalism}
\label{subsec:stabilizer-binary}

An $n$-qubit Pauli operator is a tensor product of single-qubit Pauli operators, with an overall phase.
We write $f=cf_1\otimes\cdots\otimes f_n$, where $f_i\in\{I,X,Y,Z\}$ and $c\in\{\pm1,\pm\mathrm{i}\}$, and denote their group by $\sfP_n$.
Here, $I$ is the identity, while $X$ and $Z$ produce bit and phase flips, respectively.
The support of $f$ is $\supp(f):=\{i:f_i\ne I\}$, the set of qubits on which it acts nontrivially.
Its Pauli weight is $\wt_{\sfP}(f):=|\supp(f)|$.

Let $\sfS=\langle g_1,\ldots,g_{n-k}\rangle\subseteq\sfP_n$ be an Abelian group with $n-k$ independent generators and $-I\notin\sfS$.
The associated stabilizer code is their common $+1$ eigenspace $\calQ=\{\ket{\psi}\in(\CC^2)^{\otimes n}:g_a\ket{\psi}=\ket{\psi}\text{ for }a=1,\ldots,n-k\}$~\cite{gottesman_1997_stabilizer}.
This space has dimension $2^k$ and encodes $k$ logical qubits in $n$ physical qubits.
We denote its parameters by $\qc{n,k}$.

Syndrome measurements test the stabilizer constraints.
After a Pauli error $f$ acts on a code state, measurement of $g_a$ gives $-1$ if $f$ anticommutes with $g_a$, and $+1$ otherwise.
These signs form the error syndrome, with all $+1$ outcomes corresponding to zero syndrome.
The normalizer $\sfN(\sfS)$ consists of Pauli operators that commute with every stabilizer.
Its elements preserve the code space and have zero syndrome.
Stabilizers act trivially on code states, whereas normalizer elements not proportional to a stabilizer implement nontrivial logical Pauli operations.

We now express these operator relations using binary vectors.
Ignoring an overall phase, represent $f$ by $\vb{e}=(\vb{x}^{\sfT},\vb{z}^{\sfT})^{\sfT}\in\FF_2^{2n}$ through $E(\vb{e})=\bigotimes_{i=1}^n X^{x_i}Z^{z_i}$.
The local pairs $(0,0),(1,0),(0,1),(1,1)$ represent $I,X,Z,Y$, respectively.
Pauli multiplication becomes vector addition over $\FF_2$.
Two operators commute exactly when their symplectic product $\vb{e}^{\sfT}\vb{\Lambda}\vb{e}'$ is zero, where $\vb{\Lambda}:=\left(\begin{smallmatrix}\vb{0}&\vb{I}_n\\\vb{I}_n&\vb{0}\end{smallmatrix}\right)$.
The weight notation carries over as $|\vb{e}|_{\sfP}:=\wt_{\sfP}(E(\vb{e}))=\sum_{i=1}^n\mathbf{1}[(x_i,z_i)\ne(0,0)]$.

Let $\vb{H}\in\FF_2^{(n-k)\times2n}$ be the stabilizer tableau, with the transposed binary vectors of $g_1,\ldots,g_{n-k}$ as its rows.
Independence and commutation give $\rank\vb{H}=n-k$ and $\vb{H}\vb{\Lambda}\vb{H}^{\sfT}=\vb{0}$.
Define the syndrome matrix $\vb{M}_s:=\vb{H}\vb{\Lambda}$.
The stabilizer group and its normalizer correspond to the binary spaces $S:=\rowsp(\vb{H})$ and $N:=S^{\perp_s}=\ker\vb{M}_s$, respectively, where $S^{\perp_s}$ denotes the symplectic orthogonal complement.
We have $S\subseteq N$, with dimensions $n-k$ and $n+k$.
Writing the measured signs as $(-1)^{s_a}$ gives the binary syndrome $\vb{s}=\vb{M}_s\vb{e}$.
Thus, $N\setminus S$ represents the nontrivial logical Paulis whose minimum weight defines the code distance $d_{\mathrm{code}}:=\min_{\vb{e}\in N\setminus S}|\vb{e}|_{\sfP}$ for $k\geq1$.

\subsection{Rank decomposition, rank-width, and subspace overlap}
\label{subsec:rank-width}

A rank decomposition divides a graph into parts whose interaction is measured by cut rank~\cite{oum_2005_rankwidth,oum_seymour_2006_approximating}.
Let $G=(V,E)$ be a finite simple undirected graph with adjacency matrix $\vb{A}_G$ over $\FF_2$.
For a cut $Y\mid(V\setminus Y)$, define the cut rank $\rho_G(Y):=\rank_{\FF_2}\vb{A}_G[Y,V\setminus Y]$, which counts the independent neighborhood vectors across the cut.
Symmetry of the adjacency matrix gives $\rho_G(Y)=\rho_G(V\setminus Y)$.

We define the decomposition directly for a partition $\calP$ of $V$ into nonempty blocks.
A rank decomposition over $\calP$ is a rooted binary tree $\calT$, in which every internal node has two children, together with a bijection $\mu:\leaf(\calT)\to\calP$.
Let $Y_u$ be the union of the blocks below a node $u$.
The width of this decomposition is $\omega(\calT,\mu):=\max_{u\in V(\calT)}\rho_G(Y_u)$.
The rank-width of $G$ with respect to $\calP$ is $\rw(G;\calP):=\min_{(\calT,\mu)}\omega(\calT,\mu)$.
Thus, every cut keeps each block intact.
Singleton blocks give the usual rank-width $\rw(G)$.
A partition with one block has width zero.

Binary constraints give an algebraic description of the same interaction.
Let $\vb{M}\in\FF_2^{m\times \ell}$, where $\ell\geq1$.
For $Y\subseteq[\ell]$, let $\vb{M}_Y$ contain the columns in $Y$.
If $\vb{M}\vb{e}=\vb{0}$, the contribution $\vb{M}_Y\vb{e}_Y$ must be canceled by the remaining coordinates.
The possible shared contributions form the boundary space $I_Y:=\im\vb{M}_Y\cap\im\vb{M}_{[\ell]\setminus Y}$.
Since the two column spaces sum to $\im\vb{M}$, we have $\dim I_Y=\rank\vb{M}_Y+\rank\vb{M}_{[\ell]\setminus Y}-\rank\vb{M}$.
This construction recovers the boundary spaces of a trellis when $Y$ ranges over prefixes of a linear order of the coordinates.
More generally, taking $Y$ to be the coordinates below each node of a decomposition tree gives the boundary spaces used in branch decompositions of subspace arrangements~\cite{jeong_2017_trellis}.

The graph representation follows from a column basis of $\vb{M}$.
The Tanner graph $T(\vb{M})$ of $\vb{M}$ has one variable node for each column and one check node for each row.
The check node for row $i$ is connected to the variable node for column $j$ if and only if the $(i,j)$-entry is $1$.
Choose a column basis $B_0\subseteq[\ell]$ and define $\vb{A}$ by $\vb{M}_{[\ell]\setminus B_0}=\vb{M}_{B_0}\vb{A}$.
With the basis columns ordered first, expressing the columns in this basis gives the \emph{systematic representation} $(\vb{I}_{B_0}\mid\vb{A})$.
The reduced Tanner graph $G:=T(\vb{A})$ has bipartition $(B_0,[\ell]\setminus B_0)$, with the original column indices as vertices.
With the vertices ordered by these two parts, its adjacency matrix is
\begin{align}
    \vb{A}_G=
    \begin{pmatrix}
        \vb{0} & \vb{A}\\
        \vb{A}^{\sfT} & \vb{0}
    \end{pmatrix}.
    \label{eq:reduced-tanner-adjacency}
\end{align}
The graph $G$ is a fundamental graph of the binary matroid represented by $\vb{M}$.
Oum's fundamental-graph relation gives the following identity (ref.~\citenum{oum_2005_rankwidth}, Proposition~3.1).\footnote{The matroid connectivity function in ref.~\citenum{oum_2005_rankwidth} includes an additive $1$; the overlap dimension $\dim I_Y$ does not.}

\begin{proposition}
\label{prop:overlap-cutrank}
For every $Y\subseteq[\ell]$, $\rho_G(Y)=\dim I_Y$.
\end{proposition}

The identity holds for every column basis and every union of coordinate blocks.
In particular, for a partition $\calB$ of the columns into blocks, the arrangement $(\im\vb{M}_B)_{B\in\calB}$ and the reduced Tanner graph have the same cut values on a block tree.
Its branch-width, with the overlap convention of ref.~\citenum{jeong_2017_trellis}, equals $\rw(G;\calB)$.
The graph description specifies the width, while the boundary spaces provide coordinates for the dynamic program.

\section{Code and circuit distance}
\label{sec:distance}

Code distance measures the smallest support of a nontrivial logical Pauli.
Circuit distance measures the smallest number of specified faults that produce an undetected logical error.
We formulate both problems using binary linear constraints, Hamming or Pauli weight, and a nonzero logical label.

\subsection{Code distance}
\label{subsec:minimum-distance}

Let $\calQ$ be an $\qc{n,k}$ stabilizer code with $k\geq1$.
Group each qubit's coordinates into $B_i:=\{i,n+i\}$.
Set $\calB:=\{B_i:i\in[n]\}$.

Let $\vb{P}\in\FF_2^{(n+k)\times2n}$ be a full-row-rank matrix such that $\ker\vb{P}=S$.
For $\vb{e}\in N$, $\vb{P}\vb{e}$ is nonzero when $E(\vb{e})$ is a nontrivial logical Pauli, i.e., when $\vb{e} \not\in S$.
Since $N=\ker\vb{M}_s$, the constraint $\vb{e} \in N \setminus S$ in the distance definition gives
\begin{align}
    d_{\mathrm{code}}
    =\min_{\substack{\vb{M}_s\vb{e}=\vb{0}\\\vb{P}\vb{e}\ne\vb{0}}}
        |\vb{e}|_{\sfP}.
    \label{eq:distance-constrained}
\end{align}
The first constraint imposes zero syndrome.
The second excludes every stabilizer, including the identity.
A code of distance $d_{\mathrm{code}}$ can detect any error acting on at most $d_{\mathrm{code}}-1$ physical qubits.

For CSS codes, the minimization splits into two smaller problems with unit costs.
Let the $X$-type and $Z$-type generators have binary matrices $\vb{H}_X\in\FF_2^{m_X\times n}$ and $\vb{H}_Z\in\FF_2^{m_Z\times n}$, so that $\vb{H}_X\vb{H}_Z^{\sfT}=\vb{0}$.
A $Z$-type Pauli $E(\vb{0},\vb{z})$ has zero syndrome when $\vb{H}_X\vb{z}=\vb{0}$, and it is a stabilizer when $\vb{z}\in\rowsp(\vb{H}_Z)$.
Let $\vb{L}_X\in\FF_2^{k\times n}$ satisfy $\ker\vb{H}_X\cap\ker\vb{L}_X=\rowsp(\vb{H}_Z)$.
A basis of the $X$-type logical operators, with one row for each logical qubit, has this property.
Define
\begin{align}
    d_Z:=\min_{\substack{\vb{z}\in\FF_2^n\\\vb{H}_X\vb{z}=\vb{0},\;\vb{L}_X\vb{z}\ne\vb{0}}}|\vb{z}|,
    \label{eq:css-distance}
\end{align}
and define $d_X$ in the same way with $X$ and $Z$ exchanged.

\begin{proposition}[CSS distance]
\label{prop:css-distance}
For a CSS code with $k\geq1$, $d_{\mathrm{code}}=\min\{d_X,d_Z\}$.
\end{proposition}

\begin{proof}
A nontrivial logical Pauli of $X$ or $Z$ type lies in $N\setminus S$, and its Pauli weight is the Hamming weight of its nonzero part.
Hence, $d_{\mathrm{code}}\leq\min\{d_X,d_Z\}$.
Conversely, let $\vb{e}=(\vb{x}^{\sfT},\vb{z}^{\sfT})^{\sfT}\in N\setminus S$.
Commutation with the $Z$-type and $X$-type generators gives $\vb{H}_Z\vb{x}=\vb{0}$ and $\vb{H}_X\vb{z}=\vb{0}$, so $E(\vb{x},\vb{0})$ and $E(\vb{0},\vb{z})$ lie in the normalizer.
In these coordinates, $S=\rowsp(\vb{H}_X)\times\rowsp(\vb{H}_Z)$, so at least one of the two is not a stabilizer.
Its weight is at most $|\vb{e}|_{\sfP}$ and at least $d_X$ or $d_Z$, respectively.
\end{proof}

The feasible set $\ker\vb{H}_X\setminus\rowsp(\vb{H}_Z)$ does not depend on the choice of $\vb{L}_X$.
The choice can still affect solvers that encode each row of $\vb{L}_X$ as a separate constraint.

\subsection{Circuit distance}
\label{subsec:circuit-distance}

Circuit-level noise includes faults in the gates, resets, and measurements used for syndrome extraction.
These faults can affect the data and corrupt the measurement outcomes.
We consider a fixed syndrome extraction circuit with Clifford gates and Pauli measurements, subject to a specified local Pauli fault model.

A \emph{detector} checks a parity of measurement outcomes that is deterministic in the absence of faults~\cite{mcewen_2023_relaxing}.
For example, two consecutive measurements of a stabilizer with unchanged ideal value give equal outcomes without faults, so their binary parity is zero.
A detection event occurs when a detector parity differs from its noiseless value.
The circuit also specifies measured logical observables as parities of measurement outcomes, such as a final logical $Z$ readout.

An undecomposed detector error model lists error mechanisms by the detector and logical-observable flips produced by allowed single-location faults.
Mechanisms with equal detector and observable effects are merged.
Let $\ell$ be the number of resulting mechanisms, and let $\vb{e}\in\FF_2^\ell$ indicate which mechanisms are selected.

Let $m$ be the number of detectors and $q$ the number of specified logical observables.
Define the detector matrix $\vb{D}\in\FF_2^{m\times \ell}$ by $D_{aj}=1$ if mechanism $j$ flips detector $a$, and $D_{aj}=0$ otherwise.
Define the observable matrix $\vb{O}\in\FF_2^{q\times \ell}$ by $O_{bj}=1$ if mechanism $j$ flips logical observable $b$, and $O_{bj}=0$ otherwise.
Their columns are obtained by propagating the corresponding single-location faults through the Clifford circuit~\cite{stim}.
Since these effects add over $\FF_2$, the detection-event vector $\vb{s}$ and logical-flip vector $\vb*{\eta}$ satisfy
\begin{align}
    \vb{s}=\vb{D}\vb{e}\in\FF_2^m,\qquad
    \vb*{\eta}=\vb{O}\vb{e}\in\FF_2^q.
    \label{eq:circuit-maps}
\end{align}

The Hamming weight $|\vb{e}|$ counts the selected mechanisms.
In particular, a mechanism produced by any allowed nonidentity two-qubit Pauli at one gate has weight one.

An undetected logical error leaves every detector unchanged but flips at least one specified logical observable.
Its minimum mechanism count defines the circuit distance
\begin{align}
    d_{\mathrm{circ}}
    :=\min_{\substack{\vb{D}\vb{e}=\vb{0}\\\vb{O}\vb{e}\ne\vb{0}}}
        |\vb{e}|.
    \label{eq:circuit-distance}
\end{align}
The minimum is $+\infty$ if no set of mechanisms produces an undetected logical error.
If only one logical observable is specified, this definition concerns undetected flips of that observable.

The following condition identifies $d_{\mathrm{circ}}$ with the minimum number of faulty circuit locations.
It holds for depolarizing channels, for single Pauli flips, and for measurement outcome flips.

\begin{lemma}[Mechanisms and locations]\label{lem:mechanism-count}
Suppose the allowed faults at each circuit location include the no-fault event and are closed under composition up to Pauli phase. If every allowed nonidentity fault is represented in the detector error model, with mechanisms having identical detector and logical-observable effects merged, then $d_{\mathrm{circ}}$ equals the minimum number of faulty locations that produce an undetected logical error.
\end{lemma}

\begin{proof}
Any configuration with $r$ faulty locations produces the same detector and logical effects as at most $r$ merged mechanisms, since mechanisms with identical effects cancel in pairs over $\mathbb F_2$. Hence, $d_{\mathrm{circ}}$ cannot exceed the minimum number of faulty locations.

Conversely, each selected mechanism can be realized by an allowed fault at some location. If several selected mechanisms are realized at the same location, their faults can be combined into a single allowed fault by the closure assumption. This preserves the detector and logical effects and uses no more faulty locations than selected mechanisms, giving the reverse inequality.
\end{proof}

\subsection{Unified tropical formulation}
\label{subsec:tropical-partition}

We reduce code and circuit distance to evaluations of a tropical partition function.
The distance objectives in \cref{eq:distance-constrained,eq:circuit-distance} require zero syndrome and a nonzero logical label.
Here, $\vb{M}\in\FF_2^{m\times \ell}$ imposes $m$ linear constraints, and a label matrix $\vb{R}\in\FF_2^{q\times \ell}$ specifies $q$ binary label components.
Let $\calB$ be a partition of $[\ell]$ into blocks of size at most two.
We count the blocks on which $\vb{e}$ is nonzero.
The common distance objective is
\begin{align}
    \Delta(\vb{M},\vb{R};\calB)
    :=\min_{\substack{\vb{e}\in\FF_2^\ell\\
                      \vb{M}\vb{e}=\vb{0},\;\vb{R}\vb{e}\ne\vb{0}}}
        \sum_{B\in\calB}\mathbf{1}[\vb{e}_B\ne\vb{0}].
    \label{eq:unified-distance}
\end{align}
Here, $\vb{e}_B$ denotes the restriction to $B$, and the minimum of the empty set is $+\infty$.
For the physical-qubit blocks $B_i=\{i,n+i\}$, the sum is $|\vb{e}|_{\sfP}$, so \cref{eq:distance-constrained} gives $d_{\mathrm{code}}=\Delta(\vb{M}_s,\vb{P};\calB)$.
For singleton blocks, the sum is $|\vb{e}|$, so \cref{eq:circuit-distance} gives $d_{\mathrm{circ}}=\Delta(\vb{D},\vb{O};\calB)$.
For a CSS code, \cref{eq:css-distance} gives $d_Z=\Delta(\vb{H}_X,\vb{L}_X;\calB)$ with singleton blocks; similarly, $d_X=\Delta(\vb{H}_Z,\vb{L}_Z;\calB)$.
We omit $\calB$ from $\Delta$ when it is the singleton partition, as in the main text.

Let $\TT:=\RR_{\geq0}\cup\{+\infty\}$ be the tropical semiring with operations $a\oplus b:=\min\{a,b\}$ and $a\otimes b:=a+b$.
The additive and multiplicative identities are $+\infty$ and $0$, respectively.

\begin{definition}[Tropical partition function]
\label{def:tropical-partition}
Let $\vb{M}\in\FF_2^{m\times \ell}$, let $\vb{t}\in\FF_2^m$, and let $\calB$ be a partition of $[\ell]$ into blocks of size at most two.
Define
\begin{align}
    Z_{\vb{M}}^{\trop}(\vb{t};\calB)
    &:=\bigoplus_{\substack{\vb{e}\in\FF_2^\ell\\\vb{M}\vb{e}=\vb{t}}}
        \bigotimes_{B\in\calB}\mathbf{1}[\vb{e}_B\ne\vb{0}]\nonumber\\
    &=\min_{\substack{\vb{e}\in\FF_2^\ell\\\vb{M}\vb{e}=\vb{t}}}
        \sum_{B\in\calB}\mathbf{1}[\vb{e}_B\ne\vb{0}].
    \label{eq:tropical-partition}
\end{align}
\end{definition}

To reduce $\Delta$ to the tropical partition function in \cref{def:tropical-partition}, write $\vb{r}_a^{\sfT}$ for row $a$ of $\vb{R}$.
The condition $\vb{R}\vb{e}\ne\vb{0}$ holds exactly when $\vb{r}_a^{\sfT}\vb{e}=1$ for at least one $a$.
For each $a$, define
\begin{align}
    \vb{M}^{(a)}:=\begin{pmatrix}\vb{M}\\\vb{r}_a^{\sfT}\end{pmatrix},\qquad
    \vb{t}^{(a)}:=\begin{pmatrix}\vb{0}\\1\end{pmatrix}.
\end{align}
Then,
\begin{align}
    \Delta(\vb{M},\vb{R};\calB)
    =\min_{1\leq a\leq q}
        Z_{\vb{M}^{(a)}}^{\trop}(\vb{t}^{(a)};\calB).
    \label{eq:distance-unguarded-reduction}
\end{align}
Hence, both distance problems reduce to at most $q$ tropical partition function evaluations, with the same partition $\calB$.
Section~\ref{sec:tropical-rank-dp} evaluates these partition functions by Rank DP.

The distance formulations above are minimization problems.
Searching for the minimum by testing cost thresholds leads to the corresponding decision problem.
Given a nonnegative threshold $K$, determine whether there exists $\vb{e}\in\FF_2^\ell$ such that
\begin{align}
    \vb{M}\vb{e}=\vb{0},\quad
    \vb{R}\vb{e}\ne\vb{0},\quad
    \sum_{B\in\calB}\mathbf{1}[\vb{e}_B\ne\vb{0}]\leq K.
    \label{eq:distance-threshold}
\end{align}
The answer is positive exactly when $\Delta(\vb{M},\vb{R};\calB)\leq K$.
The dense Rank DP of \cref{sec:tropical-rank-dp} computes $\Delta(\vb{M},\vb{R};\calB)$ by evaluating the tropical partition functions in \cref{eq:distance-unguarded-reduction}.
For singleton blocks, the weight-truncated Rank DP of \cref{sec:sparse-dp} and the weight-bounded search of \cref{sec:weight-search} decide whether $\Delta(\vb{M},\vb{R})\leq K$.

\section{Tropical Rank DP for distance}
\label{sec:tropical-rank-dp}

\Cref{subsec:tropical-partition} expresses code and circuit distance in terms of tropical partition functions.
We first develop the linear dense Rank DP of the main text.
We then extend the same minimum-cost invariant to rank DP on a general decomposition tree, with singleton or physical-qubit blocks.
In both cases, partial assignments with the same exterior completions are represented by their minimum cost.
We prove the state bounds in terms of cut rank and combine the evaluations through \cref{eq:distance-unguarded-reduction} to obtain code and circuit distance.

\subsection{Linear dense Rank DP}
\label{subsec:dense-rank-dp}

We begin with Hamming weight, as in the main text.
Let $\vb{M}\in\FF_2^{m\times \ell}$ and $\vb{t}\in\FF_2^m$, and fix an order of the $\ell$ columns.
Write $Z_{\vb{M}}^{\trop}(\vb{t})$ for the specialization of \cref{def:tropical-partition} with singleton blocks.
This is the function $Z_{\vb{M}}(\vb{t})$ in the main text.
Gaussian elimination on $(\vb{M}\mid\vb{t})$ first tests consistency; an inconsistent system has value $+\infty$.
Henceforth, assume that the system is consistent.

At the cut after column $j\in\{0,\ldots,\ell\}$, a prefix $\vb{e}_{\le j}\in\FF_2^j$ has partial syndrome $\vb{s}:=\vb{M}_{\leq j}\vb{e}_{\le j}$.
A suffix $\vb{e}_{>j}$ completes it exactly when $\vb{M}_{>j}\vb{e}_{>j}=\vb{t}+\vb{s}$.
Thus, the states of extendible partial assignments form the affine space
\begin{align}
    \Sigma_j:=\im\vb{M}_{\leq j}\cap\bigl(\vb{t}+\im\vb{M}_{>j}\bigr).
    \label{eq:sparse-state}
\end{align}

The dense table stores the least weight for each state:
\begin{align}
    T_j(\vb{s}):=\min\{|\vb{e}_{\le j}|:\vb{e}_{\le j}\in\FF_2^j,\ \vb{M}_{\leq j}\vb{e}_{\le j}=\vb{s}\},
    \qquad\vb{s}\in\Sigma_j.
    \label{eq:sparse-forward}
\end{align}
We use $\min\varnothing=+\infty$ and initialize $T_0(\vb{0})=0$.
Entries outside $\Sigma_j$ have value $+\infty$.
Before processing column $j$, initialize every entry of $T_j$ to $+\infty$.
Let $\vb{m}_j$ be column $j$ of $\vb{M}$.
For each $\vb{s}\in\Sigma_{j-1}$ and choice $e_j\in\{0,1\}$, let $\vb{s}'=\vb{s}+e_j\vb{m}_j$.
If $\vb{s}'\in\Sigma_j$, update
\begin{align}
    T_j(\vb{s}')\gets\min\{T_j(\vb{s}'),\ T_{j-1}(\vb{s})+e_j\}.
    \label{eq:sparse-update}
\end{align}
Otherwise, discard the extension.
The two choices of $e_j$ give the min-plus update \cref{eq:dp} of the Methods.

\begin{lemma}[States of a linear sweep]
\label{lem:sparse-states}
A prefix extends to a solution if and only if its partial syndrome lies in $\Sigma_j$, and every element of $\Sigma_j$ occurs for such a prefix.
Two extendible prefixes have the same exterior completions if and only if they have the same state.
The number of states is $|\Sigma_j|=2^{w_j}$, where
\begin{align}
    w_j:=\rank\vb{M}_{\leq j}+\rank\vb{M}_{>j}-\rank\vb{M}.
    \label{eq:dense-boundary-dimension}
\end{align}
\end{lemma}

\begin{proof}
A prefix with state $\vb{s}$ has a completion when $\vb{t}+\vb{s}\in\im\vb{M}_{>j}$.
Since its state lies in $\im\vb{M}_{\leq j}$, this is equivalent to membership in $\Sigma_j$.
Conversely, both the prefix and a completion exist for every $\vb{s}\in\Sigma_j$ by its definition.
Equal states therefore give equal completion sets.
Distinct states require different suffix syndromes, so their nonempty completion sets are disjoint.

Choose any solution $\vb{e}^{0}$ of $\vb{M}\vb{e}=\vb{t}$ and let $I_j:=\im\vb{M}_{\leq j}\cap\im\vb{M}_{>j}$.
Translating the intersection in \cref{eq:sparse-state} by $\vb{M}_{\leq j}\vb{e}^{0}_{\leq j}$ gives $\Sigma_j=\vb{M}_{\leq j}\vb{e}^{0}_{\leq j}+I_j$.
The dimension formula for the intersection of two subspaces gives $\dim I_j=w_j$, since their sum is $\im\vb{M}$.
Hence, $|\Sigma_j|=2^{w_j}$, and $w_j$ equals the cut rank by \cref{prop:overlap-cutrank}.
\end{proof}

\begin{proposition}[Linear dense Rank DP]
\label{prop:linear-dense-dp}
The preceding initialization and update compute the tables in \cref{eq:sparse-forward} and return $T_\ell(\vb{t})=Z_{\vb{M}}^{\trop}(\vb{t})$.
After preparing the state spaces and transitions, the sweep uses $O(\sum_{j=0}^{\ell}2^{w_j})$ state updates.
Predecessor records recover a minimizing assignment by traceback.
\end{proposition}

\begin{proof}
Every extendible assignment on the first $j$ columns has a unique prefix of length $j-1$ and a last bit $e_j$.
Its prefix is also extendible, and the state test rejects exactly the extensions that have no completion.
By \cref{lem:sparse-states}, prefixes with the same state admit the same extensions, and each extension adds the same cost $e_j$.
Taking the minimum in \cref{eq:sparse-update} therefore preserves the table invariant by induction.
At $j=\ell$, $\Sigma_\ell=\{\vb{t}\}$, so the sole table entry minimizes weight subject to all constraints.
There are $2^{w_j}$ states at cut $j$, and each state produces at most two transitions.
This gives the stated update count.
Recording a minimizing predecessor and its chosen bit at each entry gives the traceback.
\end{proof}

In implementation, we store the restriction of $\vb{s}$ on active rows, instead of the full vector.
A row is called \emph{active} at cut $j$ if it has nonzero entries among both the first $j$ columns and the remaining columns.
Let $A_j$ be the set of active rows.
Once all nonzero entries of a row have been processed, its parity is checked against $t_i$, and a prefix that fails this check is discarded.
A row whose nonzero entries all lie in the remaining columns has partial parity zero.
Thus, we store only the restriction $\vb{s}|_{A_j}$.
Before the sweep, we put the check rows with target zero in minimal-span form, with distinct first and distinct last nonzero column positions, and reduce the support of the logical row by adding check rows.
A prefix with no completion may be rejected only when a later row is checked.

\subsection{From a linear sweep to a decomposition tree}
\label{subsec:rank-dp}

We now return to the general instance $(\vb{M},\vb{t},\calB)$ of \cref{def:tropical-partition}, with $\ell$ coordinates.
A node of a decomposition tree represents a union of whole blocks, and a join combines the assignments of its two children.
We write $T_u$ for the DP table at node $u$, extending the notation $T_j$ from prefixes to subtrees.
The table still stores one minimum cost for each class of partial assignments with the same exterior completions.

Joining the blocks one at a time in a fixed order gives a caterpillar tree.
With singleton blocks, the nodes along its spine represent prefixes, as in \cref{subsec:dense-rank-dp}.
A general tree allows both children of a join to contain many coordinates, so its update minimizes over compatible pairs of child states.

For an instance of \cref{def:tropical-partition}, Gaussian elimination first tests whether $\vb{M}\vb{e}=\vb{t}$ is consistent.
An inconsistent system has value $+\infty$.
Henceforth, assume that the system is consistent.

Fix a rank-decomposition tree $\calT$ of a reduced Tanner graph $G$ of $\vb{M}$ over the local blocks $\calB$.
For a node $u\in V(\calT)$, let $Y_u$ be the union of its leaf blocks.
At a cut $Y=Y_u$, write $\overline{Y}:=[\ell]\setminus Y$.
We represent a partial assignment $\vb{e}_Y$ by its partial syndrome $\vb{s}:=\vb{M}_Y\vb{e}_Y$.
An exterior assignment $\vb{e}_{\overline{Y}}$ completes it exactly when $\vb{M}_{\overline{Y}}\vb{e}_{\overline{Y}}=\vb{t}+\vb{s}$.
Thus, the states of extendible partial assignments form the affine space
\begin{align}
    \Sigma_Y:=\im\vb{M}_Y\cap\bigl(\vb{t}+\im\vb{M}_{\overline{Y}}\bigr).
    \label{eq:tree-state-space}
\end{align}
We discard partial assignments whose syndromes lie outside $\Sigma_Y$.
Section~\ref{subsec:rank-dp-analysis} proves the extension criterion and the state-count bound.

Assignments with the same partial syndrome impose the same conditions on every exterior completion.
A fixed exterior completion adds the same cost to all of them, because no local block crosses the cut.
Thus, at a node $u \in V(\calT)$ with $Y:=Y_u$, we store
\begin{align}
    T_u(\vb{s})
    :=\min_{\substack{\vb{e}_Y\in\FF_2^Y\\
                      \vb{M}_Y\vb{e}_Y=\vb{s}}}
        \sum_{\substack{B\in\calB\\B\subseteq Y}}\mathbf{1}[\vb{e}_B\ne\vb{0}],
    \qquad\vb{s}\in\Sigma_Y.
    \label{eq:tropical-table}
\end{align}
For Hamming weight, a node with $Y_u=\{1,\ldots,j\}$ represents the same partial assignments as cut $j$ in \cref{subsec:dense-rank-dp}.
In this case, $\Sigma_{Y_u}=\Sigma_j$ and $T_u(\vb{s})=T_j(\vb{s})$.
At a leaf with block $B$, enumerate the local assignments and compute $\vb{s}=\vb{M}_B\vb{e}_B$.
Discard states outside $\Sigma_B$ and store the minimum value of $\mathbf{1}[\vb{e}_B\ne\vb{0}]$ for each remaining state.
Thus, a singleton leaf has assignments $0,1$ with weights $0,1$, and a qubit leaf has assignments $(0,0),(1,0),(0,1),(1,1)$ with weights $0,1,1,1$.

Let $u$ have children $u_1,u_2$, and write $Y_i:=Y_{u_i}$, so that $Y=Y_1\sqcup Y_2$.
The partial syndromes of the two child assignments add to the parent syndrome:
\begin{align}
    \vb{s}=\vb{s}_1+\vb{s}_2,\qquad\vb{s}_i\in\Sigma_{Y_i}.
    \label{eq:tree-parent-state}
\end{align}
Here, addition is over $\FF_2$, and the pair is compatible when $\vb{s}_1+\vb{s}_2\in\Sigma_Y$.

The costs of compatible child assignments add, so the parent table is obtained by minimizing over compatible pairs:
\begin{align}
    T_u(\vb{s})
    =\min_{\substack{\vb{s}_i\in\Sigma_{Y_i}\\
                     \vb{s}_1+\vb{s}_2=\vb{s}}}
        \left(T_{u_1}(\vb{s}_1)
             +T_{u_2}(\vb{s}_2)\right),
    \qquad\vb{s}\in\Sigma_Y.
    \label{eq:tropical-join}
\end{align}
At the root $u_\star$, the state space is $\Sigma_{[\ell]}=\{\vb{t}\}$.
The algorithm returns $T_{u_\star}(\vb{t})$, which minimizes the total cost subject to all the constraints.
Algorithm~\ref{alg:tropical-rank-dp} summarizes the computation.

\begin{algorithm}[htbp]
\caption{Tropical rank-decomposition dynamic programming.}
\label{alg:tropical-rank-dp}
\begin{algorithmic}[1]
\Require A consistent instance $(\vb{M},\vb{t},\calB)$ and a tree $\calT$ of its reduced graph $G$
\Ensure $Z_{\vb{M}}^{\trop}(\vb{t};\calB)$
\State Compute the state spaces $\Sigma_{Y_u}$
\For{each leaf $u$ with block $B$}
    \State Initialize $T_u(\vb{s})\gets+\infty$ for $\vb{s}\in\Sigma_B$
    \State Enumerate $\vb{e}_B\in\FF_2^B$, storing the minimum $\mathbf{1}[\vb{e}_B\ne\vb{0}]$ at each $\vb{s}=\vb{M}_B\vb{e}_B\in\Sigma_B$
\EndFor
\For{each internal node $u$ in postorder, with children $u_1,u_2$}
    \State Initialize $T_u(\vb{s})\gets+\infty$ for $\vb{s}\in\Sigma_{Y_u}$
    \For{each compatible pair $(\vb{s}_1,\vb{s}_2)\in\Sigma_{Y_1}\times\Sigma_{Y_2}$}
        \State $\vb{s}\gets\vb{s}_1+\vb{s}_2$
        \State $T_u(\vb{s})\gets\min\{T_u(\vb{s}),T_{u_1}(\vb{s}_1)+T_{u_2}(\vb{s}_2)\}$
    \EndFor
\EndFor
\State \Return the root entry $T_{u_\star}(\vb{t})$
\end{algorithmic}
\end{algorithm}

An implementation stores states in affine coordinates of $\Sigma_Y$, using $\rho_G(Y)$ bits per state by \cref{lem:cutrank-states}.
Binary linear algebra computes an offset and a basis at each node and precomputes the parent-state index for each compatible child pair.
Our software contains such an implementation for single-column blocks; the results reported here use the linear sweep of \cref{subsec:dense-rank-dp}.

\subsection{Correctness and complexity}
\label{subsec:rank-dp-analysis}

We first characterize extendible partial assignments and count the states at each cut.
These properties establish the table invariant and the complexity of the preceding algorithm.

\begin{lemma}[Boundary states and cut rank]
\label{lem:cutrank-states}
A partial assignment $\vb{e}_Y$ extends to a solution of $\vb{M}\vb{e}=\vb{t}$ if and only if $\vb{M}_Y\vb{e}_Y\in\Sigma_Y$.
Every element of $\Sigma_Y$ occurs for an extendible partial assignment, and $|\Sigma_Y|=2^{\rho_G(Y)}$.
\end{lemma}

\begin{proof}
An exterior assignment completes $\vb{e}_Y$ when $\vb{M}_{\overline{Y}}\vb{e}_{\overline{Y}}=\vb{t}+\vb{M}_Y\vb{e}_Y$.
Such an assignment exists if and only if $\vb{t}+\vb{M}_Y\vb{e}_Y\in\im\vb{M}_{\overline{Y}}$.
Since $\vb{M}_Y\vb{e}_Y\in\im\vb{M}_Y$, this is equivalent to membership in $\Sigma_Y$.
Conversely, every $\vb{s}\in\Sigma_Y$ has a partial assignment and a completion by the two defining conditions in \cref{eq:tree-state-space}.
Equal partial syndromes impose the same exterior equation, while distinct partial syndromes have disjoint completion sets.

To count the states, choose any solution $\vb{e}^{0}$ of $\vb{M}\vb{e}=\vb{t}$.
Translating the intersection in \cref{eq:tree-state-space} by its point $\vb{M}_Y\vb{e}^{0}_Y$ gives
\begin{align}
    \Sigma_Y=\vb{M}_Y\vb{e}^{0}_Y+I_Y.
    \label{eq:affine-boundary-space}
\end{align}
Hence, \cref{prop:overlap-cutrank} gives $|\Sigma_Y|=2^{\dim I_Y}=2^{\rho_G(Y)}$.
\end{proof}

By \cref{prop:overlap-cutrank}, the state count is independent of the chosen column basis.

\begin{theorem}[Tropical Rank DP]
\label{thm:tropical-rank-dp}
Let $\omega:=\max_u\rho_G(Y_u)$ be the width of the given rank-decomposition tree.
The leaf initialization and recurrence \cref{eq:tropical-join} compute the tables in \cref{eq:tropical-table}.
The root value is $Z_{\vb{M}}^{\trop}(\vb{t};\calB)$.
After structural preprocessing, the computation uses
\begin{align}
    O\!\left(|\calB|4^{\omega}\right)
    \label{eq:tropical-dp-cost}
\end{align}
cost additions and comparisons.
Its tables contain at most $(2|\calB|-1)2^{\omega}$ values.
\end{theorem}

\begin{proof}
The leaf rule is the definition of the table.
At a join, disjointness of the child coordinate sets gives the syndrome addition in \cref{eq:tree-parent-state}.
Thus, all assignments with the same child states induce the same parent state.
By \cref{lem:cutrank-states}, the compatibility condition is exactly the existence of a completion outside the parent.
Every extendible parent assignment restricts to extendible child assignments.
Conversely, assignments with compatible child states concatenate to an extendible parent assignment.
Their costs add because no block is divided by the join.
Taking minima proves \cref{eq:tropical-join} and the table invariant by induction.
At the root, $\Sigma_{[\ell]}=\{\vb{t}\}$, so its sole entry minimizes the total cost subject to $\vb{M}\vb{e}=\vb{t}$.

Each child table has at most $2^{\omega}$ entries by \cref{lem:cutrank-states}.
A join therefore examines at most $4^{\omega}$ pairs, each contributing to at most one parent entry.
The tree has $|\calB|-1$ joins and $2|\calB|-1$ nodes.
Each leaf enumerates at most four assignments, since every block has size at most two.
Together with leaf enumeration, these counts give the stated bounds.
\end{proof}

\begin{remark}[Linear updates on a tree]
\label{rem:linear-decomposition}
For a caterpillar tree obtained by joining blocks in order, the cuts consist of prefix cuts and cuts isolating individual blocks.
For singleton blocks, its width is $\max_j w_j$, the largest prefix cut rank from \cref{subsec:dense-rank-dp}: a cut isolating column $j$ has rank at most one, and if it is one, then $\vb{m}_j=\vb{a}+\vb{b}$ with $\vb{a}\in\im\vb{M}_{<j}$ and $\vb{b}\in\im\vb{M}_{>j}$, so that a nonzero $\vb{a}$ lies in $\im\vb{M}_{\leq j-1}\cap\im\vb{M}_{>j-1}$ and $w_{j-1}\geq1$, while $\vb{a}=\vb{0}$ puts $\vb{m}_j$ in $\im\vb{M}_{\leq j}\cap\im\vb{M}_{>j}$ and $w_j\geq1$.
For general blocks, the width $\omega$ of this rank decomposition is the maximum cut rank over both prefix cuts and cuts isolating individual blocks.
At a join adding a block $B$, the leaf table has at most $2^{|B|}$ entries, so the join examines at most $2^{\omega+|B|}$ pairs rather than $4^{\omega}$.
With singleton blocks, each preceding state has at most two extensions, recovering the update count in \cref{prop:linear-dense-dp}.
\end{remark}

The preceding bound assumes that a decomposition is given.
We next account for its construction using the following approximation guarantee.

\begin{proposition}[Constructing a decomposition over a partition]
\label{prop:construct-decomposition}
Let $G$ be a graph with $v\geq1$ vertices and partition $\calB$.
Set $r:=\rw(G;\calB)$ and $c:=\max_{B\in\calB}\rho_G(B)$.
A rank decomposition over $\calB$ of width at most $3r+c\leq4r$ can be constructed deterministically in $\poly(v)2^{O(r)}$ time.
\end{proposition}

This is Proposition~4.2 of ref.~\nocite{cheng_pan_2026_rankdp}\citenum{cheng_pan_2026_rankdp}, obtained from the connectivity-function form of the Oum--Seymour approximation algorithm~\cite{oum_seymour_2006_approximating}.
To apply that algorithm to a fixed partition, define $f(\calD):=\rho_G(\bigcup_{B\in\calD}B)$ for $\calD\subseteq\calB$.
This function is symmetric, integer-valued, and submodular.
Binary Gaussian elimination evaluates it in polynomial time.
Its branch-decompositions are exactly the rank decompositions of $G$ over $\calB$.
The cited algorithm, for $k\geq c$, either returns width at most $3k+c$ or certifies that the optimum exceeds $k$.
Trying successive $k$ from $c$ terminates by $k=r$ with the stated running time.
When $c=0$, no edges join distinct parts, and any tree has width zero.

The structural preprocessing uses binary linear algebra and enumeration of boundary-state pairs.
Together, \cref{prop:construct-decomposition,thm:tropical-rank-dp} give a running time polynomial in the matrix dimensions and singly exponential in $r=\rw(G;\calB)$.
Every finite table value is an integer between $0$ and $|\calB|$, so it uses $O(\log(|\calB|+1))$ bits.

A decomposition of the original Tanner graph also supplies a width bound for this algorithm.
The next lemma constructs a block tree for the reduced graph by deleting the check leaves.

\begin{lemma}[Check elimination preserves the width bound]
\label{lem:width-check-elimination}
Let $\calC$ consist of the coordinate blocks and singleton check nodes of $T(\vb{M})$.
Deleting the check leaves from a rank-decomposition tree of $T(\vb{M})$ over $\calC$, then repeatedly deleting childless internal nodes and suppressing nodes with one child, yields a tree of $G=T(\vb{A})$ over $\calB$ whose width does not increase.
In particular,
\begin{align}
    \rw(G;\calB)\leq\rw(T(\vb{M});\calC).
    \label{eq:reduced-width-bound}
\end{align}
\end{lemma}

\begin{proof}
Every cut of the resulting tree arises from a cut $X$ of the original tree.
Write $Y$ for its variable coordinates and $J$ for its check-row indices.
Let $r_0:=\rank\vb{M}$ and choose $r_0$ independent rows of $\vb{M}$, of which $a$ have indices in $J$.
Restricting to these rows preserves every column-submatrix rank.
Therefore,
\begin{align}
    \rank\vb{M}_Y&\leq a+\rank\vb{M}[\overline{J},Y],\nonumber\\*
    \rank\vb{M}_{\overline{Y}}&\leq r_0-a+\rank\vb{M}[J,\overline{Y}].
\end{align}
Here, $\overline{J}$ is taken within the set of check rows.
By \cref{prop:overlap-cutrank} and the dimension formula for $I_Y$,
\begin{align}
    \rho_G(Y)
    &\leq\rank\vb{M}[\overline{J},Y]+\rank\vb{M}[J,\overline{Y}]
     =\rho_{T(\vb{M})}(X).
\end{align}
Taking the maximum over all cuts of the resulting tree proves the statement.
\end{proof}

Consequently, the same type of complexity bound holds with $r=\rw(T(\vb{M});\calC)$.

\subsection{Code and circuit applications}
\label{subsec:distance-applications}

The reduction in \cref{eq:distance-unguarded-reduction} computes distance by evaluating the $q$ augmented systems and taking the minimum.
For each row $\vb{r}_a^{\sfT}$ of $\vb{R}$, the corresponding query imposes $\vb{M}\vb{e}=\vb{0}$ and $\vb{r}_a^{\sfT}\vb{e}=1$.
The other label components are unrestricted.
We first test each augmented system for consistency and assign $+\infty$ to an inconsistent query.
We apply \cref{thm:tropical-rank-dp} separately to the consistent affine queries.
For each such query, the root is evaluated at its target $\vb{t}^{(a)}$.

The augmented systems can all use the same block tree.
The affine state spaces and DP tables are formed separately for each query, using its augmented matrix and target.
Let $G$ be a reduced Tanner graph of $\vb{M}$, and let a tree over $\calB$ have width $\omega$.
Appending one row increases each overlap dimension by at most one.
If the row is dependent, the ranks of the full matrix and its two column submatrices are unchanged.
Otherwise, the total rank increases by one, and each restricted rank increases by at most one.
By the overlap formula in \cref{subsec:rank-width}, this tree has width at most $\omega+1$ for a reduced Tanner graph of each augmented matrix $\vb{M}^{(a)}$.

By \cref{thm:tropical-rank-dp}, the total number of cost additions and comparisons after structural preprocessing is
\begin{align}
    O\!\left(q|\calB|4^{\omega}\right).
    \label{eq:distance-query-cost}
\end{align}
The extra unit of width contributes only a constant factor to this bound.
Evaluating the queries sequentially requires $O(|\calB|2^{\omega})$ scalar table entries.
Recording the minimizing choices at leaves and joins permits traceback in a finite minimizing query.

For code distance, take $\vb{M}=\vb{M}_s$, $\vb{R}=\vb{P}$, and the physical-qubit partition from \cref{subsec:minimum-distance}.
The objective is the Pauli weight $|\vb{e}|_{\sfP}$.
Each query selects a zero-syndrome Pauli with one specified component of $\vb{P}\vb{e}$ equal to one.
Taking the minimum over the $n+k$ rows of $\vb{P}$ gives $d_{\mathrm{code}}$.
Let $G_s$ and $G_P$ be reduced Tanner graphs of $\vb{M}_s$ and $\vb{P}$, respectively.
Both graphs have vertex set $[2n]$ and use the physical-qubit partition $\calB$.
The following lemma shows that the same block tree has the same width for both graphs.

\begin{samepage}
\begin{lemma}[Stabilizer and normalizer cut ranks]
\label{lem:dual-widths}
For $U\subseteq[n]$, let $Y_U:=\bigcup_{i\in U}B_i$ and $\overline{U}:=[n]\setminus U$.
For a binary subspace $C\subseteq\FF_2^{2n}$, let $C_U:=\{\vb{e}\in C:\vb{e}_{\overline{Y_U}}=\vb{0}\}$ be the subspace supported on $Y_U$.
Define
\begin{align}
    \lambda_C(U):=\dim C-\dim C_U-\dim C_{\overline{U}}.
    \label{eq:code-connectivity}
\end{align}
Then,
\begin{align}
    \rho_{G_s}(Y_U)=\lambda_N(U)=\lambda_S(U)=\rho_{G_P}(Y_U).
    \label{eq:dual-cut-rank}
\end{align}
Consequently, $\rw(G_s;\calB)=\rw(G_P;\calB)$.
\end{lemma}
\end{samepage}

\begin{proof}
For any matrix $\vb{M}$ with $\ker\vb{M}=C$, zero extension identifies the kernel of its column submatrix $\vb{M}_{Y_U}$ with the shortened space $C_U$.
Thus, rank-nullity and \cref{prop:overlap-cutrank} give
\begin{align}
    \rho_{T(\vb{A})}(Y_U)
    &=\rank\vb{M}_{Y_U}+\rank\vb{M}_{\overline{Y_U}}-\rank\vb{M}\nonumber\\*
    &=\dim C-\dim C_U-\dim C_{\overline{U}}
     =\lambda_C(U).
    \label{eq:code-connectivity-cutrank}
\end{align}
This proves the first and last equalities of \cref{eq:dual-cut-rank}.

Let $\pi_U(S)$ denote the projection of $S$ to the $2|U|$ coordinates of $U$.
The restricted symplectic form on these coordinates is nondegenerate.
Since $N=S^{\perp_s}$, restriction to these coordinates identifies $N_U$ with the symplectic orthogonal complement of $\pi_U(S)$.
Therefore,
\begin{align}
    \dim N_U
    &=2|U|-\dim\pi_U(S)
     =2|U|-\dim S+\dim S_{\overline{U}}.
    \label{eq:shortened-normalizer}
\end{align}
The analogous identity holds for $\overline{U}$.
Substituting both identities and $\dim N=2n-\dim S$ into \cref{eq:code-connectivity} gives $\lambda_N(U)=\lambda_S(U)$.
The equality holds for every cut displayed by every tree over $\calB$.
Taking the minimum width over these trees proves the final assertion.
\end{proof}

By \cref{lem:dual-widths}, a tree supplied for either $G_s$ or $G_P$ can be used for the code-distance queries.
The equality applies to cuts that keep every physical-qubit block intact, for any column bases in the two reductions.
Using all $n+k$ rows of $\vb{P}$ in \cref{eq:distance-query-cost} gives $O(n(n+k)4^{\omega})$ cost operations after structural preprocessing.
Each scalar table entry stores a cost of at most $n$, or $+\infty$.
Traceback in a minimizing query recovers a minimum-weight nontrivial logical Pauli.

For circuit distance, let $G_D$ be a reduced Tanner graph of the detector matrix $\vb{D}$.
Take $\vb{M}=\vb{D}$, $\vb{R}=\vb{O}$, and the singleton partition of the $\ell$ mechanisms from \cref{subsec:circuit-distance}.
The objective is the Hamming weight $|\vb{e}|$ of the mechanism vector.
Each of the $q$ queries requires zero detector syndrome and a flip of one specified logical observable.
Their minimum is $d_{\mathrm{circ}}$, with the operation bound $O(q\ell 4^{\omega})$ from \cref{eq:distance-query-cost} for a tree of $G_D$ over the mechanisms.
Each scalar table entry has value in $\{0,\ldots,\ell\}\cup\{+\infty\}$.
A finite minimizing query yields a minimum set of mechanisms by traceback.
Under \cref{lem:mechanism-count}, composing representative faults at each location gives a minimum fault configuration.
The detector matrix determines the base width, and each appended observable row increases it by at most one.

\begin{corollary}[Rank-width bounds for distance]
\label{cor:final-complexity}
For the objective in \cref{eq:unified-distance}, let $G$ be a reduced Tanner graph of $\vb{M}$ and let $r:=\rw(G;\calB)$.
The value $\Delta(\vb{M},\vb{R};\calB)$ can be computed in $\poly(N_{\mathrm{in}})2^{O(r)}$ time, where $N_{\mathrm{in}}$ is the total binary encoding length of the matrices and partition.
This bound includes decomposition construction, structural preprocessing, and bit operations.
In particular, code distance is computable in $\poly(n)2^{O(r)}$ time for $r=\rw(G_s;\calB)=\rw(G_P;\calB)$.
Circuit distance has the corresponding bound for $r=\rw(G_D)$.
The same bounds hold with $r$ replaced by the rank-width of the original Tanner graph of the constraint matrix, with respect to its coordinate blocks and singleton check nodes.
\end{corollary}

\begin{proof}
Apply \cref{prop:construct-decomposition} to obtain a block decomposition of width at most $4r$.
By the overlap argument above, the same block tree has width at most $4r+1$ for a reduced Tanner graph of each augmented constraint matrix.
Apply \cref{thm:tropical-rank-dp} to each consistent affine query and take the minimum as in \cref{eq:distance-unguarded-reduction}.
The number of queries $q$ is bounded by the input length.
Every finite stored value is an integer at most $|\calB|$, so its binary encoding length is $O(\log(|\calB|+1))$.
Exact binary linear algebra and structural preprocessing contribute a polynomial factor per enumerated state or state pair.
The original Tanner graph bound follows from \cref{lem:width-check-elimination}.
For a stabilizer code, \cref{lem:dual-widths} also permits the bound to use the original Tanner graph of $\vb{P}$.
\end{proof}

Thus, rank-width logarithmic in the input length gives polynomial-time distance computation, with no bound on the number of logical classes.

\section{Weight-truncated Rank DP}
\label{sec:sparse-dp}

The weight-truncated algorithm uses the linear trellis and exact forward tables $T_j$ from \cref{subsec:dense-rank-dp}.
For the threshold decision problem $Z^{\trop}_{\vb{M}}(\vb{t})\leq K$, only states on solutions of weight at most $K$ are needed.
We first define the backward tables on the same state space, and then prove the truncation rule and its meet-in-the-middle extension.
Throughout this section, $\vb{M}\vb{e}=\vb{t}$ imposes zero syndrome and parity one for a fixed logical row, as in \cref{eq:distance-unguarded-reduction}.
We assume that this system has a solution.
We use the same column order, partial syndromes $\vb{s}$ and state spaces $\Sigma_j$ as in the linear dense sweep.

\subsection{Backward sweep and joins}
\label{subsec:sparse-states}

For a state $\vb{s}\in\Sigma_j$, define the backward cost by
\begin{align}
    B_j(\vb{s}):=\min\{|\vb{y}|:\vb{y}\in\FF_2^{\ell-j},\ \vb{M}_{>j}\vb{y}=\vb{t}+\vb{s}\}.
    \label{eq:sparse-backward}
\end{align}
By \cref{lem:sparse-states}, $B_j(\vb{s})$ is the least completion weight of every extendible prefix with state $\vb{s}$.
Its entries use the same state labels as $T_j$: a suffix is indexed by the partial syndrome $\vb{s}=\vb{t}+\vb{M}_{>j}\vb{y}$ it requires from a prefix.

At cut $\ell$, the suffix is empty and its required prefix syndrome is $\vb{t}$, so we initialize $B_\ell(\vb{t})=0$.
We then process columns $j=\ell,\ell-1,\ldots,1$.
Before processing column $j$, set every entry of $B_{j-1}$ to $+\infty$.
Let $\vb{m}_j$ be column $j$ of $\vb{M}$.
For each state $\vb{s}\in\Sigma_j$ and choice $e_j\in\{0,1\}$, let $\vb{s}'=\vb{s}+e_j\vb{m}_j$.
If $\vb{s}'\in\Sigma_{j-1}$, update
\begin{align}
    B_{j-1}(\vb{s}')\gets\min\{B_{j-1}(\vb{s}'),\ B_j(\vb{s})+e_j\}.
    \label{eq:sparse-backward-update}
\end{align}
Otherwise, discard this choice.
Prepending $e_j$ gives a suffix with syndrome $\vb{t}+\vb{s}+e_j\vb{m}_j=\vb{t}+\vb{s}'$ and increases its weight by $e_j$.
Every suffix on columns $j,\ldots,\ell$ arises from such a choice, so descending induction gives the tables in \cref{eq:sparse-backward}.

The two sweeps satisfy $T_\ell(\vb{t})=B_0(\vb{0})=Z^{\trop}_{\vb{M}}(\vb{t})$.
At every cut,
\begin{align}
    Z^{\trop}_{\vb{M}}(\vb{t})=\min_{\vb{s}\in\Sigma_j}\bigl[T_j(\vb{s})+B_j(\vb{s})\bigr].
    \label{eq:sparse-join}
\end{align}
To see this, fix a state $\vb{s}\in\Sigma_j$ and choose a prefix $\vb{e}_{\le j}$ attaining $T_j(\vb{s})$ and a suffix $\vb{e}_{>j}$ attaining $B_j(\vb{s})$.
By definition, $\vb{M}_{\le j}\vb{e}_{\le j}=\vb{s}$ and $\vb{M}_{>j}\vb{e}_{>j}=\vb{t}+\vb{s}$.
Their concatenation $\vb{e}$ therefore satisfies
\begin{align}
    \vb{M}\vb{e}
    =\vb{M}_{\le j}\vb{e}_{\le j}+\vb{M}_{>j}\vb{e}_{>j}
    =\vb{s}+(\vb{t}+\vb{s})=\vb{t}.
    \label{eq:sparse-join-concatenation}
\end{align}
The two parts occupy disjoint coordinates, so $|\vb{e}|=|\vb{e}_{\le j}|+|\vb{e}_{>j}|=T_j(\vb{s})+B_j(\vb{s})$.
Thus, $Z^{\trop}_{\vb{M}}(\vb{t})\leq T_j(\vb{s})+B_j(\vb{s})$ for every $\vb{s}\in\Sigma_j$, which gives the inequality $\leq$ in \cref{eq:sparse-join}.

Conversely, let $\vb{e}$ be any solution and set $\vb{s}:=\vb{M}_{\le j}\vb{e}_{\le j}$.
Then, $\vb{M}_{>j}\vb{e}_{>j}=\vb{t}+\vb{s}$, so $\vb{s}\in\Sigma_j$ and both parts are candidates in the minimizations defining the two tables.
Consequently, $T_j(\vb{s})+B_j(\vb{s})\leq|\vb{e}_{\le j}|+|\vb{e}_{>j}|=|\vb{e}|$.
The minimum on the right-hand side of \cref{eq:sparse-join} is therefore at most the weight of every solution.
Taking the minimum over solutions gives the opposite inequality and proves the identity.

\subsection{Truncation}
\label{subsec:sparse-truncation}

Fix a threshold $K\geq0$ and choose efficiently computable lower bounds $h_j(\vb{s})$ satisfying $0\leq h_j(\vb{s})\leq B_j(\vb{s})$.
Any admissible choice may be used, including $h_j\equiv0$.
Section~\ref{subsec:sparse-implementation} gives a specific implementation of $h_j$.
The exact backward values $B_j$ are reference quantities for the proof and need not be computed.

We run the forward sweep of \cref{subsec:dense-rank-dp} with sparse tables $\widehat{T}_j$, initialized by $\widehat{T}_0(\vb{0})=0$.
Missing entries have value $+\infty$.
At each column, we apply \cref{eq:sparse-update} to stored predecessor entries and take the minimum of candidates reaching each valid successor.
We then discard every entry satisfying
\begin{align}
    \widehat{T}_j(\vb{s})+h_j(\vb{s})>K.
    \label{eq:sparse-pruning-rule}
\end{align}
Fix a cut $j$ and a state $\vb{s}\in\Sigma_j$.
By the argument for \cref{eq:sparse-join}, $T_j(\vb{s})+B_j(\vb{s})\leq K$ holds if and only if there exists $\vb{e}\in\FF_2^\ell$ with $\vb{M}\vb{e}=\vb{t}$, $|\vb{e}|\leq K$, and $\vb{M}_{\le j}\vb{e}_{\le j}=\vb{s}$.

\begin{lemma}[Exact truncation]
\label{lem:sparse-truncation}
At each cut $j$, every state $\vb{s}\in\Sigma_j$ with $T_j(\vb{s})+B_j(\vb{s})\leq K$ is kept, with $\widehat{T}_j(\vb{s})=T_j(\vb{s})$.
Consequently, the truncated sweep returns $Z^{\trop}_{\vb{M}}(\vb{t})$ if it is at most $K$.
Otherwise, it certifies that no solution has weight at most $K$.
\end{lemma}

\begin{proof}
The truncated tables minimize over a subset of the partial assignments, so $\widehat{T}_j\geq T_j$.
We prove the claim by induction on $j$.
The initialization proves the case $j=0$.

Suppose that the claim holds at cut $j-1$, and fix a state $\vb{s}$ with $T_j(\vb{s})+B_j(\vb{s})\leq K$.
Choose a prefix $\vb{e}_{\le j}$ attaining $T_j(\vb{s})$ and let $\vb{s}':=\vb{M}_{\leq j-1}\vb{e}_{\le j-1}$ be its preceding state.
The prefix $\vb{e}_{\le j-1}$ and the completion obtained by prepending $e_j$ to an optimal suffix for $\vb{s}$ give
\begin{align}
    T_{j-1}(\vb{s}')\leq|\vb{e}_{\le j-1}|=T_j(\vb{s})-e_j,\qquad
    B_{j-1}(\vb{s}')\leq e_j+B_j(\vb{s}).
    \label{eq:sparse-predecessor-bounds}
\end{align}
Thus, $T_{j-1}(\vb{s}')+B_{j-1}(\vb{s}')\leq T_j(\vb{s})+B_j(\vb{s})\leq K$, and the induction hypothesis gives $\widehat{T}_{j-1}(\vb{s}')=T_{j-1}(\vb{s}')$.

Extending this entry by $e_j$ generates $\vb{s}$ and, before pruning, gives $\widehat{T}_j(\vb{s})\leq T_{j-1}(\vb{s}')+e_j\leq T_j(\vb{s})$.
Since $\widehat{T}_j\geq T_j$, the entry is exact.
It passes the pruning test because $\widehat{T}_j(\vb{s})+h_j(\vb{s})\leq T_j(\vb{s})+B_j(\vb{s})\leq K$.

At $j=\ell$, $B_\ell(\vb{t})=0$, so the claim gives the exact optimum when it is at most $K$.
Otherwise, every candidate for the final entry has weight greater than $K$ and is deleted.
\end{proof}

\subsection{Meet in the middle}
\label{subsec:sparse-mitm}

The single sweep of \cref{subsec:sparse-truncation} allows prefix weights up to $K$ and can therefore build large tables.
We split this weight budget between a forward sweep over prefixes and a backward sweep over suffixes, combining matching states using \cref{eq:sparse-join}.
A balanced split gives each direction a weight limit of about $K/2$.

The bidirectional weight-bounded search follows the principle of BEAST~\cite{bocharova_2004_beast}.
Here, we formulate it as a min-plus dynamic program on the syndrome trellis, storing the minimum partial weight for each state.

First, fix integers $K\geq1$ and $1\leq a\leq K$.
Compute backward tables with suffix weight at most $K-a$, then run a forward sweep with prefix weight at most $a$.
For this basic algorithm, truncation uses only the accumulated weight.
At every cut $j=0,\ldots,\ell$, test all matching entries of the two tables.
At a join, write $\widehat T_j$ and $\widehat B_j$ for these truncated tables, with $+\infty$ for absent states.
The test uses the stored costs:
\begin{align}
    \min_{\vb{s}\in\Sigma_j}\bigl[\widehat T_j(\vb{s})+\widehat B_j(\vb{s})\bigr]\leq K.
    \label{eq:sparse-truncated-join}
\end{align}
The exact tables $T_j$ and $B_j$ in \cref{eq:sparse-join} remain the reference values for the correctness proof.

\begin{lemma}[Meet-in-the-middle truncation]
\label{lemma:sparse-mitm}
For any fixed split $1\leq a\leq K$, the basic two-sweep algorithm satisfies \cref{eq:sparse-truncated-join} at some cut if and only if there exists $\vb{e}\in\FF_2^\ell$ with $\vb{M}\vb{e}=\vb{t}$ and $|\vb{e}|\leq K$.
A successful test gives such a solution by concatenating the matched prefix and suffix, with weight equal to the tested sum.
If all cuts are tested without success, no solution has weight at most $K$.
If $Z_{\vb{M}}(\vb{t})\leq K$, the least tested sum over all cuts equals $Z_{\vb{M}}(\vb{t})$.
\end{lemma}

\begin{proof}
If a prefix has weight at most the fixed cutoff, so do all its shorter prefixes.
Induction using the forward update therefore gives $\widehat T_j(\vb{s})=T_j(\vb{s})$ whenever $T_j(\vb{s})\leq a$.
The same argument from right to left gives $\widehat B_j(\vb{s})=B_j(\vb{s})$ whenever $B_j(\vb{s})\leq K-a$.
Each finite matched sum is the weight of a solution by the argument for \cref{eq:sparse-join}, so a successful test gives a valid solution within the budget.

Conversely, let $\vb{e}$ be a solution with $|\vb{e}|\leq K$.
If $|\vb{e}|\leq a$, use cut $\ell$.
Otherwise, choose a cut with $|\vb{e}_{\le j}|=a$, which exists because prefix weight increases by zero or one at each step.
In either case, $|\vb{e}_{\le j}|\leq a$ and $|\vb{e}_{>j}|\leq K-a$.
For $\vb{s}=\vb{M}_{\le j}\vb{e}_{\le j}$, the two table entries are exact and their sum is at most $|\vb{e}|\leq K$, so the test succeeds.
Applied to a minimum-weight solution, this gives a tested sum of at most $Z_{\vb{M}}(\vb{t})$, and every finite tested sum is the weight of a solution, so the least tested sum equals $Z_{\vb{M}}(\vb{t})$.
\end{proof}

\paragraph{Varying the split.}
Now let the prefix budget $a(j)\in\{1,\ldots,K\}$ be non-increasing in $j$, with suffix budget $K-a(j)$.
For weight-only truncation, the required forward invariant is $\widehat T_j(\vb{s})=T_j(\vb{s})$ whenever $T_j(\vb{s})\leq a(j)$.
To prove it, choose a prefix $\vb{e}_{\le j}$ attaining such a value and let $\vb{s}'$ be its preceding state.
\begin{align}
    T_{j-1}(\vb{s}')\leq|\vb{e}_{\le j-1}|\leq T_j(\vb{s})\leq a(j)\leq a(j-1).
    \label{eq:mitm-one-sided-budget}
\end{align}
Induction makes the predecessor entry exact, and its extension gives the exact value at $\vb{s}$, which passes the current cutoff.
The backward argument is identical because $K-a(j)$ is non-increasing in the backward sweep direction.
Thus, $\widehat B_j(\vb{s})=B_j(\vb{s})$ whenever $B_j(\vb{s})\leq K-a(j)$.

When $a(j)$ decreases, test the whole forward candidate table before applying the smaller cutoff.
To see why this is sufficient, take any solution $\vb{e}$ with $|\vb{e}|\leq K$ and let $c$ be the first cut with $|\vb{e}_{>c}|\leq K-a(c)$.
Such a cut exists at $\ell$; if $c=0$, the initial join succeeds.
Otherwise, for $j<c$, we have $|\vb{e}_{\le j}|\leq K-|\vb{e}_{>j}|<a(j)$, so the prefix reaches cut $c-1$.
The next update produces a candidate of weight at most $|\vb{e}_{\le c}|\leq a(c-1)$.
It is tested by the ordinary join if the budget is unchanged, or by the extra join before applying the smaller cutoff.
The matching backward entry has weight at most $|\vb{e}_{>c}|$, so the joined weight is at most $|\vb{e}|\leq K$.

\paragraph{Reducing forward updates and joins.}
After an entry of weight at least $a(j)$ has been joined, it need not be used in further forward updates.
Any completion of total weight at most $K$ would have suffix weight at most $K-a(j)$ and would already be found by that join.
We therefore pass only entries of weight at most $a(j)-1$ to the next cut.
At a budget decrease, the extra join is performed before this stricter cutoff.
The forward invariant then holds for $T_j(\vb{s})\leq a(j)-1$, and an entry of weight $a(j)$ is new, decreased or at a budget decrease, so it is always joined.

Between budget changes, it is also sufficient to test only new entries or decreases in their stored values.
An unchanged entry has the same full syndrome and weight as at its last tested cut.
A matching suffix at the current cut can be extended by zeros to that earlier cut without changing its weight.
The suffix budget is unchanged, so any passing join for this entry would already have passed at the earlier cut.
These two refinements preserve the decision guarantee and the least tested sum of \cref{lemma:sparse-mitm}.

\subsection{Witnesses and implementation}
\label{subsec:sparse-implementation}

\paragraph{State representation and witnesses.}
In implementation, both sweeps use the active-row encoding described in \cref{subsec:dense-rank-dp}.
The backward kernel stores the suffix parity $(\vb{M}_{>j}\vb{y})|_{A_j}=(\vb{t}+\vb{s})|_{A_j}$, whereas $B_j(\vb{s})$ is indexed by the prefix syndrome.
Thus, the kernel matches a forward key $\vb{s}|_{A_j}$ to a backward key obtained by XOR with $\vb{t}|_{A_j}$, which implements the join in \cref{eq:sparse-join}.
Any additional unextendible states stored by the trellis cannot yield a valid join.
Every reconstructed witness is checked against the original query $\vb{M}\vb{e}=\vb{t}$, and its weight must equal the reported value.

\paragraph{Look-ahead bounds.}
We compute the look-ahead bound using the covering bound of \cref{eq:covering-bound}.
For the forward sweep, choose nonnegative row weights $y_i$ such that $\sum_{i:M_{ij'}=1}y_i\leq1$ for each remaining column $j'>j$, and set $h_j(\vb{s}):=\lceil\sum_{i:s_i\ne t_i}y_i\rceil$.
Every completion must cover each unsatisfied row at least once, while each selected column carries total row weight at most one.
Its integer weight is therefore at least $h_j(\vb{s})$, proving admissibility.
For the backward sweep, apply the same construction to columns $j'\leq j$ and residual $\vb{s}$.

The weights are precomputed from the remaining-column incidences, with rounding downward where needed to preserve the column inequalities.
Look-ahead pruning compares the stored weight plus this bound with the total threshold $K$; the single-sided caps bound only the corresponding partial weights.
A candidate that belongs to a solution of total weight at most $K$ cannot fail an admissible look-ahead test, so this pruning preserves the preceding decision guarantees.

\paragraph{Parameter choices.}
The split schedule is chosen before the sweeps.
Low-weight probe sweeps estimate the table sizes in each direction, and a dynamic program selects a non-increasing schedule with the least predicted work.
The prediction guides performance; correctness holds for every schedule satisfying the conditions above.
A varying schedule can exploit small forward tables at early cuts and small backward tables at late cuts.

Sparse or dense table storage is selected according to occupancy, and checkpoints trade stored tables for recomputation during traceback.
These choices affect work and memory, while conversions and recomputation must preserve the same state-value mapping.
The distance is found by increasing $K$ from a certified lower bound; failure for every logical row certifies the next lower bound.
The first successful threshold is the distance, so that run can stop at its first solution.

\paragraph{Validation and performance.}
The kernels were tested against brute force on 272,076 random queries in both table implementations, NumPy and C (forward sweeps,
every constant split, with and without the bound $h$, and with the dense fallback), and on 66,814 queries with
random split schedules.
On every instance of \cref{tab:sparse-times}, the value equals the distance found by our other evaluators where they finish and the dense DP's value where
that one runs, and every witness checks.
The sparse sweep wins where problems are wide but light partial solutions merge, as for the superdense color-code
circuits; there it settles the circuit of distance 11, which was open.
On narrow codes the dense DP is faster, and on expanding qLDPC codes and graph-like circuits the search is.

\begin{table}[t]
\centering
\caption{Time to the certified distance of the three evaluators on one thread (seconds; server described in
\cref{sec:benchmark}). Truncated DP: the sweeps at $K=d-1$ for every logical row and the sweep at $K=d$ that returns
the witness, with the split probes. Search: every threshold from $K=1$. Dense DP: one sweep per logical row; ``too
wide'', the largest table exceeds $2^{33}$ bytes. Superdense and surface: the memory-basis sector problems of
the circuits of the prior benchmark\cite{webster_2026_distance}. $^\ddagger$A dedicated run.}
\label{tab:sparse-times}
\small
\begin{tabular}{@{}lrrrrrr@{}}
\toprule
instance & columns & cut rank & $d$ & dense DP & search & truncated DP \\
\midrule
color code, $d=21$ & 331 & 21 & 21 & \textbf{0.11} & 194 & 0.60 \\
color code, $d=27$ & 547 & 27 & 27 & \textbf{0.86} & $\geq21$ (600~s) & 6.75 \\
concatenated Steane, $L=4$ & 2401 & 12 & 81 & \textbf{0.02} & $\geq18$ (600~s) & 0.50 \\
gross code \qc{144,12,12} & 144 & 40 & 12 & too wide & \textbf{0.06} & 5.59 \\
\qc{108,8,10} & 108 & 30 & 10 & 78.5 & \textbf{0.01} & 0.24 \\
superdense circuit, $d=7$ & 1248 & 25 & 7 & 2.59 & 2.42 & \textbf{0.57} \\
superdense circuit, $d=9$ & 2788 & 39 & 9 & too wide & 378 & \textbf{10.1} \\
superdense circuit, $d=11$ & 5250 & 55 & 11 & too wide & $\geq8$ (600~s) & \textbf{772}$^\ddagger$ \\
surface circuit, $d=9$ & 1945 & 42 & 9 & too wide & \textbf{0.98} & 5.74 \\
surface circuit, $d=11$ & 3631 & 62 & 11 & too wide & \textbf{36.1} & 1664 \\
\bottomrule
\end{tabular}
\end{table}

\section{Weight-bounded search and distance certificates}
\label{sec:weight-search}

Section~\ref{sec:tropical-rank-dp} computes distance with a cost controlled by the width of a rank decomposition.
When this width is large, the boundary tables can exceed the available memory.
The threshold formulation of \cref{eq:distance-threshold} gives an alternative: search for a feasible support within a prescribed weight budget.
The search uses the residual syndrome to select the next fault columns and does not require a rank decomposition.
We first give the search algorithm and its correctness and complexity guarantees.
We then introduce reductions of the search work and use relaxed problems to obtain distance bounds.

\subsection{Weight-bounded search}
\label{subsec:threshold-search}

Throughout this section, each of the $\ell$ binary coordinates is an independently selectable fault of unit cost.
Thus, $\calB$ consists of singleton blocks and the objective is Hamming weight.
For a support $F\subseteq[\ell]$, let $\vb{1}_F$ be its indicator vector, and write $\vb{s}(F):=\vb{M}\vb{1}_F$ and $\vb*{\eta}(F):=\vb{R}\vb{1}_F$.
The distance objective becomes
\begin{align}
    \delta:=\min_{\substack{F\subseteq[\ell]\\
                            \vb{s}(F)=\vb{0},\;\vb*{\eta}(F)\ne\vb{0}}}|F|.
    \label{eq:search-distance}
\end{align}
The value is $+\infty$ when no feasible support exists.
This is the unit-cost specialization of \cref{eq:unified-distance}.
Under \cref{lem:mechanism-count}, this minimum also counts faulty circuit locations.

A threshold search asks whether $\delta\leq K$.
The empty support has zero logical label, so it suffices to consider integer thresholds $1\leq K\leq \ell$.
At a search node, let $A$ be the selected support and let $U\subseteq[\ell]\setminus A$ contain the available columns.
The remaining budget is $b:=K-|A|$.
A completion is a set $B\subseteq U$ satisfying
\begin{align}
    |B|\leq b,\qquad
    \vb{s}(B)=\vb{s}(A),\qquad
    \vb*{\eta}(B)\ne\vb*{\eta}(A).
    \label{eq:search-completion}
\end{align}
These conditions make $A\cup B$ a feasible support of weight at most $K$.

The residual syndrome determines the branching rule.
If $\vb{s}(A)\ne\vb{0}$, select a row $r$ with $s_r(A)=1$.
Every completion must contain an available column $j$ with $M_{rj}=1$.
Order these columns as $j_1,\ldots,j_t$ and give child $p$ the sets
\begin{align}
    A_p:=A\cup\{j_p\},\qquad
    U_p:=U\setminus\{j_1,\ldots,j_p\}.
    \label{eq:canonical-search-branch}
\end{align}
Each completion belongs to the child corresponding to its first column in this order.
Removing earlier candidates prevents the same completion from appearing in two children.
An unsatisfied row with no available column gives no children.

The search starts from $A=\{a\}$ and $U=\{a+1,\ldots,\ell\}$ for each anchor $a\in[\ell]$.
It stops at a nonempty zero-syndrome support: a nonzero logical label gives a solution, and a zero label rejects the branch.
It also rejects a node when the budget is exhausted and the residual syndrome is nonzero.
Section~\ref{subsec:search-certificates} proves that these rules preserve a path to a minimum support.
Algorithm~\ref{alg:certified-search} gives the complete threshold search, with $\varnothing$ denoting a negative answer.

\begin{algorithm}[htbp]
\caption{Weight-bounded search at threshold $K$.}
\label{alg:certified-search}
\makeatletter
\def\theHALG@line{\thealgorithm.\arabic{ALG@line}}
\makeatother
\begin{algorithmic}[1]
\Require Unit-cost matrices $\vb{M},\vb{R}$ and an integer $1\leq K\leq \ell$
\For{$a=1,\ldots,\ell$}
    \State $F\gets\Call{Search}{\{a\},\{a+1,\ldots,\ell\},K}$
    \If{$F\ne\varnothing$}
        \State \Return $F$
    \EndIf
\EndFor
\State \Return $\varnothing$
\Function{Search}{$A,U,K$}
    \If{$\vb{s}(A)=\vb{0}$}
        \State \Return $A$ if $\vb*{\eta}(A)\ne\vb{0}$; otherwise, return $\varnothing$
    \EndIf
    \If{$|A|=K$}
        \State \Return $\varnothing$
    \EndIf
    \State Choose a row $r$ with $s_r(A)=1$ and form the children in \cref{eq:canonical-search-branch}
    \For{each child $(A_p,U_p)$}
        \State $F\gets\Call{Search}{A_p,U_p,K}$
        \If{$F\ne\varnothing$}
            \State \Return $F$
        \EndIf
    \EndFor
    \State \Return $\varnothing$
\EndFunction
\end{algorithmic}
\end{algorithm}

To compute the distance, start from a valid integer lower bound $L\geq1$ and run successive thresholds $K=L,L+1,\ldots,\ell$.
After each completed negative search, increase the lower bound to $K+1$.
The first positive threshold determines the distance.

\subsection{Correctness and complexity}
\label{subsec:search-certificates}

Threshold correctness requires a path to a minimum support whenever $\delta\leq K$.
The search need not enumerate every feasible support.
The following property justifies stopping at each nonempty zero-syndrome support.

\begin{lemma}[Minimum supports]
\label{lem:minimal-search-support}
Let $F$ attain the finite minimum in \cref{eq:search-distance}.
No nonempty proper subset of $F$ has zero syndrome.
\end{lemma}

The proof is in Section~\ref{app:search-minimality}.
For a minimum support $F$, its smallest column supplies an anchor, and the other columns are initially available.
At every nonempty proper prefix $A\subsetneq F$, the residual syndrome is nonzero by \cref{lem:minimal-search-support}.
The first candidate from $F\setminus A$ in \cref{eq:canonical-search-branch} selects a child that still admits $F\setminus A_p$.
Thus, canonical branching preserves a path to $F$.

\begin{theorem}[Certified weight-bounded search]
\label{thm:certified-search}
For the unit-cost problem in \cref{eq:search-distance}, \cref{alg:certified-search} returns a feasible support of weight at most $K$ if $\delta\leq K$.
Otherwise, it returns $\varnothing$, proving $\delta>K$.
Starting from a valid integer lower bound and increasing $K$ by one after each negative answer yields the exact distance at the first positive threshold.
If all thresholds through $\ell$ are negative, then $\delta=+\infty$.
\end{theorem}

Section~\ref{app:search-correctness} gives the proof.
The lower bound increases only after the threshold search is complete.
An interrupted threshold leaves this bound unchanged.
A feasible support found before interruption still gives an upper bound after verification with the original matrices.

The branching factor gives a bound in terms of the input and the threshold.
Let $\beta$ be the maximum of one and all row weights of $\vb{M}$, and let $V_K$ count the search nodes at threshold $K$.
There are at most $\ell$ anchors, at most $\beta$ children per node, and at most $K-1$ branching steps after an anchor.
Hence,
\begin{align}
    V_K\leq \ell\sum_{j=0}^{K-1}\beta^j.
    \label{eq:search-node-bound}
\end{align}
Syndrome updates and candidate selection take polynomial time per node, giving a total of $\poly(m+\ell+q)V_K$ time.
For successive thresholds, sum these costs up to the first positive threshold, or through $\ell$ if no feasible support exists.
Since $\beta$ can grow with the number of columns, this bound does not imply fixed-parameter tractability in the distance alone.
The fixed-parameter guarantee in rank-width is given separately by \cref{cor:final-complexity}.

\subsection{Pruning and meet-in-the-middle completion}
\label{subsec:search-mitm}

The search work can be reduced at the anchors, at internal nodes, and near the weight threshold.
Symmetries reduce the anchor set, necessary conditions reject internal nodes, and completion tables replace short search subtrees.
Each modification must preserve the minimum-support guarantee used in \cref{thm:certified-search}.

For the anchor reduction, let $\calG$ be a group of permutations of the fault coordinates that preserves the feasible supports in \cref{eq:search-distance}.
Each permutation preserves weight.
Order the column orbits $\{g(j):g\in\calG\}$ and choose one anchor in each orbit.
For an anchor $a_i$ in orbit $i$, initialize $U$ with all columns outside earlier orbits except $a_i$.
Every minimum support has an image under $\calG$ that contains the anchor of its first occupied orbit and uses no earlier orbit.
Thus, the reduced anchor searches still contain a minimum support.
The symmetries must preserve both zero syndrome and nonzero logical action; Section~\ref{app:search-symmetry} gives the proof and a sufficient algebraic condition.

At an internal node, necessary conditions can exclude a completion before branching.
The bound $|\vb{s}(A)|>b\max_j|\vb{M}_{\{j\}}|$ rejects the node, since $b$ columns cannot cancel that many nonzero syndrome entries.
A sharper bound uses only the available columns.
Assign weights $y_r\geq0$ to the unsatisfied rows $r\in\supp(\vb{s}(A))$, the rows with $s_r(A)=1$.
The load of an available column $j\in U$ is $\sum_{r\in\supp(\vb{s}(A)):\,M_{rj}=1}y_r$.
If every available column has load at most one, then every completion $B$ satisfies
\begin{align}
    |B|\geq\sum_{r\in\supp(\vb{s}(A))}y_r,
    \label{eq:covering-bound}
\end{align}
and the node is rejected when this sum exceeds $b$.
This is the dual bound for covering the unsatisfied rows with available columns.
One feasible choice is $y_r:=1/m_r$, where $m_r$ is the largest number of unsatisfied rows met by a single available column incident to $r$; with unit costs this is the admissible heuristic of the Tesseract decoder\cite{beni_2025_tesseract}, which the ascent below raises.
If $m_r=0$, no available column reaches row $r$, and the node is rejected.
The weights can then be raised one row at a time while every load stays at most one.
Equal weights $y_r=1/\max_j|\vb{M}_{\{j\}}|$ recover the preceding bound.
A second rule uses supports with zero syndrome and zero logical label.
For every such support $Z$ and every minimum support $F$,
\begin{align}
    |F\cap Z|\leq\lfloor |Z|/2\rfloor.
    \label{eq:kernel-dominance}
\end{align}
Otherwise, the symmetric difference $F\mathbin{\triangle}Z$ would be a feasible support of smaller weight.
Thus, a partial support that violates \cref{eq:kernel-dominance} can be rejected.
In particular, a minimum support contains at most one column from any zero-syndrome, zero-label triple of distinct columns.
Section~\ref{app:search-minimality} proves these pruning rules.
Among the remaining nodes, choosing an unsatisfied row with the fewest available columns minimizes the immediate branching factor.

Near the weight threshold, a completion table replaces the remaining branching steps by a syndrome query.
Fix a table budget $k\geq0$ and enumerate supports of size at most $k$.
The table stores the exclusion-query value
\begin{align}
    Q_k(\vb{h},\vb*{\gamma})
    :=\min\{ |B|:B\subseteq[\ell],\ |B|\leq k,
                    \vb{s}(B)=\vb{h},\ \vb*{\eta}(B)\ne\vb*{\gamma}\}.
    \label{eq:completion-table}
\end{align}
At a fixed syndrome, a completion record is the pair $(\vb*{\eta}(B),|B|)$ of its logical label and cost.
For a finite set $\calR$ of such records, first keep a lowest-cost record for each label.
Let $\Ret(\calR)$ consist of the two lowest-cost records with distinct labels, or all records if fewer than two labels occur.
Ties may be resolved arbitrarily.

\begin{lemma}[Two-label representatives]
\label{lem:two-labels}
For a finite record set $\calR$ and a forbidden label $\vb*{\gamma}$, define
\begin{align}
    \Psi_{\calR}(\vb*{\gamma})
    :=\min\{c:(\vb*{\eta},c)\in\calR,\ \vb*{\eta}\ne\vb*{\gamma}\},
    \label{eq:exclusion-query}
\end{align}
with $\min\varnothing=+\infty$.
Then, $\Psi_{\calR}(\vb*{\gamma})=\Psi_{\Ret(\calR)}(\vb*{\gamma})$ for every $\vb*{\gamma}$.
\end{lemma}

\begin{proof}
If fewer than two labels occur, $\Ret$ keeps a minimum-cost record for every available label, and the assertion is immediate.
Forbidding one label removes at most one of the two chosen labels.
If the label of a cheapest chosen record is allowed, that record attains the constrained minimum.
Otherwise, the second chosen record has minimum cost among the remaining labels.
\end{proof}

Applying \cref{lem:two-labels} to the records at each exact syndrome preserves every value of $Q_k$.
The table does not encode the available set $U$.

When $b\leq k$, a value $Q_k(\vb{s}(A),\vb*{\eta}(A))>b$ rejects the node.
A value at most $b$ requires reconstruction, since the minimizing record may use unavailable columns.
Reconstruction searches the full remaining budget $b$ for a set $B\subseteq U$ satisfying \cref{eq:search-completion}.
Exact subset enumeration is one valid procedure.
Every returned completion is checked with the original syndrome and logical-label maps.
Tables that omit logical labels or use finite syndrome fingerprints can also serve as filters if they preserve every true completion.
Their positive queries require the same full-budget reconstruction and verification.
Section~\ref{app:search-fingerprints} gives the detailed conditions and the fingerprint counterexample.

These modifications preserve the conclusion of \cref{thm:certified-search}.
The anchor reduction preserves a minimum support up to symmetry, the pruning rules preserve its canonical path, and a completion table cannot reject its remaining support.
Any further pruning test must preserve such a path whenever $\delta\leq K$.
Section~\ref{app:search-correctness} extends the completeness argument to these modifications.

The relaxation arguments of \cref{subsec:search-reductions} require all minimizers, not only one.
In the recording variant of \cref{alg:certified-search}, \textsc{Search} records each support with zero syndrome and nonzero label and then continues.
Each positive table query reconstructs all completions within the full remaining budget.

\begin{corollary}[Enumeration]
\label{cor:search-enumeration}
At threshold $K$, the recording variant without the dominance bound \cref{eq:kernel-dominance} records every feasible support $F$ with $|F|\leq K$ that has no nonempty proper zero-syndrome subset.
With the dominance bound, it records every minimum support.
With the anchor reduction, it records at least one member of each symmetry orbit of these supports.
\end{corollary}

By \cref{lem:minimal-search-support}, minimum supports have no nonempty proper zero-syndrome subset, so the recording variant at $K=\delta$ lists all of them.
Section~\ref{app:search-correctness} gives the proof.

Completion tables trade construction and storage costs for shorter searches.
The number of supports enumerated for a table of budget $k$ is
\begin{align}
    N_k:=\sum_{j=0}^{\min\{k,\ell\}}\binom{\ell}{j}.
    \label{eq:search-table-size}
\end{align}
Construction takes $\poly(m+\ell+q)N_k$ time and storage.
Let $V_K$ now count the nodes visited with these modifications, and let $C_K$ count the subsets inspected during exact reconstruction.
With explicit syndrome keys and polynomial-time pruning tests, the total work at threshold $K$, including table construction, is at most
\begin{align}
    \poly(m+\ell+q)\bigl(N_k+V_K+C_K\bigr).
    \label{eq:search-work}
\end{align}
The node bound \cref{eq:search-node-bound} still applies.
If reconstruction uses subset enumeration only when $b\leq k$, then $C_K\leq V_KN_k$.
For successive thresholds, construct the table once and sum the branching and reconstruction costs.
The costs of finding and verifying symmetries and dominance constraints must also be included.

\subsection{Distance certification through relaxations}
\label{subsec:search-reductions}

The same search can be applied to derived problems to bound the original distance.
A relaxed problem supplies a lower bound $L$, while a verified feasible support supplies an upper bound $d_{\mathrm{up}}$.
The bounds determine the distance when $L=d_{\mathrm{up}}$.

A row projection relaxes the syndrome constraints.
For a binary matrix $\vb{T}$, keep the fault coordinates, partition, and label map, and replace $\vb{M}$ by $\vb{T}\vb{M}$.
Every original feasible support remains feasible, so
\begin{align}
    \Delta(\vb{T}\vb{M},\vb{R})
    \leq\Delta(\vb{M},\vb{R}).
    \label{eq:search-relaxation-bound}
\end{align}
The same inequality holds with any fixed partition $\calB$ in \cref{eq:unified-distance}.
Deleting detector rows and adding rows within specified groups are examples.
Time collapse adds all detector rows for the same check.
Time folding adds the rows in each specified time class.
For a memory circuit with identical syndrome rounds that carry every Pauli error on the data qubits through
unchanged, as in all circuits here, and a final transversal measurement of the data qubits, the
sum of a check's detectors is the parity of its value in that measurement with its reference value, so the
collapsed column of a fault is the memory-basis syndrome and the logical flip of the data error that the fault
leaves at readout.
Since later rounds pass that error on unchanged, it depends on the fault's place within a round but not on the
round, so after merging the collapsed problem is the same for every number of rounds, and
\cref{eq:search-relaxation-bound} bounds all of them.
A round that permutes the data qubits, for example, would break this argument.

Column reduction can simplify the projected problem before a search.
A joint column combines the corresponding columns of $\vb{M}$ and $\vb{R}$.
In the unit-cost model, removing zero joint columns and merging identical joint columns preserve the distance of that problem.
Section~\ref{app:search-relaxations} proves these statements and gives a sufficient lifting condition for a detector-sector projection to preserve the original distance.

Restricting the allowed columns provides the complementary upper bound.
Every feasible support of the restricted problem is also feasible in the original problem.
For example, a time window can keep only columns whose detector support lies within the selected rounds.
A search on this restriction can supply a witness and improve $d_{\mathrm{up}}$.
A negative search on the restriction does not raise the lower bound for the original problem.

Lifting can close the gap between the relaxed lower bound and the original upper bound.
Suppose that a relaxed search gives the finite minimum $L$.
If a relaxed minimizer lifts to an original feasible support of weight $L$, then the distance is $L$.
To raise the lower bound to $L+1$, one must exclude every original support of weight $L$.
It suffices to exhaust all relaxed minimizers and all their original preimages of that weight.
\Cref{cor:search-enumeration}, applied to the relaxed problem at threshold $L$, lists the relaxed minimizers.
Failure to lift a single candidate does not establish this exclusion.
Symmetry representatives suffice only when the symmetries also act on the original problem and respect the projection.
Section~\ref{app:search-relaxations} proves this certification rule.

The same reasoning can raise the lower bound by more than one.
Call original columns with the same nonzero projected joint column a class, and call a column with zero projected joint column image-free.
Every original support $F$ maps to the set $\kappa(F)$ of classes that it meets an odd number of times.

\begin{proposition}[Lifting through weight $W$]
\label{prop:lifting-W}
Let $\calO$ be a set of relaxed feasible supports that contains a member of the orbit of every relaxed feasible support of weight at most $W$, under a group that acts on both problems and commutes with $\kappa$.
Suppose that, for each $O\in\calO$, the original problem has no feasible support $F$ with $\kappa(F)=O$ and $|F|\leq W$.
Then the original distance is at least $W+1$.
\end{proposition}

For $W=L$, a support with $\kappa(F)=O$ and $|F|\leq L$ takes exactly one column from each class of $O$, which recovers the rule above.
For larger $W$, $\calO$ must also contain relaxed supports that are not minimum.
The next lemma shows that \cref{cor:search-enumeration} still finds them, up to a simple correction at weight $L+3$.

\begin{lemma}[Feasible supports near the minimum]
\label{lem:near-minimum}
Suppose that no joint column is zero, no two joint columns are equal, and $\delta\geq4$ is finite.
A feasible support of weight at most $\delta+2$ has no nonempty proper zero-syndrome subset.
A feasible support of weight $\delta+3$ that has such a subset is the disjoint union of a minimum support and a zero-syndrome, zero-label set of three columns.
\end{lemma}

After column reduction, the relaxed problem satisfies these conditions.
Let $L\geq4$ and $W\leq L+3$.
At threshold $W$, the recording variant without the dominance bound lists every relaxed feasible support of weight at most $W$, except those of weight $L+3$ with a nonempty proper zero-syndrome subset.
By \cref{lem:near-minimum}, the exceptions are unions of a relaxed minimizer with a disjoint zero-syndrome, zero-label triple, and they can be added directly from the listed minimizers.
The search for supports with $\kappa(F)=O$ and $|F|\leq W$ is \cref{alg:certified-search} on the original columns with one more necessary condition.
The number of selected columns minus the number of classes of $O$ that they meet cannot exceed $W-|O|$.
Section~\ref{app:search-relaxations} proves \cref{prop:lifting-W,lem:near-minimum} and the completeness of these restricted searches.

The projected constraints also give completion filters inside the search of \cref{subsec:search-mitm}.
Every original completion must cancel the relaxed residual $\vb{T}\vb{s}(A)$ with at most the same cost and the same logical label.
If no such relaxed completion exists within budget $b$, the original node can be rejected.
A truncated table supports this test only when its table budget is at least $b$.
Each auxiliary search or table contributes its construction and query costs, together with the cost of constructing and verifying the projection.

\subsection{Proofs}
\label{app:search-proofs}

\subsubsection{Minimum supports and canonical branching}
\label{app:search-minimality}

\begin{proof}[Proof of \cref{lem:minimal-search-support}]
Suppose that a nonempty proper subset $A\subset F$ has $\vb{s}(A)=\vb{0}$.
If $\vb*{\eta}(A)\ne\vb{0}$, then $A$ is a feasible support of smaller weight.
Otherwise, $F\setminus A$ has zero syndrome and logical label $\vb*{\eta}(F)\ne\vb{0}$, again with smaller weight.
Both cases contradict the minimality of $F$.
\end{proof}

For the dominance bound \cref{eq:kernel-dominance}, let $F$ be a minimum support and let $Z$ have zero syndrome and zero logical label.
Write $F\mathbin{\triangle}Z$ for their symmetric difference.
Its syndrome and logical label equal those of $F$, since both maps vanish on $Z$.
Moreover,
\begin{align}
    |F\mathbin{\triangle} Z|=|F|+|Z|-2|F\cap Z|.
\end{align}
If $|F\cap Z|>|Z|/2$, this gives a feasible support of smaller weight.
Thus, \cref{eq:kernel-dominance} holds.

\Cref{lem:minimal-search-support} also implies that a minimum support is connected in the graph that joins two fault columns whenever they share a nonzero detector row.
Distinct connected components of a selected support affect disjoint sets of detector rows.
If the total syndrome vanishes, each component has zero syndrome.
A minimum support therefore has only one component.
A column with zero detector part and nonzero logical part is a feasible support of weight one and is tested as an anchor.

We next show that canonical branching preserves a path to each minimum support available at a node.
Let $F$ be a minimum support with $\varnothing\ne A\subsetneq F$ and $F\setminus A\subseteq U$.
The residual syndrome is nonzero by \cref{lem:minimal-search-support}.
For a selected row $r$ with $s_r(A)=1$, the set $F\setminus A$ contains an odd number of columns incident to $r$.
In particular, it contains at least one candidate.
Let $j_p$ be its first member in the candidate order.
No earlier candidate belongs to $F\setminus A$.
Hence, the child in \cref{eq:canonical-search-branch} still satisfies $A_p\subseteq F$ and $F\setminus A_p\subseteq U_p$.
Induction gives a path to $F$.
The choice of the first candidate also makes this path unique among the children of the current node.

The additional elementary pruning rules preserve this path.
The remaining part of $F$ has weight at most $K-|A|$ whenever $|F|\leq K$.
It supplies a candidate at every unsatisfied row and cannot violate \cref{eq:kernel-dominance}.
Also, the Hamming weight of a sum of $b$ detector columns is at most $b$ times their maximum Hamming weight.
Thus, the detector-count bound cannot reject a node on this path.

For the covering bound \cref{eq:covering-bound}, let $B\subseteq U$ be a completion and let the weights satisfy the load condition.
Each unsatisfied row meets $B$ in an odd, hence positive, number of columns.
Therefore,
\begin{align}
    \sum_{r\in\supp(\vb{s}(A))}y_r
    \leq\sum_{j\in B}\,\sum_{\substack{r\in\supp(\vb{s}(A))\\M_{rj}=1}}y_r
    \leq|B|.
\end{align}
The remaining part of $F$ is such a completion, so the covering bound cannot reject a node on this path.
The choice $y_r=1/m_r$ satisfies the load condition.
Indeed, if an available column $j$ meets $o_j\geq1$ unsatisfied rows, then each of these rows has $m_r\geq o_j$, so the load of $j$ is at most $o_j/o_j=1$.
Raising a weight only while every load stays at most one preserves the condition.
The argument uses only $B\subseteq U$ and $\vb{s}(B)=\vb{s}(A)$, not minimality.

\subsubsection{Completion tables and finite fingerprints}
\label{app:search-fingerprints}

The table in \cref{eq:completion-table} minimizes over a superset of the available completions.
If $B\subseteq U$ satisfies \cref{eq:search-completion} and $b\leq k$, then
\begin{align}
    Q_k(\vb{s}(A),\vb*{\eta}(A))\leq |B|\leq b.
\end{align}
Therefore, a negative table test cannot remove such a completion.
For each syndrome, \cref{lem:two-labels} preserves this scalar query using records $(\vb*{\eta}(B),|B|)$.
Restrictions on the support of $B$ must be enforced during reconstruction or included in the table index.

Exact subset enumeration is one valid reconstruction procedure.
Residual-syndrome branching also suffices if it preserves every minimum support through the current node.
Reconstruction may instead use an enlarged available set that contains $U$ and excludes the selected columns.
Every returned support must still satisfy the original syndrome, logical-label, and weight conditions.
The enlargement can find a solution from another branch, but it cannot exclude an available completion.

A fingerprint table gives the same rejection guarantee by minimizing over a larger set.
Let $\phi$ be a deterministic map from detector syndromes to finite keys, and define
\begin{align}
    \widetilde Q_k(\vb{h},\vb*{\gamma})
    :=\min\{ |B|: B\subseteq[\ell],\ |B|\leq k,\
          \phi(\vb{s}(B))=\phi(\vb{h}),\
          \vb*{\eta}(B)\ne\vb*{\gamma}\}.
    \label{eq:fingerprint-filter}
\end{align}
Logical labels in this definition are exact.
Every exact syndrome match is a fingerprint match, so $\widetilde Q_k(\vb{h},\vb*{\gamma})\leq Q_k(\vb{h},\vb*{\gamma})$.
When $b\leq k$, a filtered value greater than $b$ certifies the absence of an available completion.
A filtered value at most $b$ requires exact reconstruction within the full budget $b$.
If the filter omits logical labels, its candidate set only grows, so negative tests remain valid.

An apparent fingerprint cost is not a safe reconstruction limit.
Consider
\begin{align}
    \vb{M}=\begin{pmatrix}1&1\end{pmatrix},\qquad
    \vb{R}=\begin{pmatrix}1&0\\0&1\end{pmatrix}.
    \label{eq:fingerprint-counterexample}
\end{align}
The unique feasible support is $\{1,2\}$, so $\delta=2$.
Let $k=1$ and suppose that $\phi(0)=\phi(1)$.
For either one-column prefix $A$, the empty support has the same fingerprint as $\vb{s}(A)$ and a different logical label.
Thus, the apparent completion cost is zero, although the other column is needed to cancel the exact syndrome.
Reconstruction limited to the apparent cost finds no completion.
Rejecting the anchors on this basis would incorrectly exclude a feasible support of weight two.

The implementation used for this paper reconstructs within the full remaining budget after every positive query, in both of its table modes (\texttt{cdist/gmitm.c}).
An earlier version limited reconstruction in its labeled mode to the cost returned by the table.
A forced-collision test of \cref{eq:fingerprint-counterexample} shows that this path can report exhaustion wrongly; the test is part of the test suite (\texttt{tests/test\_collision.py}).
In a completed negative threshold, every reconstruction that was started has failed, since a successful one would have returned a solution, and if none was started the cost limit never affected the search.
All benchmark runs used label-free tables, which always reconstructed within the full budget.
For every other value reported in the paper, either the stored number of failed reconstructions is zero at every negative threshold, which certifies the value by the argument above, or the value was recomputed with the corrected kernel; all recomputations reproduced the reported values (\cref{sec:software}).

\subsubsection{Symmetry reduction}
\label{app:search-symmetry}

To justify the anchor reduction, let $F$ be a minimum support and let orbit $i$ be its first occupied orbit.
Choose $j\in F$ in that orbit and an element $g\in\calG$ that sends $j$ to the selected anchor $a_i$.
The support $g(F)$ has the same weight and is feasible by the assumed symmetry.
Every group element maps each orbit to itself, so $g(F)$ has no column in an earlier orbit.
Its remaining columns are therefore available in the search rooted at $a_i$.
The canonical branching argument then preserves a path to this minimum support.

The following identities give a sufficient condition for a column permutation to preserve feasibility.
Let $\vb{\Pi}$ be its permutation matrix.
Suppose that there are matrices $\vb{U},\vb{V},\vb{W}$ such that $\vb{U}$ is invertible and
\begin{align}
    \vb{M}\vb{\Pi}=\vb{U}\vb{M},\qquad
    \vb{R}\vb{\Pi}=\vb{V}\vb{R}+\vb{W}\vb{M}.
    \label{eq:search-symmetry-condition}
\end{align}
The first identity makes $\ker\vb{M}$ invariant under $\vb{\Pi}$.
Both identities imply
$\vb{\Pi}(\ker\vb{M}\cap\ker\vb{R})\subseteq\ker\vb{M}\cap\ker\vb{R}$.
The inclusion is equality because $\vb{\Pi}$ is invertible and preserves dimension.
Thus, $\vb{\Pi}$ preserves $\ker\vb{M}\setminus(\ker\vb{M}\cap\ker\vb{R})$, which is precisely the feasible set.
Coordinate permutations also preserve Hamming weight.
The matrix $\vb{V}$ need not be invertible on the full logical-label space.

\subsubsection{Projection, column reduction, and lifting}
\label{app:search-relaxations}

Since $\vb{M}\vb{e}=\vb{0}$ implies $\vb{T}\vb{M}\vb{e}=\vb{0}$, keeping the label map and costs fixed gives \cref{eq:search-relaxation-bound}.
For a partial support $A$, every available completion maps to a completion of the relaxed residual $\vb{T}\vb{s}(A)$ with the same label and at most the same cost.
Consequently, rejection by an exact relaxed completion test is sound.
A table truncated at weight $k$ is used for this rejection only when the remaining budget is at most $k$.

To prove that column merging preserves the unit-cost distance, partition the columns into classes with identical joint columns.
Map a support to the nonzero classes that it meets an odd number of times.
This map preserves the syndrome and logical label and does not increase weight.
Conversely, selecting one original representative of each selected class lifts a reduced support with the same syndrome, label, and weight.
These two maps prove equality of the distances before and after merging.
They apply to the projected joint columns as well.
Restricting the allowed columns gives an upper bound on the original distance, since extension by zeros preserves the syndrome, label, and weight of each restricted solution.

A detector-sector projection is exact under a stronger lifting condition.
Suppose that the projection deletes a set of detector rows.
Assume that each joint column of the reduced problem has an original representative with the same logical part and zero detector entries on all deleted rows.
Selecting these representatives lifts every reduced feasible support to an original feasible support of the same weight.
Hence, the original and reduced distances are equal.
The condition is the one checked by the sector-reduction test in the implementation.
A row projection without such a lifting argument supplies only a lower bound.

To justify an improved lower bound from lifting, let the relaxed minimum be a finite integer $L$.
If an original feasible support of weight $L$ exists, its projected and merged image has weight at most $L$.
The definition of the relaxed minimum forces that image to have weight exactly $L$.
It is therefore a relaxed minimizer.
Consequently, excluding all original preimages of weight $L$ for every relaxed minimizer proves that the original distance is at least $L+1$.
Orbit representatives suffice only for symmetries that also act on the original problem and respect the projection.
Partial lists of relaxed minimizers or incomplete lifting searches do not establish this conclusion.

\begin{proof}[Proof of \cref{prop:lifting-W}]
Suppose that the original problem has a feasible support of weight at most $W$, and let $F$ have minimum weight among them.
The class map $\kappa$ is the column-merging map above after the projection, so $\kappa(F)$ is relaxed feasible with $|\kappa(F)|\leq|F|\leq W$.
Choose $g$ in the group with $g(\kappa(F))\in\calO$.
Then $g(F)$ is feasible, has weight $|F|$, and satisfies $\kappa(g(F))=g(\kappa(F))$.
This contradicts the assumption for $O=g(\kappa(F))$.
\end{proof}

The proof only needs the restricted searches to be complete for minimum original supports.
Let $F$ be a minimum support with $\kappa(F)=O$ and $|F|\leq W$.
Every class of $O$ meets $F$, so the number of selected columns minus the number of classes of $O$ that they meet equals $|F|-|O|\leq W-|O|$ at $F$.
Adding a column raises this quantity by zero or one, so the necessary condition holds along every prefix of $F$.
Since every such support meets each class of $O$, the anchors can be restricted to the members of one class of $O$, placed first in the column order.
The canonical path argument of Section~\ref{app:search-minimality} and the completion-table argument of Section~\ref{app:search-fingerprints} then apply unchanged.

\begin{proof}[Proof of \cref{lem:near-minimum}]
Let $F$ be feasible with a nonempty proper subset $A$ of zero syndrome.
Then $F\setminus A$ also has zero syndrome, and the labels of $A$ and $F\setminus A$ add to $\vb*{\eta}(F)\ne\vb{0}$.
If both labels are nonzero, both parts are feasible, so $|F|\geq2\delta\geq\delta+4$.
Otherwise, one part $Z$ has zero label, and the other is feasible, with weight at least $\delta$.
A nonempty set with zero syndrome and zero label has at least three columns, since one column would be a zero joint column and two would be equal joint columns.
Hence $|F|\geq\delta+3$, with equality only if the feasible part is a minimum support and $|Z|=3$.
\end{proof}

\subsubsection{Search completeness and certificates}
\label{app:search-correctness}

\begin{proof}[Proof of \cref{thm:certified-search}]
Every returned support satisfies the original syndrome, logical-label, and weight tests.
Thus, a positive result at threshold $K$ implies $\delta\leq K$.

Conversely, suppose that $\delta\leq K$, and choose a minimum support $F$.
Its smallest column is an allowed anchor, and all other columns of $F$ are initially available.
The argument in Section~\ref{app:search-minimality} gives a canonical path whose selected set remains a subset of that support.
The residual cannot vanish before the entire minimum support is selected.
The weight budget cannot terminate this path before $F$ is reached, since $|F|\leq K$.
Thus, a completed search cannot report a negative result when $\delta\leq K$.
This proves the threshold claim.

Let the initial lower bound be $L_0\leq\delta$.
Each completed negative threshold raises the lower bound by one.
At the first positive threshold $K$, the earlier bounds imply $\delta\geq K$, while the verified support implies $\delta\leq K$.
Therefore, $\delta=K$.
Every support has size at most $\ell$, so a completed negative search through threshold $\ell$ proves that no feasible support exists.
An interrupted threshold establishes neither its negative decision nor the corresponding lower-bound update.
\end{proof}

The modifications in \cref{subsec:search-mitm} preserve this argument.
With verified symmetry reduction, Section~\ref{app:search-symmetry} supplies a minimum support in an allowed anchor search.
The dominance, detector-count and covering bounds preserve its canonical path by Section~\ref{app:search-minimality}.
Any further permitted pruning test must preserve such a path.
At a table query, the remaining part of the minimum support is a valid completion, so the rejection rules of Section~\ref{app:search-fingerprints} cannot exclude it.
Exact reconstruction within the full remaining budget either finds it or finds another feasible support within the threshold.
Therefore, the threshold and distance guarantees remain valid with these modifications.

\begin{proof}[Proof of \cref{cor:search-enumeration}]
Let $F$ be a feasible support with $|F|\leq K$ and no nonempty proper zero-syndrome subset.
The canonical path argument of Section~\ref{app:search-minimality} applies to $F$, since it uses only that every nonempty proper prefix has a nonzero residual syndrome.
The detector-count and covering bounds use only that $F\setminus A$ is an available completion within the budget, so they preserve this path.
The dominance bound preserves the paths of minimum supports by Section~\ref{app:search-minimality}.
At a table query on the path, $F\setminus A$ is an available completion, so the query does not reject the node.
Reconstruction by exact subset enumeration, or by residual-syndrome branching over a set that contains $F\setminus A$, reaches $F$ and records it, since no nonempty proper prefix of $F$ has zero syndrome.
With the anchor reduction, Section~\ref{app:search-symmetry} supplies an image of $F$ in an allowed anchor search.
The image has the same weight and also has no nonempty proper zero-syndrome subset, because a symmetry satisfying \cref{eq:search-symmetry-condition} maps zero-syndrome sets to zero-syndrome sets.
\end{proof}

\FloatBarrier
\section{Benchmark}
\label{sec:benchmark}

\subsection{Instances}
\label{subsec:bench-instances}

The benchmark has 74 instances: 47 memory circuits and 27 code-distance problems
(\cref{tab:bench-circuits,tab:bench-codes}).
The circuits are the standard depth-seven circuits of bivariate bicycle codes from \qc{72,12,6} to
\qc{288,12,18}\cite{bravyi_2024_bb} in the Stim files of ref.~\nocite{tesseract_code}\citenum{tesseract_code}, two depth-seven schedules found in this work
(\cref{sec:design}), the published circuits of the Cornucopia\cite{lu_2026_cornucopia},
mitten\cite{bhardwaj_2026_mitten}, routing\cite{zhang_2026_routing}, shift-automorphism\cite{kim_2026_shift},
STAR\cite{ismail_2026_star}, Walking Cat\cite{tripier_2026_walkingcat} and left-right\cite{strikis_2026_leftright}
codes, circuits designed here for self-dual\cite{liang_2025_selfdual} and weight-eight\cite{lu_2026_weight8}
bivariate bicycle codes, Stim's rotated surface-code memories from $d=3$ to $25$ and color-code memories from
$d=3$ to $11$\cite{stim}, and the superdense color-code circuits of the prior benchmark\cite{webster_2026_distance}.
The codes are bivariate bicycle, Kasai\cite{okada_2026_pair,lee_2026_ultrahigh}, balanced-product\cite{hong_2026_rate}, multicycle\cite{mian_2026_multicycle},
LLM-searched bivariate bicycle\cite{cruzbenito_2026_llm}, weight-eight bivariate bicycle, GALA\cite{yang_2026_gala},
Cornucopia and self-dual bivariate bicycle codes, together with codes chosen to test the Rank DP: 6.6.6 color codes
of distance 19 to 31, concatenated Steane codes, a many-hypercube code\cite{goto_2024_hypercube} and planar tile codes\cite{gu_2026_planar}.
They have up to 2,401 qubits, and the circuits up to $3.3\times10^5$ fault mechanisms before reduction.

Every solver receives the same pair of matrices $(\vb{M},\vb{R})$.
For a CSS memory circuit this is its sector problem, which already contains the exact reduction of
\cref{subsec:search-reductions} and is the best case for the other solvers; the exporter checks the lifting
condition of \cref{app:search-relaxations} for every circuit.
For Stim's color-code circuits, which are not CSS memory experiments in this sense, it is the full detector error
model.
For a code it is one CSS sector, $d_X$. The other sector was computed separately and has the same distance wherever \cref{tab:bench-codes-both} gives both, which is therefore $d_{\mathrm{code}}$ by \cref{prop:css-distance}; for the self-dual codes the two are equal by construction.
The four tile codes are the exception: we computed $d_X$ only, and the planar paper reports $d_Z$ of 4, 5, 4 and 12, so these instances are sector distances, not code distances.
The detector error models are Stim's, without decomposition, with equal mechanisms merged; all noise models are
built from depolarizing channels, single Pauli flips and measurement flips, so by \cref{lem:mechanism-count} the
unit-cost distance is $d_{\mathrm{circ}}$.
The ground truth of every instance comes from a certified record or, for the surface codes, from Stim's graph-like
distance; distances marked $^\dagger$ in the tables were not certified before and are certified by the runs.

\subsection{Solvers and protocol}
\label{subsec:bench-protocol}

The solvers and their configurations are listed in \cref{tab:solvers}.
On the codes, the integer-programming and SAT solvers were run with two logical bases of the same logical space, a generic one
and, for all codes but the tile code \qc{269,4}, a sparse one, and are credited with either.
dist-m4ri was run in its two modes, on our matrices and on Stim's detector error model, with and without its model
lower bound: with the bound, our adapter calls dist-m4ri's own \texttt{compute\_model\_lower\_bound} within half of
the time budget and starts the binary from that bound, and its bracketing mode runs without codeword export, which
would otherwise continue to enumerate codewords after its bounds meet.
Heuristic methods supply upper bounds only and are not counted as certifying an instance.

Each run is pinned to its own cores of one socket of an AMD EPYC 9654 processor.
In the first tier, each solver runs with one thread, 600~s and 16~GB.
In the second tier, each solver runs again on the instances it did not certify, for one hour, with 32 threads and
90~GB for solvers that use threads and one thread and 16~GB otherwise.
Within each family, instances run in order of size, and after two uncertified instances in a row (one in the second
tier) the larger ones are skipped for that solver.
A run counts as exact only if the solver certifies a lower bound equal to its upper bound and the harness verifies
the witness against the matrices; Qubitserf returns values without a witness, so its values are checked only against the other bounds.
The harness also checks every bound against the ground truth and against the bounds of every other run; no record
contradicts another.
Our method runs in one configuration throughout (\texttt{dp=2}, default settings).

\begin{table}[t]
\centering
\caption{Solvers in the benchmark.}
\label{tab:solvers}
\small
\begin{tabular}{@{}>{\raggedright\arraybackslash}p{0.19\linewidth}>{\raggedright\arraybackslash}p{0.35\linewidth}>{\raggedright\arraybackslash}p{0.40\linewidth}@{}}
\toprule
method & solver & configurations \\
\midrule
connected clusters & dist-m4ri 0.14.7, commit bed621e\cite{dumer_2017_verification,distm4ri_code} & cluster enumeration, with and without random-window upper bounds; our matrices or Stim's model; with and without its model lower bound \\
 & Qubitserf 0.1.0\cite{qubitserf_code} & connected clusters, Brouwer--Zimmermann and meet in the middle, all on codes only \\
integer programming & HiGHS (SciPy 1.16.3)\cite{huangfu_2018_highs} & 1 thread \\
 & CP-SAT (OR-Tools 9.15)\cite{cpsat_code} & 1 or 32 workers \\
 & Gurobi 13.0.3\cite{gurobi_manual} & 1 or 32 threads \\
SAT and MaxSAT & CaDiCaL (PySAT 1.8)\cite{biere_2024_cadical,ignatiev_2018_pysat} & incremental cardinality bound \\
 & RC2\cite{ignatiev_2019_rc2} & own encoding; Stim's encoding \\
 & CASHWMaxSAT-CorePlus\cite{maxsat_2023} & own encoding; Stim's encoding \\
 & enMaxCDCL\cite{zhang_2026_enmaxcdcl,enmaxcdcl_code} & own encoding; Stim's encoding \\
heuristics (upper bounds) & BP-OSD (ldpc 2.4.1)\cite{roffe_2020_decoding}, information-set decoding, Stim's undetectable-error search, dist-m4ri random windows & witnesses only \\
\bottomrule
\end{tabular}
\end{table}

\subsection{Results}
\label{subsec:bench-results}

\begin{figure}[t]
\centering
\includegraphics{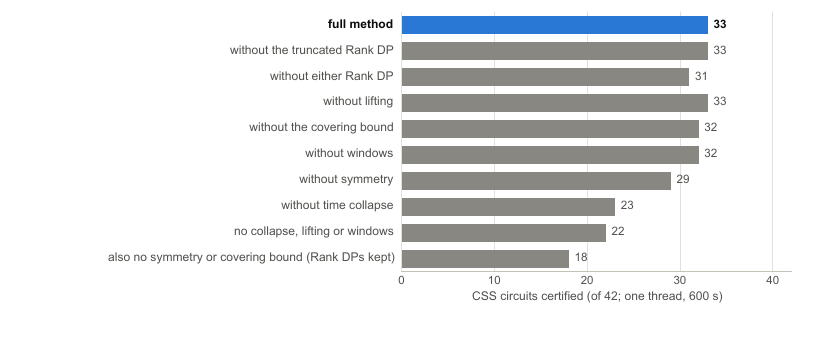}
\caption{Ablation of our method on the 42 CSS memory circuits of the benchmark (sector problem; one
thread, 600~s): the full method, one part switched off at a time, the method without time collapse, lifting and
windows, and the method without symmetry and covering bound as well, which keeps the two Rank DPs. ``Without either Rank DP'' switches off both
Rank DP evaluators, which loses the surface-code circuits of distance 23 and 25.}
\label{fig:si-ablation}
\end{figure}

With one thread and 600~s our method certifies 59 of the 74 instances, and 70 over both
tiers; all other solvers together certify 38 and 53.
On the circuits the counts are 38 and 45 against 25 and
34, and on the codes 21 and 25 against 13 and
19.
18 instances are certified by ours alone and 1 by another solver alone, the many-hypercube
code \qc{1296,256,16}, which SAT certifies in about 20 minutes and for which every column order we tried has width 85.
The three other instances that we do not certify are the mitten \qc{200,40,12} circuit, for which the first tier
certifies $d_{\mathrm{circ}}\geq9$ and which the give-up rule skips in the second (its value 9 comes from the audit
runs, \cref{sec:audit}), the mitten \qc{300,60,14} circuit, for which we certify $d_{\mathrm{circ}}\geq11$ within the
hour (its value 12 comes from a longer run, \cref{subsec:extended}), and the planar \qc{537,14} code, for which the
bounds are 20 and 30.
Per solver, over the two tiers: dist-m4ri 35 and 47 (its configurations without the model bound
33 and 42, the earlier build 0.9.0 17 and 26), Gurobi 12 and 22, CaDiCaL 12 and 20, RC2
11 and 15, CP-SAT 7 and 14, HiGHS 12 and 13,
CASHWMaxSAT 11 and 13, enMaxCDCL 11 and 11, and Qubitserf's
connected-cluster method 6 and 10.
\Cref{tab:bench-circuits,tab:bench-codes} give, for every instance, our time and that of the fastest other exact
solver in each tier.
The ablation of \cref{fig:si-ablation} switches off one part of our method at a time on the 42 CSS
circuits; the full method certifies 33 of them in the first tier.

\begin{table}[tp]
\centering
\caption{Circuits of the benchmark: columns of the problem given to every solver, circuit distance $d_{\mathrm{circ}}$, and the time
to the certified distance of our method and of the fastest other exact solver. Tier 1: one thread, 600~s; tier 2:
one hour, 32 threads where the solver uses them. An empty cell: not certified in that tier, not run because the
instance was certified in the first tier, or skipped by the give-up rule.}
\label{tab:bench-circuits}
\resizebox{\linewidth}{!}{\begin{tabular}{@{}lrrllll@{}}
\toprule
instance & columns & $d$ & ours, tier 1 & ours, tier 2 & best other, tier 1 & best other, tier 2 \\
\midrule
BB $[[108,8,10]]$, best depth-7 schedule & 3,348 & 10 & 13.7\,s &  &  & 184\,s (dist-m4ri, model bound) \\
BB $[[144,12,12]]$, best depth-7 schedule & 4,464 & 11 & 83.9\,s &  &  & 3,117\,s (dist-m4ri, model bound) \\
BB $[[72,12,6]]$, 6 rounds & 2,232 & 6 & 0.3\,s &  & 0.1\,s (dist-m4ri) & 534\,s (Gurobi) \\
BB $[[90,8,10]]$, 10 rounds & 4,590 & 8 & 0.5\,s &  & 33.5\,s (dist-m4ri, model bound) &  \\
BB $[[108,8,10]]$, 10 rounds & 5,508 & 8 & 0.6\,s &  & 37.2\,s (dist-m4ri, model bound) &  \\
BB $[[144,12,12]]$, 12 rounds & 8,784 & 10 & 1.6\,s &  &  & 348\,s (dist-m4ri, model bound) \\
BB $[[288,12,18]]$, 18 rounds & 26,208 & 16 &  & 220\,s &  &  \\
weight-8 BB $[[72,14]]$ & 3,096 & 8 & 5.0\,s &  & 205\,s (dist-m4ri) &  \\
weight-8 BB $[[144,16]]$ & 6,192 & 10 & 185\,s &  &  &  \\
Cornucopia $[[252,130]]$ & 8,568 & 5 & 2.4\,s &  & 0.4\,s (dist-m4ri) &  \\
Cornucopia $[[576,292]]$ & 19,584 & 8 & 435\,s &  &  & 214\,s (dist-m4ri, model bound) \\
Cornucopia $[[900,454]]$ & 30,600 & 9 &  & 509\,s &  &  \\
Cornucopia $[[1044,526]]$ & 35,496 & 10 &  & 927\,s &  &  \\
left-right gross $[[144,12]]$ & 4,464 & 11 & 22.3\,s &  &  & 706\,s (dist-m4ri, model bound) \\
mitten $[[150,30]]$ & 9,750 & 8 & 153\,s &  &  & 21.0\,s (dist-m4ri, model bound) \\
mitten $[[300,60]]$ & 11,820 & 12 &  &  &  &  \\
mitten $[[200,40]]$ & 15,560 & 9 &  &  &  &  \\
routing $[[54,8]]$ & 1,998 & 6 & 1.4\,s &  & 0.1\,s (dist-m4ri) & 206\,s (Gurobi) \\
routing $[[90,8]]$ & 3,330 & 9 & 6.1\,s &  & 374\,s (dist-m4ri, model bound) & 17.0\,s (dist-m4ri, model bound) \\
routing $[[110,8]]$ & 4,070 & 10 & 41.0\,s &  &  & 252\,s (dist-m4ri, model bound) \\
routing $[[140,8]]$ & 5,180 & 12 &  & 88.8\,s &  &  \\
routing $[[200,24]]$ & 11,700 & 10 &  & 509\,s &  &  \\
self-dual BB $[[64,8]]$ & 2,752 & 8 & 6.4\,s &  & 99.6\,s (dist-m4ri, model bound) &  \\
self-dual BB $[[120,8]]$ & 5,160 & 12 &  & 1,174\,s &  &  \\
shift automorphism $[[120,8]]$ & 3,720 & 8 & 12.9\,s &  & 2.3\,s (dist-m4ri, model bound) &  \\
STAR $[[42,6]]$ & 1,806 & 6 & 1.2\,s &  & 0.2\,s (dist-m4ri) & 389\,s (Gurobi) \\
color code, $d=3$ & 72 & 2$^\dagger$ & 0.1\,s &  & 0.0\,s (dist-m4ri) &  \\
color code, $d=5$ & 1,104 & 3$^\dagger$ & 0.9\,s &  & 0.0\,s (dist-m4ri) &  \\
color code, $d=7$ & 3,651 & 4$^\dagger$ & 1.8\,s &  & 0.0\,s (dist-m4ri) & 379\,s (CP-SAT) \\
color code, $d=9$ & 7,683 & 5$^\dagger$ & 38.8\,s &  & 0.4\,s (dist-m4ri) &  \\
color code, $d=11$ & 16,422 & 6$^\dagger$ & 280\,s &  & 19.7\,s (dist-m4ri, DEM input, model bound) &  \\
superdense color code, $d=7$ & 1,248 & 7 & 0.7\,s &  & 0.3\,s (dist-m4ri) & 37.7\,s (Gurobi) \\
superdense color code, $d=9$ & 2,788 & 9 & 17.0\,s &  & 245\,s (dist-m4ri, model bound) &  \\
superdense color code, $d=11$ & 5,250 & 11 &  & 1,793\,s &  &  \\
surface code, $d=3$ & 55 & 3 & 0.1\,s &  & 0.0\,s (dist-m4ri) &  \\
surface code, $d=5$ & 301 & 5 & 0.1\,s &  & 0.0\,s (dist-m4ri) &  \\
surface code, $d=7$ & 883 & 7 & 0.2\,s &  & 0.0\,s (dist-m4ri) & 82.5\,s (CP-SAT) \\
surface code, $d=9$ & 1,945 & 9 & 0.2\,s &  & 0.3\,s (dist-m4ri, model bound) & 815\,s (Gurobi) \\
surface code, $d=11$ & 3,631 & 11 & 0.2\,s &  & 0.5\,s (dist-m4ri, model bound) &  \\
surface code, $d=13$ & 6,085 & 13 & 0.3\,s &  & 2.0\,s (dist-m4ri, model bound) & 1,287\,s (dist-m4ri) \\
surface code, $d=15$ & 9,451 & 15 & 0.8\,s &  & 60.9\,s (dist-m4ri, model bound) &  \\
surface code, $d=17$ & 13,873 & 17 & 1.9\,s &  &  & 9.7\,s (dist-m4ri, model bound) \\
surface code, $d=19$ & 19,495 & 19 & 1.7\,s &  &  & 188\,s (dist-m4ri, model bound) \\
surface code, $d=21$ & 26,461 & 21 & 2.1\,s &  &  &  \\
surface code, $d=23$ & 34,915 & 23 & 2.9\,s &  &  &  \\
surface code, $d=25$ & 45,001 & 25 & 8.6\,s &  &  &  \\
Walking Cat $[[54,2]]$ & 2,322 & 8 & 8.1\,s &  & 104\,s (dist-m4ri, model bound) &  \\
\bottomrule
\end{tabular}
}
\end{table}

\begin{table}[t]
\centering
\caption{Codes of the benchmark, sector $d_X$; columns as in \cref{tab:bench-circuits}.}
\label{tab:bench-codes}
\resizebox{\linewidth}{!}{\begin{tabular}{@{}lrrllll@{}}
\toprule
instance & columns & $d$ & ours, tier 1 & ours, tier 2 & best other, tier 1 & best other, tier 2 \\
\midrule
BB $[[72,12,6]]$ & 72 & 6 & 0.2\,s &  & 0.0\,s (dist-m4ri) &  \\
BB $[[90,8,10]]$ & 90 & 10 & 0.2\,s &  & 0.0\,s (dist-m4ri) &  \\
BB $[[108,8,10]]$ & 108 & 10 & 0.2\,s &  & 0.0\,s (dist-m4ri) &  \\
BB $[[144,12,12]]$ & 144 & 12 & 0.2\,s &  & 0.1\,s (dist-m4ri) & 59.5\,s (CP-SAT) \\
BB $[[288,12,18]]$ & 288 & 18 & 1.8\,s &  & 115\,s (dist-m4ri) & 2,006\,s (Gurobi) \\
color $[[271,1,19]]$ & 271 & 19 & 1.4\,s &  & 75.3\,s (dist-m4ri) & 200\,s (Gurobi, sparse L) \\
color $[[397,1,23]]$ & 397 & 23 & 2.2\,s &  &  & 1,132\,s (dist-m4ri) \\
color $[[547,1,27]]$ & 547 & 27 & 2.4\,s &  &  &  \\
color $[[721,1,31]]$ & 721 & 31 & 351\,s &  &  &  \\
concatenated Steane $[[343,1,27]]$ & 343 & 27 & 1.3\,s &  & 3.6\,s (SAT) &  \\
concatenated Steane $[[2401,1,81]]$ & 2,401 & 81 & 1.2\,s &  &  & 977\,s (SAT) \\
Cornucopia $[[1044,526]]$ & 1,044 & 12$^\dagger$ & 2.9\,s &  & 133\,s (dist-m4ri) & 1,633\,s (Qubitserf CC) \\
thin tile $[[259,2]]$, $d_X=36$ & 259 & 36 & 1.2\,s &  & 2.5\,s (CASHWMaxSAT, sparse L) & 2.3\,s (CP-SAT, sparse L) \\
thin tile $[[269,4]]$, $d_X=28$ & 261 & 28 & 1.2\,s &  & 507\,s (CASHWMaxSAT) & 683\,s (CASHWMaxSAT) \\
thin tile $[[365,10]]$, $d_X=31$ & 365 & 31 & 13.1\,s &  &  & 816\,s (SAT) \\
thin tile $[[537,14]]$ & 537 & $\geq$20 &  &  &  &  \\
weight-8 BB $[[144,6,15]]$ & 144 & 15 & 1.4\,s &  & 86.5\,s (dist-m4ri) & 419\,s (Gurobi, sparse L) \\
multivariate multicycle $[[324,12,18]]$ & 324 & 18 & 2.9\,s &  &  & 216\,s (dist-m4ri) \\
LLM-searched BB $[[360,12,24]]$ & 360 & 24 &  & 66.7\,s &  &  \\
balanced product $[[390,78,16]]$ & 390 & 16 & 42.6\,s &  &  &  \\
balanced product $[[625,125,20]]$ & 625 & 20 &  & 59.1\,s &  &  \\
GALA $[[132,30,12]]$ & 132 & 12 & 1.4\,s &  & 30.5\,s (dist-m4ri) & 191\,s (Gurobi, sparse L) \\
many-hypercube $[[1296,256,16]]$ & 1,296 & 16 &  &  &  & 1,225\,s (SAT, sparse L) \\
Kasai $[[372,130,16]]$ & 372 & 16 & 120\,s &  &  & 1,817\,s (dist-m4ri) \\
Kasai $[[444,154,18]]$ & 444 & 18 &  & 74.3\,s &  &  \\
Kasai $[[492,170,20]]$ & 492 & 20 &  & 1,054\,s &  &  \\
self-dual BB $[[200,12,12]]$ & 200 & 12 & 0.3\,s &  & 1.2\,s (dist-m4ri) & 1,204\,s (CP-SAT) \\
\bottomrule
\end{tabular}
}
\end{table}

\begin{table}[t]
\centering
\caption{Both CSS distances of the codes of the benchmark. The sector $d_Z$ was computed with the label-free search,
with the covering bound and the verified symmetries of each code.}
\label{tab:bench-codes-both}
\small
\begin{tabular}{@{}lrrrr@{}}
\toprule
code & $n$ & $k$ & $d_X$ & $d_Z$ \\
\midrule
BB $[[72,12,6]]$ & 72 & 12 & 6 & 6 \\
BB $[[90,8,10]]$ & 90 & 8 & 10 & 10 \\
BB $[[108,8,10]]$ & 108 & 8 & 10 & 10 \\
GALA $[[132,30,12]]$ & 132 & 30 & 12 & 12 \\
BB $[[144,12,12]]$ & 144 & 12 & 12 & 12 \\
weight-8 BB $[[144,6,15]]$ & 144 & 6 & 15 & 15 \\
self-dual BB $[[200,12,12]]$ & 200 & 12 & 12 & 12 \\
thin tile $[[259,2]]$, $d_X=36$ & 259 & 2 & 36 &  \\
thin tile $[[269,4]]$, $d_X=28$ & 269 & 4 & 28 &  \\
color $[[271,1,19]]$ & 271 & 1 & 19 &  \\
BB $[[288,12,18]]$ & 288 & 12 & 18 & 18 \\
multivariate multicycle $[[324,12,18]]$ & 324 & 12 & 18 & 18 \\
concatenated Steane $[[343,1,27]]$ & 343 & 1 & 27 &  \\
LLM-searched BB $[[360,12,24]]$ & 360 & 12 & 24 & 24 \\
thin tile $[[365,10]]$, $d_X=31$ & 365 & 10 & 31 &  \\
Kasai $[[372,130,16]]$ & 372 & 130 & 16 & 16 \\
balanced product $[[390,78,16]]$ & 390 & 78 & 16 & 16 \\
color $[[397,1,23]]$ & 397 & 1 & 23 &  \\
Kasai $[[444,154,18]]$ & 444 & 154 & 18 & 18 \\
Kasai $[[492,170,20]]$ & 492 & 170 & 20 & 20 \\
thin tile $[[537,14]]$ & 537 & 14 & $\geq20$ &  \\
color $[[547,1,27]]$ & 547 & 1 & 27 &  \\
balanced product $[[625,125,20]]$ & 625 & 125 & 20 & 20 \\
color $[[721,1,31]]$ & 721 & 1 & 31 &  \\
Cornucopia $[[1044,526]]$ & 1044 & 526 & 12 & 12 \\
many-hypercube $[[1296,256,16]]$ & 1296 & 256 & 16 &  \\
concatenated Steane $[[2401,1,81]]$ & 2401 & 1 & 81 &  \\
\bottomrule
\end{tabular}

\end{table}

\subsection{The prior benchmark}
\label{subsec:bench-prior}

The most recent benchmark of distance methods\cite{webster_2026_distance} has 126 CSS codes and circuits: surface-,
color- (midout and superdense) and bivariate bicycle circuits, hyperbolic surface and color codes, lifted-product,
bivariate bicycle and quantum Tanner codes.
We ran our method on that study's own files, converted to the same matrix format, under that study's limits (one
core, 8~GB and eight hours per instance, on all 126 instances) and in our two tiers (the first had a memory cap of
16~GB but used at most 2.5~GB).
The study's own results come from its published spreadsheets: an instance counts for one of its five exact methods
(Gurobi, SCIP, CLISAT, dist-m4ri's connected-cluster mode and Magma) when that method reports a distance within its
eight-hour limit.
Under that study's limits our method certifies 95 instances, against 73 for its five methods together (dist-m4ri 71,
Gurobi 70, CLISAT 67, SCIP 64, Magma 39); with one thread and 600~s it certifies 86, and with 32 threads and an
hour 97.
Every value of the five methods agrees with ours.
Of the 53 instances without an exact value from that study's methods, 24 are certified here; 23 confirm its best
heuristic value, and one, the \qc{288,12,18} circuit, has distance 16 instead of 24.
The superdense circuit of distance 11, open for every method there, has distance 11; it was settled by the
truncated Rank DP (\cref{sec:sparse-dp}) in the second tier.
\Cref{tab:bench-prior} gives the counts by family.

\begin{table}[t]
\centering
\caption{The 126 instances of the prior benchmark\cite{webster_2026_distance}, by family: instances certified by our
method (one thread, 600~s; under that study's limits; with 32 threads and an hour), by our runs of the current
dist-m4ri and Gurobi (one thread, 600~s), and by that study's five exact methods, per method (Gurobi, SCIP, CLISAT,
dist-m4ri, Magma) and together.}
\label{tab:bench-prior}
\small
\begin{tabular}{@{}lrrrrrrll@{}}
\toprule
 & & \multicolumn{3}{c}{this method} & dist-m4ri & Gurobi & \multicolumn{2}{c}{ref.~\citenum{webster_2026_distance}} \\
\cmidrule(lr){3-5}\cmidrule(lr){8-9}
family & inst. & 600\,s & 8\,h & +32 thr. & 600\,s & 600\,s & per method & any \\
\midrule
surface-code circuits & 5 & 5 & 5 & 5 & 5 & 3 & 3/3/2/4/2 & 4 \\
color-code circuits, midout & 5 & 5 & 5 & 5 & 5 & 5 & 5/4/5/5/4 & 5 \\
color-code circuits, superdense & 6 & 5 & 5 & 6 & 5 & 3 & 4/4/4/4/3 & 5 \\
BB circuits & 7 & 4 & 5 & 5 & 3 & 0 & 1/0/0/1/0 & 1 \\
hyperbolic surface codes & 21 & 21 & 21 & 21 & 21 & 17 & 21/21/21/21/14 & 21 \\
hyperbolic color codes & 21 & 21 & 21 & 21 & 21 & 16 & 20/17/20/21/9 & 21 \\
lifted-product codes & 19 & 14 & 19 & 19 & 6 & 5 & 6/6/6/6/3 & 6 \\
BB codes & 20 & 6 & 8 & 8 & 6 & 5 & 5/5/5/5/1 & 5 \\
quantum Tanner codes & 22 & 5 & 6 & 7 & 4 & 4 & 5/4/4/4/3 & 5 \\
\midrule
all & 126 & 86 & 95 & 97 & 76 & 58 & 70/64/67/71/39 & 73 \\
\bottomrule
\end{tabular}

\end{table}

\subsection{Longer runs outside the benchmark}
\label{subsec:extended}

To close instances that a few server-days can settle, we ran the same pipeline (\texttt{dp=2}) with more threads and with limits of up to seven days (\cref{tab:extended}).
These results lie outside the two tiers and do not enter the counts above.
Where a range of thresholds ran separately, the lower bound holds together with the earlier record of the same problem.
The upper bounds are explicit errors checked against the matrices, found by information-set decoding where the search had not reached them.
The prior benchmark's Tanner instances are $Z$ distances, and these codes have no qubit permutation that exchanges $X$ and $Z$, so we ran the $X$ distances as well.
The \qc{396,4} code shows why this matters: its $Z$ distance is 27, the prior benchmark's best known value, but its $X$ distance is 26, so its code distance is 26.

\begin{table}[t]
\centering
\caption{Longer runs outside the benchmark. Weights in the last column are the thresholds of each run.}
\label{tab:extended}
\small
\begin{tabular}{@{}>{\raggedright\arraybackslash}p{4.0cm}l>{\raggedright\arraybackslash}p{9.0cm}@{}}
\toprule
instance & value & runs \\
\midrule
Tanner \qc{288,2} & 24 & $Z$: 10.5 h on 32 threads; $X$: 22.9 h on 12 threads, with a logical of weight 24 \\
Tanner \qc{324,4} & 25 & $Z$: 8.8 h on 32 threads; $X$: up to weight 23, 16.6 h on 12 threads, weight 24, 8.3 h on 40 threads, with a logical of weight 25 \\
Tanner \qc{360,4} & 24 & $Z$: 3.1 h on 32 threads; $X$: 18.0 h on 6 threads, with a logical of weight 24 \\
Tanner \qc{396,4} & 26 & $Z$ distance 27: up to weight 25, 23.1 h on 32 threads, weight 26, 13.7 h on 89 threads, with a logical of weight 27; $X$ distance 26: 58.7 h on 50 threads, with a logical of weight 26 \\
bivariate bicycle \qc{648,4} & 32 & up to weight 30, 14.7 h on 32 threads; weight 31, 20.5 h on 57 threads; a logical of weight 32 \\
bivariate bicycle \qc{756,16} & 34 & up to weight 32, 50.9 h on 57 threads; weight 33, 96.0 h on 134 threads; a logical of weight 34 \\
many-hypercube $L=4$ & 16 & 5.3 h on 32 threads \\
mitten \qc{300,60,14} circuit & 12 & up to weight 11, 22.5 h on 32 threads; the 3-round fault set of weight 12 replayed in the 6-round circuit \\
Cornucopia \qc{2304,1156,16} circuit, $Z$ basis, 8 cycles & 14 & time-collapsed problem up to weight 12, last weight 25.3 h on 64 threads; weight 13 split over four servers, 2.4 days; a fault set of weight 14 \\
\bottomrule
\end{tabular}

\end{table}

\FloatBarrier
\section{Audit of published distances}
\label{sec:audit}

\subsection{Scope and protocol}
\label{subsec:audit-protocol}

We included papers from 2023 to 2026 that state distances of codes or circuits that we could reproduce: 26 papers
and one data page.
Provenance is the authors' own circuit or data files (A), public circuit files of the paper's schedule from another source (P), or a rebuild from the paper's text (R).
Every rebuild was first checked against values that the paper reports as exact: the rebuilt planar codes give all
of that paper's integer-programming distances and Table~III's lengths and routing counts, the rebuilt Cornucopia
circuits have the check matrices and the gate orders of the authors' repository, and every rebuilt code family has
the parameters $\qc{n,k}$ that its paper prints, some after a reading stated in \cref{subsec:audit-notes}.
Two rebuilds that fail this check are not used: the hexagonal directional tile codes of ref.~\citenum{gu_2026_planar},
whose sizes the paper does not give and whose distances our sizes do not reproduce, and the amended \qc{314,12,14}
example of the bivariate bicycle surface codes\cite{wang_2026_bbsurface}, whose text is inconsistent and whose rebuild
has $d_Z=12$.
Fault models are the papers' own, as used in their simulations: two-qubit depolarizing faults after the entangling
gates, with preparation and measurement faults; the row marked P uses the SI1000 noise of its files, and
\cref{subsec:ibm-distance} gives the authors' own circuits with their noise.
Where a paper's distance estimator used another fault model, both were computed (\cref{subsec:audit-notes}).

A published value counts as too high when the paper states it as the distance or gives it as an estimate, an expectation or a
bound that a reader takes as the distance, and a certified value is lower. A value that is not closed counts when the whole certified interval lies below it.
A value that depends on a convention the paper does not print, such as the numbering of the terms of a check, counts only
if every choice gives a lower value, and a schedule's name, such as hook-free, is not a published value. The seven values
that these two rules leave out are listed in \cref{tab:audit} but not counted.
Misprints in tables are not counted.
The prior benchmark's 27 for the Tanner code \qc{396,4} is not counted either: its instance asks for the $Z$ distance, which is 27 (\cref{subsec:extended}).
Where a value could not be decided, we give the interval between the certified lower bound and the lightest error
found.
Every exact value has a certified lower bound, with all lighter errors excluded, and an explicit error that is
checked against the matrices and, for circuits, replayed in Stim.
\Cref{tab:audit} lists the 69 audited values. Fig.~3 of the main text shows the 22 that are too high and those that other papers left open. The \qc{288,12,18} circuit appears twice among the 22, with two published values, so they cover 21 distinct codes and circuits.

\begingroup\small
\begin{longtable}{@{}>{\raggedright\arraybackslash}p{2.4cm}>{\raggedright\arraybackslash}p{2.9cm}>{\raggedright\arraybackslash}p{2.5cm}>{\raggedright\arraybackslash}p{1.35cm}c>{\raggedright\arraybackslash}p{4.1cm}@{}}
\caption{Every audited value of \cref{sec:audit}, by paper. Published: the value and how the paper states it. Certified: the exact value (bold where it corrects the published value or settles one left open), or the interval between the proved lower bound and the lightest error found. Values whose note begins with ``not counted'' are listed but not counted as too high (\cref{subsec:audit-protocol}). Prov.: A, the authors' circuit or data files; P, public circuit files of the paper's schedule from another source; R, rebuilt from the paper's text and checked against values it reports as exact.}\label{tab:audit}\\
\toprule paper & object & published (how stated) & certified & prov. & note \\ \midrule \endfirsthead
\toprule paper & object & published (how stated) & certified & prov. & note \\ \midrule \endhead
\bottomrule \endlastfoot
\multicolumn{6}{@{}l}{\textit{Circuit distances}} \\[2pt]
Mitten codes~\cite{bhardwaj_2026_mitten} & \qc{200,40,12} circuit & $\le$12 (estimated upper bound) & \textbf{9} & A & version 2 of the paper, whose Table 8 gives the circuit values as upper bounds estimated with single-qubit faults and whose text presents the schedules as distance-preserving; the certified values hold under that fault model too \\
 & \qc{300,60,14} circuit & $\le$14 (estimated upper bound) & \textbf{12} & A &  \\
 & \qc{500,100,16} circuit & $\le$16 (estimated upper bound) & \textbf{9} & A & 9 in the X basis; Z basis 11--16 \\
 & \qc{540,108,18} circuit & $\le$18 (estimated upper bound) & 10--13 & A & interval over both bases \\
 & \qc{630,126,20} circuit & $\le$20 (estimated upper bound) & 9--12 & A & interval over both bases \\
 & \qc{780,156,22} circuit & $\le$22 (estimated upper bound) & 9--16 & A & interval over both bases \\
 & \qc{975,195,24} circuit & $\le$24 (estimated upper bound) & 9--17 & A & interval over both bases \\
 & \qc{150,30,10} circuit & 8 (stated as the distance) & 8 & A & value of version 2 (version 1 implied 10) \\
\addlinespace[2pt]
Mitten codes (repository)~\cite{bhardwaj_2026_mitten} & \qc{330,66,12} circuit & none (schedule called hook-free) & 7 & A & not counted: no circuit distance printed; the repository calls the schedule hook-free, as it does for all eight mitten schedules, including that of \qc{150,30,10}, whose circuit distance the paper gives as 8; code distance exactly 12 \\
 & \qc{600,120,14} circuit & none (schedule called hook-free) & 10 & A & not counted: no circuit distance printed; Z basis 11, X basis 10; code distance exactly 14 \\
 & \qc{600,120,16} circuit & none (schedule called hook-free) & 10 & A & not counted: no circuit distance printed; Z basis 10, X basis 11; code distance exactly 16; 16 rounds, 501,960 fault mechanisms in the Z basis \\
 & \qc{840,168,18} circuit & none (schedule called hook-free) & 11--14 & A & not counted: no circuit distance printed; weight 10 excluded; fault sets of weight 14 in both bases \\
\addlinespace[2pt]
IBM bivariate bicycle codes~\cite{bravyi_2024_bb} & \qc{288,12,18} circuit & $\le$18 (upper bound) & \textbf{16} & P & standard depth-seven schedule; 16 in both bases in the public Stim files of the schedule (tesseract-decoder; CZ gates, SI1000 noise) and in the memory instruction of the architecture study; with the authors' public code and noise at 18 cycles, exactly 16 in both bases; the authors call the bound unlikely to be tight and the architecture study assumes 18 \\
 & \qc{72,12,6} circuit & $\le$6 (upper bound) & 6 & P & the published upper bound is exact \\
 & \qc{90,8,10} circuit & $\le$8 (upper bound) & 8 & P & the published upper bound is exact \\
 & \qc{108,8,10} circuit & $\le$8 (upper bound) & 8 & P & the published upper bound is exact \\
 & \qc{144,12,12} circuit & $\le$10 (upper bound) & 10 & P & the published upper bound is exact \\
\addlinespace[2pt]
Distance-finding benchmark~\cite{webster_2026_distance} & \qc{288,12,18} circuit & $\le$24 (best known value) & \textbf{16} & A & the study's own circuit file; best known value from heuristics; the same circuit as the row above \\
 & superdense circuit, $d=$11 & 8--11 (open) & \textbf{11} & A & open for every method; settled by the truncated Rank DP \\
\addlinespace[2pt]
Cornucopia codes~\cite{lu_2026_cornucopia} & \qc{900,454,10} circuit & $\le$10 (upper bound) & \textbf{9} & R & matrices and gate orders equal to the authors' repository; same value in their one-cycle setting \\
 & \qc{1044,526,12} circuit & $\le$11 (upper bound) & \textbf{10} & R &  \\
 & \qc{2304,1156,16} circuit & $\le$16 (upper bound) & \textbf{14} & R & Z basis, 8 cycles; a fault set of weight 14 also in the authors' one-cycle setting \\
 & \qc{2844,1426,18} circuit & $\le$18 (upper bound) & 12--16 & R & Z basis; 8 cycles; 12--17 in one cycle \\
 & \qc{252,130,6} circuit & $\le$5 (upper bound) & 5 & R & the published upper bound is exact \\
 & \qc{576,292,8} circuit & $\le$8 (upper bound) & 8 & R & the published upper bound is exact \\
\addlinespace[2pt]
Routing codes~\cite{zhang_2026_routing} & \qc{200,24,14} circuit & $\le$14 (upper bound) & \textbf{10} & R & weight-11 checks; the rebuilt code has distance 14 \\
 & \qc{200,16,17} circuit & $\le$17 (upper bound) & 11--14 & R &  \\
\addlinespace[2pt]
Shift automorphism circuits~\cite{kim_2026_shift} & gross code, dynamic circuit & 10 (estimate) & \textbf{8} & R & 8 under all 12 valid numberings; the static circuit with the same gate order also has 8 \\
 & \qc{120,8,12} code, dynamic circuit & 10 (estimate) & 8 & R & not counted: terms numbered as printed (the paper does not give the numbering): Z basis 8, X basis 10; 7 of the 12 valid numberings give 10 in both bases \\
\addlinespace[2pt]
Walking Cat~\cite{tripier_2026_walkingcat} & \qc{70,6,9} circuit & 9 (stated as the distance) & \textbf{8} & R & terms numbered as printed; 4 of the 144 numberings give 9 in the Z basis, none in both bases \\
 & \qc{102,22,9} circuit & 9 (stated as the distance) & 8 & R & not counted: terms numbered as printed (the paper does not give the numbering); 8 of the 576 numberings give 9 in both bases \\
 & \qc{54,2,10} circuit & 9 (stated as the distance) & 8 & R & not counted: terms numbered as printed; 2 of the 576 numberings give 9 in both bases \\
\addlinespace[2pt]
Left-right circuits~\cite{strikis_2026_leftright} & gross code, three-color circuit & 11 (conjecture) & \textbf{11} & R & conjectured; 11 in both bases \\
\addlinespace[2pt]
Vine codes~\cite{nixon_2026_vine} & four patches of distance 10 & 10 (conjecture) & \textbf{10} & A & conjectured and stated as unverified; the authors' Stim files \\
\addlinespace[2pt]
Amplified qLDPC codes~\cite{liang_2026_amplified} & \qc{90,8,8} level-1 circuit & 6--8 (range) & \textbf{8} & A & certified at least 6 in the paper; the authors' builder \\
\addlinespace[2pt]
\midrule
\multicolumn{6}{@{}l}{\textit{Code distances}} \\[2pt]
Tile codes~\cite{steffan_2025_tile} & \qc{512,18} code & 23 (expected) & \textbf{21} & R & weight-10 code, upper bound 23 expected to be exact; layout inferred as the only square layout with $n=$512, which reproduces the authors' exact \qc{512,18,19} \\
\addlinespace[2pt]
Nearest-neighbor planar codes~\cite{gu_2026_planar} & \qc{259,2} code, X distance & $\le$37 (upper bound) & \textbf{36} & R & one below the published upper bound; the rebuild reproduces the paper's integer-programming distances \\
 & \qc{537,14} code, X distance & $\le$35 (upper bound) & 20--30 & R & lower bound from a 32-thread run of one hour \\
 & \qc{269,4} code, X distance & $\le$28 (upper bound) & 28 & R & the published upper bound is exact \\
 & \qc{365,10} code, X distance & $\le$31 (upper bound) & 31 & R & the published upper bound is exact \\
\addlinespace[2pt]
Multivariate multicycle codes~\cite{mian_2026_multicycle} & \qc{324,12} code & 20 (randomized search) & \textbf{18} & R & the code as printed; distance from a randomized method; several other table rows have visible misprints \\
\addlinespace[2pt]
Mitten codes (repository)~\cite{bhardwaj_2026_mitten} & \qc{840,168,18} code & 18 (estimate) & \textbf{17} & A & sampled estimate; X distance 17, Z distance 17--18 \\
 & \qc{300,60,9} code & 9 (estimate) & \textbf{10} & A & the certified value is higher; 9 is the circuit distance of its schedule \\
 & \qc{330,66,12} code & 12 (estimate) & 12 & A & the sampled estimate is exact \\
 & \qc{600,120,14} code & 14 (estimate) & 14 & A & the sampled estimate is exact \\
 & \qc{600,120,16} code & 16 (estimate) & 16 & A & the sampled estimate is exact \\
 & \qc{560,112,14} code & 14 (estimate) & 14 & A & abelian lifted-product code; attains the paper's proven ceiling \\
\addlinespace[2pt]
Weight-8 bivariate bicycle codes~\cite{lu_2026_weight8} & \qc{144,6} code & $\ge$15 (lower bound) & \textbf{15} & R & published as at least 15 \\
 & \qc{144,10} code & $\ge$11 (lower bound) & \textbf{14} & R & published as at least 11 \\
\addlinespace[2pt]
Modular hyperbolic codes~\cite{higgott_2026_hyperbolic} & \qc{1600,104} code & 14--16 (range) & \textbf{16} & R & rebuilt; the paper's range closes at its upper end \\
 & \qc{2048,36} code & 20--24 (range) & \textbf{24} & R & rebuilt; the paper's range closes at its upper end \\
 & \qc{5376,88} code & 26--28 (range) & \textbf{28} & R & rebuilt; the paper's range closes at its upper end \\
 & \qc{1344,340} code & 10--12 (range) & \textbf{12} & R & rebuilt; the paper's range closes at its upper end \\
 & \qc{5376,340} code & 18--20 (range) & \textbf{20} & R & rebuilt; the paper's range closes at its upper end \\
 & \qc{31200,1954} code & 22--24 (range) & \textbf{24} & R & rebuilt; the paper's range closes at its upper end \\
 & \qc{15600,3904} code & 10--14 (range) & \textbf{14} & R & rebuilt; the paper's range closes at its upper end \\
 & \qc{6912,220} code & 20--24 (range) & \textbf{24} & R & rebuilt; the paper's range closes at its upper end \\
 & \qc{10976,396} code & 22--24 (range) & \textbf{24} & R & rebuilt; the paper's range closes at its upper end \\
 & \qc{12288,260} code & 26--28 (range) & \textbf{28} & R & rebuilt; the paper's range closes at its upper end \\
 & \qc{23040,676} code & 24--28 (range) & \textbf{28} & R & rebuilt; the paper's range closes at its upper end \\
 & \qc{2688,1636} code & 6--8 (range) & \textbf{8} & R & rebuilt; the paper's range closes at its upper end \\
 & \qc{4896,2044} code & 8--10 (range) & \textbf{10} & R & rebuilt; the paper's range closes at its upper end \\
 & \qc{12180,4876} code & 8--10 (range) & \textbf{10} & R & rebuilt; the paper's range closes at its upper end \\
 & \qc{13680,4108} code & 8--12 (range) & \textbf{12} & R & rebuilt; the paper's range closes at its upper end \\
\addlinespace[2pt]
IBM bivariate bicycle codes~\cite{bravyi_2024_bb} & \qc{360,12} code & $\le$24 (upper bound) & 24 & A & the published upper bound is exact \\
\addlinespace[2pt]
Balanced-product codes~\cite{hong_2026_rate} & \qc{390,78} code & $\le$16 (upper bound) & 16 & R & the published upper bound is exact \\
 & \qc{525,105} code & $\le$18 (upper bound) & 18 & R & the published upper bound is exact \\
 & \qc{550,110} code & $\le$18 (upper bound) & 18 & R & the published upper bound is exact \\
 & \qc{625,125} code & $\le$20 (upper bound) & 20 & R & the published upper bound is exact \\
\addlinespace[2pt]
\end{longtable}

\endgroup

\subsection{Further values confirmed}
\label{subsec:audit-confirmed}

Besides the values of \cref{tab:audit}, the following published values were reproduced exactly.
All code distances that the papers state as exact are among them.
\begin{itemize}
\item Routing codes\cite{zhang_2026_routing}: the BP-OSD bounds of all eight weight-seven circuits are exact (6, 7,
  8, 9, 9, 10, 11, 12); the value 7 printed for \qc{80,8} in one table is a misprint, as the paper's other tables
  give 8.
\item Cornucopia codes\cite{lu_2026_cornucopia}: both CSS distances of all seven codes.
\item STAR architecture\cite{ismail_2026_star}: the paper's histogram of the distances of the 212 valid depth-eight
  schedules of its \qc{42,6,6} code (192 at 4, 20 at 5), reproduced exactly, in the family of schedules that we
  infer from its description, under the two index conventions that give it (\cref{subsec:design-other}).
\item Time-dynamic shift circuits\cite{kim_2026_shift}: the estimates of the other two codes.
\item Self-dual bivariate bicycle codes\cite{liang_2025_selfdual}: all 57 code distances; GALA
  codes\cite{yang_2026_gala}: all 11 codes over abelian groups; pair-partition codes\cite{okada_2026_pair,lee_2026_ultrahigh} from K.~Kasai's data page: 10 of
  10 stated distances.
\item Multivariate multicycle codes\cite{mian_2026_multicycle}: 214 of 215 codes computed exactly, 201 equal to the
  printed value; the others differ in table rows, several of which carry visible misprints, so only the code of \cref{tab:audit} is counted.
\item Weight-eight bivariate bicycle\cite{lu_2026_weight8}, clustered lifted-product\cite{gu_2026_qgpu}, balanced-product\cite{hong_2026_rate} and LLM-guided
  bivariate bicycle codes\cite{cruzbenito_2026_llm}: 35 codes, all consistent with the published values.
\item Planar code families (tile\cite{bae_2026_tile}, barbell\cite{choe_2026_barbell}, bivariate bicycle
  surface\cite{wang_2026_bbsurface} and space-group codes\cite{xu_2026_spacegroups}): about 200 reported instances, all consistent with the published values apart from the amended example above; for the bivariate bicycle surface codes, whose
  randomized bounds come without region sizes, some size attains the reported value in 87 of 88 rows, and exactly one size in 86, which identifies the authors' size.
\item Vine codes\cite{nixon_2026_vine}: the circuit distances of the paper's Fig.~7.
\item The benchmark of distance methods\cite{webster_2026_distance}: all 73 values certified by its five exact
  methods agree with ours.
\item IBM bivariate bicycle codes\cite{bravyi_2024_bb}: the bounds 6, 8, 8 and 10 of the \qc{72,12,6},
  \qc{90,8,10}, \qc{108,8,10} and \qc{144,12,12} circuits are exact.
\item Modular hyperbolic codes\cite{higgott_2026_hyperbolic}: the circuits are not published, and our reconstructed schedules match the graph-like upper bounds of all 64 surface-code
  circuits computed, which are exact for them.
\end{itemize}

\subsection{Notes on individual values}
\label{subsec:audit-notes}

\begin{figure}[t]
\centering
\fullwidthfigure{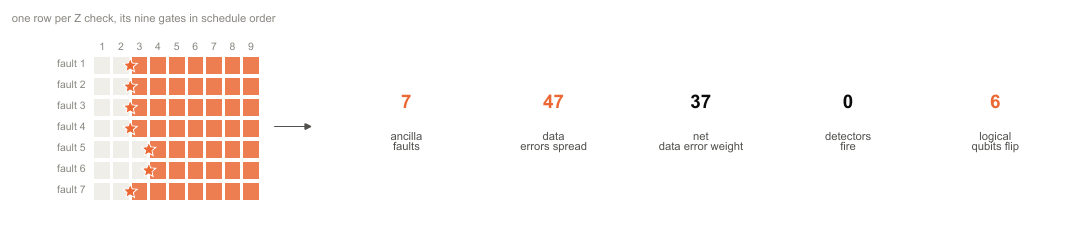}
\caption{A minimum fault set of the structured mitten \qc{330,66,12} circuit with its published schedule ($X$
memory basis). Seven ancilla faults, each after the second or third of the nine gates of a $Z$ check (stars),
spread $Z$ errors to the data qubits of the remaining gates (orange). The 47 data errors leave a net error of weight
37 that no detector sees and that flips six logical qubits.}
\label{fig:si-hook}
\end{figure}

\paragraph{Mitten codes.}
The circuit values are those of version 2 of the paper\cite{bhardwaj_2026_mitten}, which states 8 for the
\qc{150,30,10} circuit, as we find.
Its text presents the schedules as preserving the circuit-level distance, and its Table~8 gives the circuit values as
upper bounds estimated with single-qubit faults, so we list them as estimated upper bounds.
The paper's estimator used single-qubit Pauli faults, while its simulations used two-qubit depolarizing faults
after every gate.
Single-qubit faults on both qubits of every gate give exactly the same problems as ours, for all codes in both
bases.
Single-qubit faults after the gates alone give the same $Z$-basis problems and lack one column per data qubit in
the $X$ basis; every fault of this model is a fault of ours, and every recorded $X$-basis fault set uses only faults
of this model, so every value and interval holds in it too.
Idle faults can only lower a distance.
The structured-mitten circuits of the repository have no circuit distance printed. The repository calls their
schedules hook-free, but it gives the same name to all eight schedules of the paper, including that of the
\qc{150,30,10} circuit, so the name does not state a distance, and these circuits are not counted; the code
distances of the \qc{330,66,12}, \qc{600,120,14} and \qc{600,120,16} codes are exactly 12, 14 and 16, and their
circuit distances 7, 10 and 10.
\Cref{fig:si-hook} shows a minimum fault set of the structured \qc{330,66,12} circuit.

\paragraph{Cornucopia codes and the modular architecture.}
The circuit distance of \qc{2304,1156,16} is 14 (8 cycles, $Z$ basis).
Its search at weight 13, the last threshold of the time-collapsed problem, consists of 251,040 independent subsearches;
we split them over four servers, each recording the subsearches it finished, and a last run over the union of these
records found all of them finished, so weight 13 is excluded (\cref{tab:extended}).
The \qc{900,454,10} value and the weight-14 fault set of \qc{2304,1156,16} hold also in the authors' one-cycle setting
with CNOT faults only.
For the \qc{2844,1426,18} circuit, the interval 12 to 16 ($Z$ basis, 8 cycles; 12 to 17 in one cycle) implies that the leading
order of the logical error rate is at most $p^8$, below the exponent $d/2=9$ with which the Cornucopia paper
extrapolates a rate of $2.6\times10^{-16}$ per qubit and cycle\cite{lu_2026_cornucopia}, on which a modular architecture
builds\cite{lu_2026_modular}; how far that value moves depends on the number of minimum fault sets, which we did not count.

\paragraph{Index conventions.}
For the Walking Cat circuits\cite{tripier_2026_walkingcat}, the paper does not print how the terms of the checks are
numbered; with the terms numbered as printed, all three circuits have distance 8. Of the 576 numberings, 8 give 9 for \qc{102,22,9} and 2 give 9 for \qc{54,2,10}, in both bases, so these two values are not counted, and of the 144 numberings of \qc{70,6,9}, 4 give 9 in the $Z$ basis only, so its distance is 8 under every numbering.
For the shift circuits\cite{kim_2026_shift}, the lattices of the \qc{120,8,12} and \qc{170,16,10} codes as printed give
$k=0$; we use those of the construction the paper cites, which have the same polynomials and give the stated
parameters.
The paper does not print the numbering of the terms either; the terms numbered as printed
give 8 in the $Z$ basis and 10 in the $X$ basis for \qc{120,8,12}, and 7 of its 12 valid numberings give 10 in both, so this value is not counted;
for the gross code the value 8 holds under all 12 valid numberings, and the static circuit
with the same gate order also has 8, so the loss comes from the gate order rather than the dynamics.

\paragraph{Planar and tile codes.}
For the planar \qc{259,2} code\cite{gu_2026_planar}, the published value is an upper bound from random information
sets, and the exact value is one below it; with 36, the code's ranking value in that paper drops from 1.14 to 1.11.
For the tile code\cite{steffan_2025_tile} the layout is inferred as the only square layout with $n=512$; the same
layout reproduces the authors' exact \qc{512,18,19} code.
For the planar \qc{537,14} code, the benchmark instance records the lower bound 19 known when it was exported
(\cref{tab:bench-codes}); a 32-thread run of one hour raised it to 20.

\begin{figure}[t]
\centering
\includegraphics{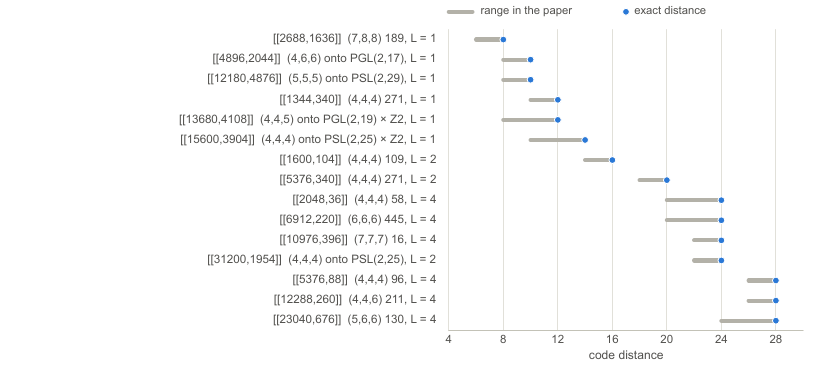}
\caption{Fifteen of the modular hyperbolic codes whose distances ref.~\citenum{higgott_2026_hyperbolic} could only
bracket: the bracket of that paper (gray) and the exact distance (blue), with the triangle group of the tiling, the normal subgroup (by its number in the list of the GAP package LINS, or by the group it gives after ``onto'') and the number
of layers $L$ of each code. Every bracket closes at its upper end.}
\label{fig:si-hyperbolic}
\end{figure}

\paragraph{Modular hyperbolic codes.}
Fifteen of the codes whose distances ref.~\citenum{higgott_2026_hyperbolic} could only bracket were rebuilt from its construction; \cref{fig:si-hyperbolic} shows each bracket with the exact distance, which is its upper end in every case.

\paragraph{Open values.}
The audit leaves open the mitten circuits from $n=540$ (intervals in \cref{tab:audit}), the structured
\qc{840,168,18} circuit (11 to 14) and the code distance of the structured \qc{1200,240,20} code, which appears only
in the authors' repository (15 to 20), the \qc{1764,886,14} (12 to 14) and
\qc{2844,1426,18} (12 to 16) Cornucopia circuits, the amplified level-2 \qc{468,16,16} circuit\cite{liang_2026_amplified}
(12 to 16), the routing \qc{200,16,17} circuit (11 to 14), IBM's \qc{756,16,34} circuit, whose code distance 34 we
certify but for which only the circuit lower bound 15 is certified, and the planar \qc{537,14} code (20 to 30).
Five further papers could not be tested: their constructions need data that we did not obtain or were not attempted.

\FloatBarrier
\section{Design studies}
\label{sec:design}

\begin{figure}[tbp]
\centering
\fullwidthfigure{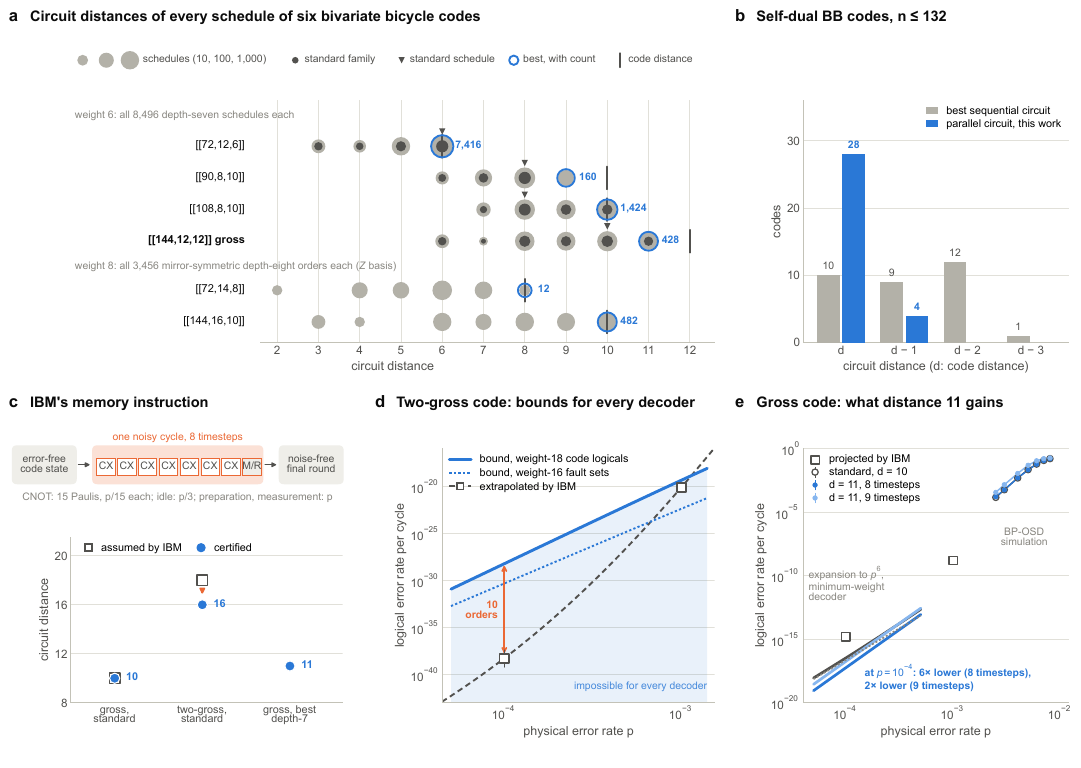}
\caption{\textbf{Certified design of syndrome-extraction circuits, and the memory of a planned fault-tolerant
processor.} \textbf{a}, Exact circuit distances of every schedule of six bivariate bicycle codes: all 8,496 valid uniform
depth-seven one-ancilla schedules of four weight-six codes\cite{bravyi_2024_bb} (the smaller of the two memory
bases) and all 3,456 mirror-symmetric depth-eight gate orders of two weight-eight codes published without
circuits\cite{lu_2026_weight8} ($Z$ basis). Marker area grows with the logarithm of the number of schedules at each
distance (light gray, all; dark gray, the 936 of the standard family); triangles mark the standard schedule, blue rings the
best schedules with their number, and bars the code distance. \textbf{b}, The 32 self-dual bivariate bicycle
codes\cite{liang_2025_selfdual} with $n\le132$: how far the circuit distance falls below the code distance $d$, for
the best sequential circuit (an $X$-check round followed by a $Z$-check round, every check with the same gate order;
optimum over all 5,040 orders) and for the parallel depth-eight circuit designed here; \cref{tab:design-sdbb} lists every code. \textbf{c}, The memory instruction of IBM's bicycle
architecture\cite{yoder_2025_tour}: one noisy syndrome cycle of eight timesteps between an error-free code state and
a noise-free final round, under circuit noise of strength $p$; below, its circuit distances assumed in
ref.~\citenum{yoder_2025_tour} and certified here, the same in both memory bases. \textbf{d}, Two-gross code with the standard schedule, per cycle in the setting of
ref.~\citenum{yoder_2025_tour} (18 cycles, failure probability divided by 18), summed over the two memory bases: lower
bounds on the logical error rate that hold for every decoder (rates in the shaded region are impossible), from the 336
logical operators of the code of weight 18 of each type (solid line) and from the fault sets of weight 16 (dotted line);
the extrapolation of ref.~\citenum{yoder_2025_tour}, computed from its published fit parameters (dashed line), and its
rates at $p=10^{-4}$ and $10^{-3}$ with their uncertainties (squares). The arrow gives the orders of magnitude between its
rate and the higher bound at $p=10^{-4}$. \textbf{e}, Gross code, per cycle in the setting of
ref.~\citenum{yoder_2025_tour} (10 cycles, failure probability divided by 10), summed over the two memory bases: logical
error rates with the standard schedule and with the best schedules of distance 11 of eight and of nine timesteps per
cycle, from an estimate of the low-noise expansion to order $p^6$ for a minimum-weight decoder using fault sets of up to 12 faults
(lines; dotted, the part of the standard schedule from failures of five faults, which distance 11 removes), simulated
with a BP-OSD decoder (points, with one-standard-deviation intervals), and the rates projected in
ref.~\citenum{yoder_2025_tour} for the standard schedule (squares).}
\label{fig:si-design}
\end{figure}

\subsection{All depth-seven schedules of weight-six bivariate bicycle codes}
\label{subsec:design-bb6}

A weight-six bivariate bicycle code\cite{bravyi_2024_bb} has $X$ checks of the form $A\,|\,B$ and $Z$ checks of the
form $B^{\sfT}\,|\,A^{\sfT}$, with three terms in each of the left and right blocks.
A uniform depth-seven schedule with one ancilla per check assigns each of the six terms of the $X$ checks, and of the $Z$
checks, to one of seven CNOT layers, the same assignment for every check of a type.
A schedule is valid when $X$- and $Z$-check gates never act on the same block in a layer, so that no qubit is used
twice in a layer, and when every overlapping pair of an $X$ and a $Z$ check is crossed consistently, so that the
measured operators commute through the circuit and the detectors are deterministic.
There are 8,496 valid schedules.
The standard family has 936 of them: those in which the $X$ checks idle in the first layer and the $Z$ checks in the
last, as in the schedule of ref.~\citenum{bravyi_2024_bb}.

For each schedule and each of the codes \qc{72,12,6}, \qc{90,8,10}, \qc{108,8,10} and \qc{144,12,12}, we built the
memory circuit with six cycles in both bases, with two-qubit depolarizing faults after every CNOT, single-qubit
depolarizing faults on every idle qubit in every layer and flips after reset and before measurement, and computed
both circuit distances exactly; the schedule's distance is the smaller of the two. Stim confirms that the detectors
of every circuit built are deterministic.
One best schedule each of \qc{108,8,10} and the gross code was checked again with six and twelve cycles in both bases
and with an independent pipeline without symmetry reduction, which reproduced every value.
\Cref{tab:design-bb6} gives the distributions, and Fig.~4a of the main text shows their ranges.

\begin{table}[t]
\centering
\caption{Exact circuit distances of all 8,496 valid depth-seven schedules (all) and of the 936 schedules of the
standard family (std.), by code: the number of schedules at each distance.}
\label{tab:design-bb6}
\small
\begin{tabular}{@{}lcrrrrrrr@{}}
\toprule
code & schedules & 3 & 4 & 5 & 6 & 7 & 8 & 9--11 \\
\midrule
\qc{72,12,6} & all & 48 & 24 & 1008 & 7416 & & & \\
 & std. & 24 & 12 & 108 & 792 & & & \\
\qc{90,8,10} & all & & & & 32 & 416 & 7888 & 160 at 9 \\
 & std. & & & & 16 & 64 & 856 & \\
\qc{108,8,10} & all & & & & & 68 & 4932 & 2072 at 9, 1424 at 10 \\
 & std. & & & & & 10 & 630 & 166 at 9, 130 at 10 \\
\qc{144,12,12} & all & & & & 48 & 4 & 2364 & 1420 at 9, 4232 at 10, 428 at 11 \\
 & std. & & & & 24 & 2 & 310 & 114 at 9, 436 at 10, 50 at 11 \\
\bottomrule
\end{tabular}
\end{table}

The standard schedule itself has circuit distance 6, 8, 8 and 10 for the four codes.
It is optimal only for \qc{72,12,6}.
The best schedules reach 9, 10 and 11 for the other three; for \qc{90,8,10} they lie outside the standard family.
Because every schedule was computed, these optima are certified for the family.

\subsection{Weight-eight bivariate bicycle codes}
\label{subsec:design-bb8}

The weight-eight bivariate bicycle codes of ref.~\citenum{lu_2026_weight8} were published without circuits.
We designed parallel depth-eight circuits in which $X$ and $Z$ checks share the eight layers with one ancilla per
check: the $X$ order has a palindromic pattern of the two blocks and the $Z$ order is its mirror image, which gives
3,456 valid orders.
For \qc{72,14,8} and \qc{144,16,10} we computed the exact $Z$-basis circuit distance of all 3,456 orders (six
cycles): 12 orders reach the code distance 8 and 482 orders the code distance 10 (Supplementary Fig.~\ref{fig:si-design}a), and four
of the best orders of each code, checked in both bases, have the full distance in both.
For \qc{144,10,14}, two orders with the best time-collapsed bound were evaluated; one has circuit
distance 12 in both bases, and the other is not decided; this is a design, not the optimum of the family.

\subsection{Self-dual bivariate bicycle codes}
\label{subsec:design-sdbb}

For the self-dual bivariate bicycle codes of ref.~\citenum{liang_2025_selfdual}, also published without circuits, we
studied two families.
In the sequential family, an $X$-check round is followed by a $Z$-check round and every check uses the same gate
order; the search covers all 5,040 inequivalent orders, and its best value is the optimum of the family for the 42
of the 57 codes for which it finished.
In the parallel family, $X$ and $Z$ checks share eight layers as above; the order is chosen among the 3,456 mirror
orders by the time-collapsed bound, and the circuit distance of the best few is computed exactly, which gives a
circuit and its distance but not the optimum of the family, except for the small codes for which all orders were
computed.
\Cref{tab:design-sdbb} lists both for the codes with a parallel circuit.
For the 32 codes with $n\leq132$ that have a parallel circuit, all but \qc{108,20,6} and \qc{126,22,6}, the parallel
circuit reaches the code distance for 28 and the best sequential circuit for 10.

\begin{table}[t]
\centering
\caption{Self-dual bivariate bicycle codes: code distance $d$, the best sequential circuit (optimum over all 5,040
orders; $\ge$, a lower bound where that search did not finish) and the parallel depth-eight circuit designed here
(an interval where it is not decided).}
\label{tab:design-sdbb}
\small
\begin{tabular}{@{}lrrr@{\hspace{2.5em}}lrrr@{}}
\toprule
code & $d$ & sequential & parallel & code & $d$ & sequential & parallel \\
\midrule
\qc{16,4,4} & 4 & 2 & 4 & \qc{80,10,8} & 8 & 6 & 7 \\
\qc{24,8,4} & 4 & 2 & 4 & \qc{84,6,10} & 10 & 8 & 10 \\
\qc{30,6,5} & 5 & 4 & 5 & \qc{90,18,6} & 6 & 6 & 6 \\
\qc{32,12,4} & 4 & 2 & 4 & \qc{96,12,8} & 8 & 7 & 8 \\
\qc{36,10,4} & 4 & 4 & 4 & \qc{98,14,6} & 6 & 6 & 6 \\
\qc{40,6,6} & 6 & 5 & 6 & \qc{100,12,8} & 8 & 7 & 8 \\
\qc{42,6,6} & 6 & 5 & 6 & \qc{102,6,10} & 10 & 9 & 10 \\
\qc{48,16,4} & 4 & 2 & 4 & \qc{104,6,12} & 12 & 9 & 11 \\
\qc{50,10,5} & 5 & 5 & 5 & \qc{110,10,8} & 8 & 8 & 8 \\
\qc{54,10,6} & 6 & 5 & 6 & \qc{112,6,12} & 12 & 10 & 11 \\
\qc{56,6,8} & 8 & 6 & 8 & \qc{114,6,10} & 10 & 10 & 10 \\
\qc{60,12,5} & 5 & 5 & 5 & \qc{120,8,12} & 12 & 10 & 12 \\
\qc{64,8,8} & 8 & 6 & 8 & \qc{128,16,8} & 8 & 8 & 8 \\
\qc{66,6,8} & 8 & 7 & 8 & \qc{130,10,10} & 10 & 9 & 10 \\
\qc{70,10,6} & 6 & 6 & 6 & \qc{132,8,12} & 12 & 10 & 11 \\
\qc{72,12,6} & 6 & 6 & 6 & \qc{152,6,16} & 16 & $\ge$11 & 12--13 \\
\qc{78,6,10} & 10 & 8 & 10 & \qc{160,8,16} & 16 & $\ge$11 & 12 \\
\bottomrule
\end{tabular}

\end{table}

\subsection{Other designs}
\label{subsec:design-other}

\paragraph{GALA \qc{132,30,12}.}
The paper\cite{yang_2026_gala} prints only the order of the twelve permutation layers; in our gate-level rebuild,
this order applied to simultaneous $X$ and $Z$ extraction does not give deterministic measurements, and run as an $X$
round followed by a $Z$ round it gives a circuit of distance 8.
Among 600 random palindromic orders of a parallel 12-layer circuit, 147 have time-collapsed bound 8; of the 14 of
them evaluated exactly, five have circuit distance 10 in both bases, for example the order F0 F1 G5 G3 G1 F5 F3 G0
G4 G2 F2 F4.
The best of 2,024 sequential orders has distance 8.

\paragraph{STAR \qc{42,6,6}.}
The STAR paper\cite{ismail_2026_star} reports 212 valid depth-eight schedules for its \qc{42,6,6} code, 192 with
distance 4 and 20 with distance 5, the chosen one among the latter.
It does not give its schedule grammar in full; with the internal order of the $A$ and of the $B$ terms of the $X$
schedule fixed, our enumeration has 212 valid schedules, the paper's count, and we take these as its family.
It does not print which monomial each term of the second block is either, so we ran all 24 conventions; each keeps the
212 schedules valid.
Two conventions give exactly the paper's histogram, with the chosen schedule at 5.
The convention $B_i=A_i^{\sfT}$ gives 192 at 4, 12 at 5 and 8 at 6, with the chosen schedule at 6; schedules of
distance 6, the code distance, exist under 5 of the 24 conventions.

\FloatBarrier
\section{The memory instruction of IBM's bicycle architecture}
\label{sec:ibm}

\subsection{Setting}
\label{subsec:ibm-setting}

The bicycle architecture of ref.~\citenum{yoder_2025_tour} stores logical qubits in modules of the gross code
\qc{144,12,12}, or of the two-gross code \qc{288,12,18} for larger computations, and specifies every operation as
an instruction with a logical error rate; the resource estimates add these rates over the instructions of a
computation\cite{yoder_2025_tour}.
The paper publishes no circuits, so we built its memory instruction from its description
(\texttt{papers/ibm\_bicycle.py}):
\begin{itemize}
\item the syndrome cycle of ref.~\citenum{bravyi_2024_bb}, or any other valid depth-seven schedule
  (\cref{subsec:design-bb6}), with one check qubit per check: seven CNOT layers, then the measurement and reset of
  all check qubits in one timestep, so that $C$ cycles take $8C+1$ timesteps;
\item an error-free code state: a noise-free preparation of the data qubits in the memory basis followed by a
  noise-free syndrome round, which fixes the reference outcomes;
\item $C$ noisy cycles, then a noise-free syndrome round and a noise-free measurement of the data qubits in the
  memory basis, from which the 12 logical observables of that basis are read;
\item circuit noise of strength $p$: every CNOT is followed by one of the 15 non-identity two-qubit Paulis with
  probability $p/15$, every qubit idle in a timestep by $X$, $Y$ or $Z$ with probability $p/3$, and preparations and
  measurements fail with probability $p$.
\end{itemize}
Detectors compare consecutive outcomes of each check, and the final data measurement gives one more detector for
each check of the memory basis; Stim confirms that all of them are deterministic.
We certify the instruction with one cycle, $C=1$; \cref{subsec:ibm-fit} treats the setting of the rates in the paper's
Table~2, which are defined over many cycles.
Its detector error model has 432 detectors and 6,336 distinct fault mechanisms for the gross code (864 columns in
the sector of one memory basis) and 864 detectors and 12,672 mechanisms for the two-gross code (1,728 in the
sector); the lattice translations give 72 and 144 symmetries.

The paper obtained its logical error rates by simulating each instruction with the Relay-BP
decoder\cite{mueller_2025_relay} at physical error rates where failures can be sampled, and extrapolated them to
$p=10^{-3}$ and $10^{-4}$ with a failure-spectrum ansatz: the fraction of failing fault configurations of weight $w$
is set to zero for $w<\lceil d^*/2\rceil$ and fitted with two parameters above.
The assumed circuit distances were $d^*=10$ for the gross code and $d^*=18$ for the two-gross code, for which the
paper notes that the true value is not known and might be lower\cite{yoder_2025_tour}.

\subsection{Certified distances}
\label{subsec:ibm-distance}

\Cref{tab:ibm-distance} lists the certified circuit distances of the instruction.
For the gross code with the standard schedule it is 10 in both bases, as assumed: the time-collapsed bound gives 8
($X$ basis) and 9 ($Z$ basis), the sector search and lifting raise it to 10, and the search finds a fault set of
weight 10 within seconds.
For the two-gross code it is 16 in both bases, two below the assumed value, certified by the pipeline of the main
text (\texttt{dp=2}, 32 threads, on instances exported in the benchmark format by
\texttt{papers/ibm\_bicycle\_export.py}): in the $X$ basis in 35 minutes; in the $Z$ basis the lifting argument gives
the lower bound 16 in about five minutes, and the search at weight 16 finds a fault set after 2.3 hours (no fault set of
weight 16 or 17 has all its detectors in the first round in this basis). Both fault sets of weight 16 were replayed in Stim.
For the gross code we also computed the instruction distance of each of the 428 depth-seven schedules whose
six-cycle memory circuits have distance 11 (\cref{subsec:design-bb6}), one single-threaded run per schedule and basis
(\texttt{papers/ibm\_bicycle\_scan.py}): all 428 have distance 11 in both bases.
Since the instruction's boundaries are noise-free, its distance need not equal that of the memory circuit; for these
schedules the two agree.

\paragraph{The circuits of the public code.}
A study of rare failures by IBM\cite{beverland_2025_failfast} analyzes the memory circuits generated by the public
code\cite{bravyi_bbcode} of ref.~\citenum{bravyi_2024_bb}: the cycle above with the $X$ checks prepared during its first CNOT
layer and the $Z$ checks measured during its last, $d$ noisy cycles followed by noise-free ones, and the circuit noise of
\cref{subsec:ibm-setting}.
For $Z$-type noise its Table~2 gives $D\le10$ for the gross code and $D\le18$ for the two-gross code, the latter from
at least 1,981,650 (compressed) logical operators of weight 18 found by search, and it counts the minimum-weight failures from these operators.
We rebuilt the circuits in Stim (\texttt{papers/bravyi\_repo\_circuit.py}).
Their detector error models have exactly the columns of the decoding matrices that the public code computes by its own
fault simulation, in both bases, for the gross code at three cycles and the two-gross code at two
(\texttt{papers/bravyi\_repo\_check.py}).
With 12 cycles the gross code's circuit has distance 10 in both bases, as stated.
With 18 cycles the two-gross code's circuit has distance 16 for both types of noise.
For $Z$-type noise, the search of a one-round window finds a fault set of weight 16, replayed in Stim, and lifting
through weight 15 (\cref{prop:lifting-W}) certifies the lower bound: the 796,704 solutions of weight up to 15 of the
time-collapsed problem, whose distance is 12, fall into 5,546 orbits of the 144 lattice translations, and none lifts
to a fault set of the circuit of weight 15 or less (2.6~hours on 48 threads, nearly all of it for the enumeration).
For $X$-type noise the distance 16 was certified in 1.7~hours on 24 threads.
Its minimum-weight failures therefore have eight faults, not nine.

\begin{table}[t]
\centering
\caption{Certified circuit distances of the memory instruction ($C=1$) in both memory bases.}
\label{tab:ibm-distance}
\small
\begin{tabular}{@{}llccl@{}}
\toprule
code & schedule & $d$ ($Z$ basis) & $d$ ($X$ basis) & assumed in ref.~\citenum{yoder_2025_tour} \\
\midrule
gross \qc{144,12,12} & standard\cite{bravyi_2024_bb} & 10 & 10 & 10 \\
 & 428 depth-seven schedules & 11 & 11 & -- \\
two-gross \qc{288,12,18} & standard\cite{bravyi_2024_bb} & 16 & 16 & 18 \\
 & best schedules (\cref{subsec:ibm-scan}) & 18 & 18 & -- \\
\bottomrule
\end{tabular}
\end{table}

\subsection{All schedules of the two-gross code}
\label{subsec:ibm-scan}

The 8,496 schedules of \cref{subsec:design-bb6} in two memory bases give 16,992 distance problems for the two-gross code. Each is far larger than a
problem of the smaller codes, so we first reduced their number.

\paragraph{Symmetries.}
A group automorphism $\varphi$ of the code's group $\mathbb{Z}_{l}\times\mathbb{Z}_{m}$ for which $\varphi(A)$ and $\varphi(B)$ are translates of $A$ and $B$, of $B$ and $A$,
of $-A$ and $-B$ or of $-B$ and $-A$ relabels the schedule and carries the circuit to the circuit of the relabeled schedule; the last two kinds exchange $X$ and $Z$.
Reversing the layers of a schedule keeps the basis.
The two-gross code has an automorphism of order 3, $x\mapsto xy^{9}$, $y\mapsto x^{9}y^{10}$, that cycles the terms of $A$ and of $B$, so its group is larger than the gross code's.
The problems form 708 orbits, against 354 for \qc{72,12,6}, 1,062 for \qc{90,8,10} and 2,124 for \qc{108,8,10} and the gross code.
The schedules form 390 classes, each holding one or two problems.
All generators reproduce the 16,992 exact values of each of the four smaller codes.
On the two-gross code itself we built the circuits of 6 schedules in both bases with one cycle and found the detector map that carries one list of columns to the other, up to an invertible change of the logical basis plus detector terms, in 144 of 144 cases.

\paragraph{One-round windows.}
At one cycle every column of the sector has its detectors in rounds 1, 2 or both.
A one-round window, the columns whose detectors all lie in one round, holds the single-qubit columns and possibly one family of hook columns.
Hashed by content, the windows of all 708 problems are only seven (\cref{tab:ibm-scan-windows}).
Identical windows have identical columns, so a fault set found once holds in every problem with that window, and we replayed it in each problem's own circuit (588 of 588).
A problem with a hook window therefore has distance at most 16, and this caps 7,056 of the 8,496 schedules. They include all 1,872 schedules that fit eight timesteps per cycle in the timing of ref.~\citenum{yoder_2025_tour}, so no such schedule exceeds the standard schedule's 16.
A pass up to weight 12 over all 708 problems finds 4 problems at exactly 12, four classes with 48 schedules, and lower bound 13 for all the others.
A weight-18 logical operator of the code at one time step fires no detector, so every schedule has distance at most 18.

\begin{table}[t]
\centering
\caption{The one-round windows of the 708 problems of the two-gross code at one cycle, named by the start of their hash. The first has single-qubit columns only, the others also a family of hook columns. Occurrences count both windows of all 708 problems.}
\label{tab:ibm-scan-windows}
\small
\begin{tabular}{@{}lrrl@{}}
\toprule
one-round window & columns & occurrences & lightest fault set \\
\midrule
\texttt{6ad9} & 288 & 828 & the code's own logical operators, weight 18 \\
\texttt{9da4} & 432 & 288 & weight 16, replayed in every occurrence \\
\texttt{a7bf} & 432 & 144 & weight 16, replayed in every occurrence \\
\texttt{f570} & 432 & 144 & weight 16, replayed in every occurrence \\
\texttt{a263} & 432 & 4 & weight 16, replayed in every occurrence \\
\texttt{2900} & 432 & 4 & weight 16, replayed in every occurrence \\
\texttt{0b59} & 432 & 4 & weight 12, replayed in every occurrence \\
\bottomrule
\end{tabular}

\end{table}

\paragraph{Exhaustive search.}
That leaves 78 classes, 1,440 schedules and 120 problems without a hook window.
For each class we ran the pipeline (\texttt{dp=2}) on the class's first problem, on the image under layer reversal whose first window is the larger, and on its second problem only below the first's value.
Since the plain window has no low-weight fault set, a fault set of weight 17 can only come from the sector search, not from the window or lifting stages.
The run stops at the first fault set, so classes at 17 end far sooner than the exhaustive search that a class at 18 needs, roughly 100 to 170 thread-hours per problem.
Every fault set was replayed in Stim.
\Cref{tab:ibm-scan-values} gives the result.
We did not split the capped schedules into 15 and 16, since the design question does not need it.
The values are not a general rule about hook families: on \qc{72,12,6}, \qc{108,8,10} and the gross code the best value is also reached with such a family, on \qc{90,8,10} and the two-gross code only without one.

\begin{table}[t]
\centering
\caption{All 8,496 schedules of the two-gross code at one cycle.}
\label{tab:ibm-scan-values}
\small
\begin{tabular}{@{}lrr@{}}
\toprule
schedules & number & circuit distance at one cycle \\
\midrule
with a family of hook columns in one round & 7,056 & at most 16, of which 48 exactly 12 \\
without such a family, 34 classes & 600 & 15 \\
without such a family, 7 classes & 144 & 16 \\
without such a family, 31 classes & 612 & 17 \\
without such a family, 6 classes & 84 & 18 \\
\bottomrule
\end{tabular}

\end{table}

\paragraph{The best schedules.}
\Cref{tab:ibm-scan-best} lists the classes certified at 18. For the first, a direct run excludes weight 17 in the $Z$ basis in 3.6~hours on 40 threads, and a direct run excludes weight 16 in the $X$ basis, whose weight 17 follows from the symmetry that carries $X$ to $Z$.
All 336 logical operators of weight 18 of each type are fault sets of its circuit, with 1,008 copies per basis replayed in Stim in the setting of \cref{subsec:ibm-fit}, and they give the bound per cycle, both bases, of $10^{-28.40}$ at $p=10^{-4}$ and $10^{-19.52}$ at $p=10^{-3}$. On the standard schedule's circuit the same operators give $10^{-28.22}$ and $10^{-19.34}$.

\begin{table}[t]
\centering
\caption{The classes of schedules of the two-gross code certified at 18 at one cycle, in the slot order of \cref{subsec:design-bb6}, with the first schedule of each class, the number of schedules in it, and the timesteps per cycle in the timing of ref.~\citenum{yoder_2025_tour}.}
\label{tab:ibm-scan-best}
\small
\begin{tabular}{@{}llrr@{}}
\toprule
$p_X$ & $p_Z$ & schedules & timesteps per cycle \\
\midrule
(0,4,6,1,3,5) & (5,3,1,6,2,0) & 12 & 9 \\
(0,4,6,3,5,1) & (3,1,5,6,2,0) & 12 & 9 \\
(0,2,6,1,5,3) & (5,1,3,6,4,0) & 12 & 9 \\
(0,3,6,1,5,4) & (4,1,2,6,3,0) & 24 & 9 \\
(0,3,6,1,5,4) & (5,1,2,6,3,0) & 12 & 9 \\
(0,3,6,2,5,4) & (4,1,2,6,3,0) & 12 & 9 \\
\bottomrule
\end{tabular}

\end{table}

\paragraph{More cycles.}
A value at one cycle is an upper bound for more cycles.
To see how much that matters we ran every orbit of problems of the four smaller codes at one and two cycles in the matched circuit, 11,328 runs, all exact and replayed, against the six-round tables of \cref{subsec:design-bb6} (\cref{tab:cycles}).
Two cycles equal six rounds on all 67,968 problems.
One cycle equals six rounds on 67,352, and on the other 616 it is exactly one too high.
Without a hook window, one cycle is too high on only 32 of 11,520 problems, and at each code's top value on none of 3,560.
With a hook window, the top value does drop: on 128 of 2,688 problems of \qc{108,8,10} and on 128 of 1,472 of the gross code.
For the first schedule above, the instruction with two cycles has a sector of $720\times3{,}168$, the same columns as the matched circuit in both bases, and weights 13 to 16 are excluded in the $Z$ basis in 7.3~hours on 32 threads and weight 17 in 32.4~hours on 48 threads.
With three cycles, weights up to 16 are excluded in 19.4~hours on 32 threads.
At two and at three cycles the symmetry that carries $X$ to $Z$ holds as at one cycle.
With the logical operators of weight 18 as upper bound, the circuit distance of this schedule is therefore 18 at two cycles and 17 or 18 at three, and in these instructions a suitable decoder can correct any eight faults, which is what the fit of ref.~\citenum{yoder_2025_tour} assumes (its failure spectrum vanishes below nine faults).

\begin{table}[t]
\centering
\caption{Number of cycles on the four smaller codes: the problems (schedule and basis) on which one or two cycles give the same circuit distance as six rounds.}
\label{tab:cycles}
\small
\begin{tabular}{@{}lrrrr@{}}
\toprule
code & problems & orbits run & one cycle equals six & two cycles equal six \\
\midrule
\qc{72,12,6} & 16,992 & 354 & 16,944 & 16,992 \\
\qc{90,8,10} & 16,992 & 1,062 & 16,736 & 16,992 \\
\qc{108,8,10} & 16,992 & 2,124 & 16,864 & 16,992 \\
\qc{144,12,12} & 16,992 & 2,124 & 16,808 & 16,992 \\
\bottomrule
\end{tabular}

\end{table}

\subsection{Decoder-independent lower bounds on the logical error rate}
\label{subsec:ibm-bound}

\begin{proposition}[Unavoidable failures]
\label{prop:ibm-bound}
Consider a detector error model whose mechanisms $j$ occur independently with probabilities $p_j$, with detector
matrix $\vb{D}$ and observable matrix $\vb{O}$, and a decoder, that is, any map from detector outcomes to a
prediction of the observable flips.
\begin{enumerate}
\item If two sets of mechanisms $A$ and $A'$ satisfy $\vb{D}\vb{1}_A=\vb{D}\vb{1}_{A'}$ and
  $\vb{O}\vb{1}_A\neq\vb{O}\vb{1}_{A'}$, the decoder fails on at least one of the events ``exactly the mechanisms
  of $A$ occur'' and ``exactly the mechanisms of $A'$ occur''.
\item Let $F_1,\ldots,F_m$ be sets of mechanisms with $\vb{D}\vb{1}_{F_i}=0$ and $\vb{O}\vb{1}_{F_i}\neq0$, split as
  $F_i=E_i\disjoint E_i'$ so that the $2m$ sets $E_1,E_1',\ldots,E_m,E_m'$ are pairwise distinct, and let
  $U\supseteq F_1\cup\cdots\cup F_m$. Then
  \begin{equation}
  P_{\mathrm{fail}}\ \ge\ \sum_{i=1}^{m}\min\bigl(P_U(E_i),P_U(E_i')\bigr),\qquad
  P_U(E)=\prod_{j\in E}p_j\prod_{j\in U\setminus E}(1-p_j).
  \label{eq:si-ibm-bound}
  \end{equation}
\end{enumerate}
\end{proposition}

\begin{proof}
(1) The two events produce the same detector outcomes, so the decoder makes the same prediction in both; the true
observable flips differ, so the prediction is wrong in at least one of them.
(2) For every set $S$ of mechanisms outside $U$, the sets $E_i\cup S$ and $E_i'\cup S$ satisfy the conditions of (1),
since $\vb{D}\vb{1}_{E_i}=\vb{D}\vb{1}_{E_i'}$ and $\vb{O}\vb{1}_{E_i}+\vb{O}\vb{1}_{E_i'}=\vb{O}\vb{1}_{F_i}\neq0$.
All these sets are distinct, because their intersections with $U$ are the distinct sets $E_i$ and $E_i'$, so the
events that exactly their mechanisms occur are disjoint, and the failure probability is at least
$\sum_i\sum_S\min\bigl(P(E_i\cup S),P(E_i'\cup S)\bigr)$, where $P(A)$ is the probability that exactly the
mechanisms of $A$ occur. Since $P(E\cup S)=P_U(E)\,Q(S)$ with
$Q(S)=\prod_{j\in S}p_j\prod_{j\notin U\cup S}(1-p_j)$, and the $Q(S)$ sum to one, \cref{eq:si-ibm-bound} follows.
\end{proof}

We apply the proposition to minimum fault sets, $|F_i|=d$, split into all parts $E$ of size $\lfloor d/2\rfloor$ and
their complements; for even $d$ each unordered pair is counted once.
If two minimum fault sets share fewer than $\lfloor d/2\rfloor$ mechanisms, no part of one equals a part of the
other, so the parts of a family of minimum fault sets with pairwise intersections below $\lfloor d/2\rfloor$ are
pairwise distinct.
For the gross code, the search enumerates all minimum fault sets at the certified distance (enumeration mode of
\cref{cor:search-enumeration}, one set per symmetry orbit, then expanded); for the two-gross code we use the orbit of
the fault set found by the certified run, so that the bound sums over a subset of the minimum fault sets and remains
valid.
The family is chosen greedily from these sets, and $U$ is the union of its sets.
The search runs on the sector problem, whose columns are classes of mechanisms with the same detectors in the sector;
each column of a minimum fault set is realized by the mechanism of the full model that fires no detector outside the
sector, so that $\vb{D}\vb{1}_F=0$ holds for all detectors of the full model.
The probabilities $p_j$ are those of the full detector error model at each $p$, with equal mechanisms merged.
The bound holds for every decoder that acts on the instruction's detector outcomes, Relay-BP among them.
A logical error of the instruction in either basis is a failure of the instruction, so its logical error rate is at
least the larger of the two bounds (\texttt{papers/ibm\_bicycle\_bound.py}).

\Cref{tab:ibm-bound} lists the bounds.
For the gross code with the standard schedule they lie below the rates of ref.~\citenum{yoder_2025_tour}.
For the two-gross code the one-cycle bound at $p=10^{-3}$ is below the paper's rate, but at $p=10^{-4}$ it is
$10^{-30.9}$, above the extrapolated $10^{-38.3}$, and it rises as $p^{8}$ at low noise; \cref{subsec:ibm-fit} makes
this comparison in the paper's own setting, where the bound exceeds the paper's rates at both noise levels.
In each basis the bound uses one orbit of 36 minimum fault sets, that of the certified run's fault set, and the two
bases give the same values, as the symmetry between the $X$ and $Z$ checks of the code suggests. In the $X$ basis a
complete enumeration of the fault sets of weight 16 whose detectors lie in one round finds no other; more minimum fault sets would only raise the
bound.

\begin{table}[t]
\centering
\caption{Lower bounds on the logical error rate of the memory instruction with one cycle for every decoder
(\cref{prop:ibm-bound}) and the rates of ref.~\citenum{yoder_2025_tour} (Table~2, in the setting of
\cref{subsec:ibm-fit}), at $p=10^{-4}$ and $10^{-3}$. Minimum fault sets: the
number of sets used (all minimum fault sets; \textsuperscript{a}the orbit of the certified run's fault set) and the
size of the family with pairwise intersections below $\lfloor d/2\rfloor$. The best schedule of the gross code is
the nine-timestep schedule of \cref{subsec:ibm-leading}.}
\label{tab:ibm-bound}
\footnotesize
\setlength{\tabcolsep}{4pt}
\begin{tabular}{@{}lcrrrrrrr@{}}
\toprule
 & & & \multicolumn{2}{c}{minimum fault sets} & \multicolumn{2}{c}{bound, any decoder} & \multicolumn{2}{c}{ref.~\citenum{yoder_2025_tour}} \\
\cmidrule(lr){4-5}\cmidrule(lr){6-7}\cmidrule(l){8-9}
code, schedule & basis & $d$ & used & family & $p=10^{-4}$ & $10^{-3}$ & $10^{-4}$ & $10^{-3}$ \\
\midrule
gross, standard & $Z$ & 10 & 144 & 72 & $10^{-17.6}$ & $10^{-12.6}$ & $10^{-14.8}$ & $10^{-8.8}$ \\
 & $X$ & 10 & 288 & 144 & $10^{-17.6}$ & $10^{-12.7}$ &  &  \\
\addlinespace[2pt]
gross, best & $Z$ & 11 & 3,960 & 660 & $10^{-20.7}$ & $10^{-15.0}$ &  &  \\
 & $X$ & 11 & 936 & 316 & $10^{-20.8}$ & $10^{-14.9}$ &  &  \\
\addlinespace[2pt]
two-gross, standard & $Z$ & 16 & 36\textsuperscript{a} & 36 & $10^{-30.9}$ & $10^{-22.9}$ & $10^{-38.3}$ & $10^{-20.1}$ \\
 & $X$ & 16 & 36\textsuperscript{a} & 36 & $10^{-30.9}$ & $10^{-22.9}$ &  &  \\
\bottomrule
\end{tabular}

\end{table}

\subsection{The published extrapolation}
\label{subsec:ibm-fit}

The rates in Table~2 of ref.~\citenum{yoder_2025_tour} are not those of one cycle.
The idle rate is the failure probability of a circuit of $C=d^*$ cycles, 18 for the two-gross code, divided by $C$.
Preparations and measurements take one timestep each and are staggered as in its Fig.~4, every fault is an independent
event of probability $p_{\mathrm{el}}=p/15$, with an event of probability $mp/15$ replaced by $m$ such faults, and the experiment
prepares Bell pairs, so that it detects $X$ and $Z$ logical errors together (its Section~2.6 and Appendix~A.7).
We built the circuit in that setting (\texttt{papers/ibm\_bicycle\_matched.py}).
For $C=1$ its detector error model has the same mechanisms as that of the instruction above, so the certified
distances carry over.
The 18-cycle circuit has 576 qubits, 5,760 detectors and 213,408 mechanisms, and every fault set of the one-cycle
instruction has a copy in each cycle.
All 648 copies per basis of the 36 fault sets of weight 16 fire no detector in its detector error model and replay in
Stim with the expected logical flips (\texttt{papers/ibm\_bicycle\_matched\_bound.py}).

\Cref{prop:ibm-bound} applies to each cycle on its own. Its proof holds with $U_c$, the mechanisms used in cycle $c$, in
place of $U$, the sets $S$ then including all mechanisms of the other cycles, so the failure probability restricted to the
event $H_c$ that the mechanisms of $U_c$ that occur form a half of cycle $c$ is at least the sum $B_c$ of the proposition
with $U=U_c$. The $U_c$ are disjoint, so the $H_c$ are independent, and by the Bonferroni inequality
\begin{equation}
P_{\mathrm{fail}}\ \ge\ \sum_{c}B_c-\sum_{c<c'}P(H_c)P(H_{c'}).
\label{eq:si-ibm-cycles}
\end{equation}
The last term is of second order ($10^{-54}$ against $10^{-27}$ at $p=10^{-4}$, and $10^{-36}$ against $10^{-18}$ at
$p=10^{-3}$). The same holds for the two types of logical error in the paper's Bell-pair experiment. The mechanisms we use
for $X$-type and for $Z$-type errors fire detectors of one memory basis only, so they are disjoint, and the joint law of
the detectors and of the logical flips of one type depends only on the same merged mechanisms as in the memory experiment
of that basis. The bounds of the two bases therefore add, up to a term of the same second order, and
\cref{eq:si-ibm-cycles} divided by 18 gives the bound per cycle in \cref{tab:ibm-fit} (\texttt{papers/ibm\_bicycle\_code\_logicals.py}). Fault sets that span several cycles are not
used and could only raise the bound; we did not compute the distance of the 18-cycle circuit.

The paper's rates follow from the failure spectrum $f(w)$, the fraction of configurations of $w$ faults that fail, as
\begin{equation}
P(p_{\mathrm{el}})=\sum_{w\ge w_0}f(w)\binom{N}{w}p_{\mathrm{el}}^{w}(1-p_{\mathrm{el}})^{N-w},\qquad
f(w)=a\Bigl[1-\exp\Bigl(-\frac{f_0}{a}\Bigl(\frac{w}{w_0}\Bigr)^{\gamma}\Bigr)\Bigr],\qquad a=1-2^{-K_{\mathrm{log}}},
\label{eq:si-ibm-fit}
\end{equation}
where $N$ is the number of faults of the circuit, $w_0=\lceil d^*/2\rceil$, and $f_0$ and $\gamma$ are fitted (its
Eqs.~(82) and (83)).
With the parameters of its Table~6, for the two-gross idle instruction $f_0=e^{-95.83}$, $\gamma=25.78$, $w_0=9$,
$K_{\mathrm{log}}=24$ and $N=762{,}912$, the sum reproduces the idle and automorphism rates of both codes in its Table~2 to within 0.13 in the
exponent (\texttt{papers/ibm\_bicycle\_fit.py}).
It does so without a further division by $C$. We take the rates as per cycle, as the paper states; if they were per
circuit, the gaps below would grow by $\log_{10}18\approx1.3$ orders of magnitude.

At low weight the ansatz is the power law $f_0(w/w_0)^{\gamma}$, anchored by failures sampled at roughly 200 to 300
faults (its Fig.~7).
At $p=10^{-4}$, where $N p_{\mathrm{el}}=5.1$, the largest term of the sum has $w=19$ and 90\% of the sum comes from $w$ between 15 and
24, while the term $w=9$ contributes $1.7\times10^{-5}$ of it and leads only below $p\approx1.3\times10^{-5}$.
At $p=10^{-3}$ the largest term has $w=72$.
Between the two noise levels the sum falls by 18.2 orders of magnitude.
The curve itself does not depend on $w_0$, which $f_0$ absorbs, and extending it to $w=8$ or to $w=1$ changes the rate
at $p=10^{-4}$ by less than $10^{-6}$ orders of magnitude.
The assumed distance therefore has almost no influence on these rates, and a fit with $d^*=16$ to the same data would
give the same values.

The gap to the bound therefore does not come from the distance.
The code's logical operators of weight 18 are fault sets of 18 data-qubit errors in one round. We enumerated them in the
window of one round of the sector problem and found 336 of each type, the number the paper gives for one type.
They have 6,048 copies per basis in the 18-cycle circuit, all of which fire no detector in its detector error model, and
the 1,008 copies in three of the cycles also replay in Stim.
They overlap in up to 15 qubits, so we use not a family with small intersections but a greedy matching over their splits in
which no half is used twice, which \cref{prop:ibm-bound} allows; it keeps 8,166,816 of the 8,168,160 splits per cycle.
Their mechanisms have probabilities $14p_{\mathrm{el}}$ and $19p_{\mathrm{el}}$ in the bulk, so the bound per cycle at $p=10^{-4}$ is about
$8\times10^{6}$ pairs times $(1.1\times10^{-4})^{9}$ per basis, and in full $10^{-28.22}$ for the two bases. At that noise
level this is about a hundred times the bound from weight 16, which gives the higher bound only below $p\approx10^{-6}$.
For one noisy cycle the mechanisms are less likely: the noise-free boundary rounds remove the faults of the neighboring
cycle, 7 instead of 14 faults of probability $p_{\mathrm{el}}$ for many mechanisms, which is why the instruction with $C=1$ understates
the bulk.
In the whole circuit the ansatz allows about $6\times10^{5}$ failing configurations of 9 faults, whereas these sets alone
supply about $4\times10^{19}$.
The bound from weight 18 meets the curve near $p=1.1\times10^{-3}$ and the one from weight 16 near $p=6.7\times10^{-4}$.
At $p=7.3\times10^{-4}$, where the paper's TFIM example with the two-gross code becomes feasible, the curve gives
$10^{-23.0}$ and the bound $10^{-20.5}$.
The bound does not by itself change the paper's error budget: the paper assumes $10^{-11}$ and $10^{-9}$ at $p=10^{-3}$ and
$10^{-20}$ and $10^{-18}$ at $p=10^{-4}$ for the two-gross code's in-module and inter-module measurements, far above it.
Supplementary Fig.~\ref{fig:si-design}d shows both bounds on the standard schedule's circuit against the curve.
At $p=10^{-4}$ the bound from weight 18 lies 10.1 orders of magnitude above the paper's central value and 9.2 above the
top of its stated uncertainty; on the circuit of a schedule certified at 18 at one cycle the margins are 9.9 and 9.0,
so the higher distance leaves the gap almost unchanged.

\begin{table}[t]
\centering
\caption{Two-gross code in the setting of ref.~\citenum{yoder_2025_tour}, per cycle: its Table~2 with the stated
uncertainties, the sum of \cref{eq:si-ibm-fit} with its fit parameters, and lower bounds for every decoder
(\cref{eq:si-ibm-cycles}), summed over the two memory bases, from the 36 fault sets of weight 16 and from the 336
logical operators of the code of weight 18 of each type, with copies in all 18 cycles. Last column: the higher bound
minus the sum, in orders of magnitude.}
\label{tab:ibm-fit}
\footnotesize
\begin{tabular}{@{}lrrrrr@{}}
\toprule
 & \multicolumn{2}{c}{ref.~\citenum{yoder_2025_tour}} & \multicolumn{2}{c}{bound per cycle, any decoder} & \\
\cmidrule(lr){2-3}\cmidrule(lr){4-5}
$p$ & Table 2 & sum of its eq.~82 & weight 16 & weight 18 & higher bound minus sum \\
\midrule
$10^{-4}$ & $10^{-38.3\pm0.9}$ & $10^{-38.27}$ & $10^{-30.32}$ & $10^{-28.22}$ & $+10.05$ \\
$2\times10^{-4}$ &  & $10^{-33.60}$ & $10^{-27.91}$ & $10^{-25.52}$ & $+8.08$ \\
$5\times10^{-4}$ &  & $10^{-26.35}$ & $10^{-24.75}$ & $10^{-21.98}$ & $+4.37$ \\
$7.322\times10^{-4}$ &  & $10^{-22.98}$ & $10^{-23.44}$ & $10^{-20.52}$ & $+2.46$ \\
$10^{-3}$ & $10^{-20.1\pm0.5}$ & $10^{-20.09}$ & $10^{-22.38}$ & $10^{-19.34}$ & $+0.75$ \\
\bottomrule
\end{tabular}

\end{table}

\subsection{Low-weight failures of a minimum-weight decoder}
\label{subsec:ibm-leading}

A decoder that returns a least-weight set of mechanisms consistent with the detector outcomes, with ties broken
toward the more likely set, fails on a configuration $C$ contained in an undetectable logical fault set $G$ when $C$
is heavier than its complement $G\setminus C$, which fires the same detectors and whose logical effect differs by
that of $G$, and on the less likely of the two when they are equally heavy.
A failing configuration of weight $w$ therefore lies in a logical fault set of weight at most $2w$, and the logical
fault sets of weight at most $2w$ give every failing configuration of weight at most $w$.
For a circuit of distance $d$ the lowest-order failures have weight $\lceil d/2\rceil$: for even $d$ they are halves of
minimum fault sets; for odd $d$ they are the larger parts of minimum fault sets together with the halves of logical
fault sets of weight $d+1$, which fail at the same order.

For the gross code we obtained the undetectable logical fault sets of weight at most 12 of the instruction in the
setting of \cref{subsec:ibm-fit}, with $C=10$ cycles, with the standard schedule and with the two schedules of distance 11
defined below, in both bases (\texttt{papers/ibm\_bicycle\_matched\_spectrum.py}).
The search ran in enumeration mode without its dominance exclusions (\cref{subsec:search-mitm}), which keep every
fault set of minimum weight but may drop heavier ones; enumerations up to weights 11 and 12 give the same sets of
weight 11.
A lookup table of three columns does not fit in memory for 10 cycles, but the fault sets are local in time: every translate
in time of a fault set of a shorter circuit whose mechanisms exist in the longer circuit is a fault set of the longer
circuit, since its detectors and observables are the same up to the shift.
We therefore enumerated 3 and 4 cycles and translated the sets to 10 cycles.
The translates of the sets of 3 cycles reproduce the enumeration of 4 cycles exactly, except for one orbit of 72 sets of
weight 12 in the $X$ basis of the two schedules of distance 11, out of $1.7\times10^{6}$, so sets spanning more
cycles should not change the rates at the precision given.
For the standard schedule in the $Z$ basis, a direct enumeration with 10 cycles and a lookup table of two columns finds
exactly the translated sets at all three weights.
\Cref{tab:ibm-spectrum} lists their numbers.
From them, the failing configurations of weight 5 and 6 give the logical error rate of the decoder to order $p^{6}$
(\texttt{papers/ibm\_bicycle\_matched\_mwd.py}): each configuration contributes $\prod_{j\in C}p_j$, the probability that its
mechanisms occur whatever else occurs.
Whether a configuration of weight 5 fails is decided exactly. With distance 10, the other explanations of its detectors
have weight 5: of the other logical class, its complements in the sets of weight 10 that contain it; of its own class,
the parts $G'\setminus B$ of the sets $G'$ of weight 10 that contain such a complement $B$ and have the logical class of
the configuration and $B$ together (an explanation of its own class and $B$ form a logical fault set, which has weight
10 only if they are disjoint). It fails when the likeliest of these is of the other class, ties counting as failures;
explanations of its own class rescue 720 of 178,200 configurations in the $Z$ basis of the standard schedule and 12,456
of 351,000 in the $X$ basis.
Configurations of weight 5 are counted once each, although one can be the failing half of two sets (2\% to 3\% of
them are), and two that share four mechanisms both occur when their union of weight 6 does, so each such union is
subtracted once for every failing configuration of weight 5 it holds beyond the first (inclusion--exclusion at order
$p^6$; about $3\times10^{-4}$ of the term of weight 5 at $p=10^{-4}$).
Configurations of weight 6 from sets of weight 10 and 11 are counted once, and those that contain a failing
configuration of weight 5 are left out.
A split of a set of weight 12 in which one half contains a reducible pair, two mechanisms whose joint effect is that
of a single mechanism of the model, is left out: that half has a lighter explanation in its own logical class, and
the failure on the other half is counted among the configurations of sets of weight 11. This removes 9\% to 31\% of
the splits; the halves of the remaining $1.4\times10^{9}$ to $2.8\times10^{9}$ splits are counted without checking
for repeats.
Configurations of weight 6 are counted as failing whenever they are heavier than their complement in the set, or the
less likely half of a set of weight 12, without checking for lighter or likelier explanations of their own logical
class, and every configuration of weight 6 that contains a failing one of weight 5 counts as failing, so the term of
order $p^6$ is an upper estimate.
Only mechanisms that fire detectors of the memory basis alone enter, as in the bound, so configurations involving the
other mechanisms are not counted; the expansion thus describes a minimum-weight decoder that uses all detectors,
whereas the simulations of \cref{subsec:ibm-sim} decode the detectors of the memory basis alone.
Terms of weight 7 are not included. Extrapolating the numbers of fault sets to weight 14 with a growth of 50 to 70 per
unit of weight, as from weight 10 to 12 for the standard schedule, suggests that they add about a tenth to the standard schedule's rate at
$p=10^{-4}$ and between half and all of the rate of the schedules of distance 11, whose rates start at order $p^6$; at
$p=5\times10^{-4}$ the terms of weight 6 already exceed those of weight 5, and the expansion does not converge.

\begin{table}[t]
\centering
\caption{Undetectable logical fault sets of weight 10 to 12 of the gross code's instruction with 10 cycles, and the
logical error rate per cycle of a minimum-weight decoder at $p=10^{-4}$ to order $p^6$: failing configurations of weight 5
(halves of the sets of weight 10), of weight 6 from the sets of weight 10 and 11, and of weight 6 from halves of the sets
of weight 12. The sets come from enumerations of 3 and 4 cycles translated to 10 cycles; a direct enumeration with 10
cycles confirms them for the standard schedule in the $Z$ basis, and the distance 11 of the nine-timestep schedule is
also certified directly at 10 cycles.}
\label{tab:ibm-spectrum}
\footnotesize
\setlength{\tabcolsep}{4pt}
\begin{tabular}{@{}lcrrrrrrrr@{}}
\toprule
 & & & \multicolumn{3}{c}{logical fault sets of weight} & \multicolumn{4}{c}{failures of weight, per cycle at $p=10^{-4}$} \\
\cmidrule(lr){4-6}\cmidrule(l){7-10}
schedule & basis & $d$ & 10 & 11 & 12 & 5 & 6 (sets 10, 11) & 6 (sets 12) & total \\
\midrule
standard (8 timesteps) & $Z$ & 10 & 1,440 & 141,912 & 7,497,120 & $10^{-16.9}$ & $10^{-18.1}$ & $10^{-17.3}$ & $10^{-16.8}$ \\
 & $X$ & 10 & 2,880 & 199,944 & 8,875,332 & $10^{-16.9}$ & $10^{-18.2}$ & $10^{-17.4}$ & $10^{-16.8}$ \\
best of 8 timesteps & $Z$ & 11 & 0 & 52,776 & 3,743,616 & -- & $10^{-18.7}$ & $10^{-17.6}$ & $10^{-17.5}$ \\
 & $X$ & 11 & 0 & 27,504 & 4,860,852 & -- & $10^{-19.0}$ & $10^{-17.6}$ & $10^{-17.5}$ \\
best of 9 timesteps & $Z$ & 11 & 0 & 42,840 & 4,176,420 & -- & $10^{-18.3}$ & $10^{-17.1}$ & $10^{-17.0}$ \\
 & $X$ & 11 & 0 & 27,504 & 4,860,852 & -- & $10^{-18.5}$ & $10^{-17.1}$ & $10^{-17.0}$ \\
\bottomrule
\end{tabular}

\end{table}

For the standard schedule at $p=10^{-4}$, the halves of the minimum fault sets contribute $10^{-16.9}$ per cycle and
basis, and the halves of the sets of weight 12 add $10^{-17.3}$ ($Z$) and $10^{-17.4}$ ($X$); with the configurations of weight 6
from sets of weight 10 and 11, the rate is $10^{-16.48}$ per cycle summed over the two bases.
For the schedules of distance 11, the configurations of weight 5 are gone, but the halves of sets of weight 12 remain.
We chose two of the 428 schedules by the configurations from minimum fault sets alone, in the instruction with one cycle.
Summed over the two bases, these give $10^{-19.8}$ at $p=10^{-4}$ for 16 of the schedules, among them $p_X=(1,0,6,4,3,2)$,
$p_Z=(3,4,2,5,6,0)$ in the slot order of \cref{subsec:design-bb6}, against $10^{-19.3}$ for the median schedule (the numbers
of minimum fault sets range from 936 to 51,192 per basis).
The check qubits of both types of this schedule act in slots 0 and 6, so with one timestep each for preparation and
measurement it needs nine timesteps per cycle, which adds idle noise.
Among the 100 schedules that fit eight timesteps, as the standard schedule does, the same measure is best, at $10^{-19.4}$,
for $p_X=(5,6,1,2,3,4)$, $p_Z=(3,2,4,0,1,5)$.
With these two schedules the rate per cycle at $p=10^{-4}$, summed over the bases, is $10^{-17.24}$ for eight timesteps and
$10^{-16.74}$ for nine.
The eight-timestep schedule thus lowers the logical error rate of a minimum-weight decoder sixfold to order $p^6$, or about
threefold allowing for the terms of higher weight, and the nine-timestep schedule twofold, much less than the roughly 90-fold and 30-fold that the
minimum fault sets alone would suggest in this setting; since their rates are of order $p^6$ and the standard schedule's of
order $p^5$, the gain grows toward lower noise, to 43-fold and 13-fold at $p=10^{-5}$.
Ranking all 428 schedules by the full expansion would require their sets of weight 12.

For the standard schedule, the rate of $10^{-16.5}$ at $p=10^{-4}$ lies about one and a half orders of magnitude below the
$10^{-14.8}$ that ref.~\citenum{yoder_2025_tour} extrapolated with Relay-BP.
The expansion holds for a minimum-weight decoder, which is not efficient in general, so it does not bound the rate of
Relay-BP; it suggests that the projection for the gross code leaves room for better decoding, whereas the projection for
the two-gross code lies below what any decoder can reach.

\subsection{Simulations}
\label{subsec:ibm-sim}

We sampled the matched circuit of \cref{subsec:ibm-fit} with 10 cycles in Stim and decoded the outcomes of the detectors of the memory basis with
BP-OSD\cite{roffe_2020_decoding} (ldpc's \texttt{BpOsdDecoder}: min-sum belief propagation with 100 iterations and
scaling factor 0.625, then combination-sweep ordered-statistics decoding of order 7; \texttt{papers/ibm\_bicycle\_sim.py}),
on the detector error model restricted to these detectors with mechanisms merged by their effect, for the three schedules
in both bases at $p=2.5\times10^{-3}$ to $8\times10^{-3}$, until at least 100 failures or $1.5\times10^{6}$ shots per point.
A test with the standard schedule at $p=5\times10^{-3}$ gave the same rate with 1,000 iterations (89 failures in 800 shots,
against 86 with 100 iterations).
\Cref{tab:ibm-sim} lists the rates.
The eight-timestep schedule agrees with the standard one within statistical error at every noise level, the largest
difference being 13\%.
Up to $p=5\times10^{-3}$ the nine-timestep schedule fails 2.2 to 2.5 times as often; above that the failure
probability of the 10 cycles approaches saturation for both, and the ratio falls to 1.1 at $p=8\times10^{-3}$.
In this regime failures of many faults dominate, so the rates reflect the numbers of fault sets of all weights and the
noise per cycle more than the distance; the rates of ref.~\citenum{yoder_2025_tour} were extrapolated from simulations in
this regime.
The simulations with BP-OSD show no gain of the eight-timestep schedule at $p\ge2.5\times10^{-3}$; the expansion of
\cref{subsec:ibm-leading}, for a minimum-weight decoder, gives three- to sixfold at $10^{-4}$ and 43-fold at $10^{-5}$.

\begin{table}[t]
\centering
\caption{Failure probabilities of the 10 cycles of the gross code's instruction simulated with BP-OSD, in the unit of each
row, and in parentheses the number of failures, for each schedule and memory basis.}
\label{tab:ibm-sim}
\footnotesize
\begin{tabular}{@{}lccccccc@{}}
\toprule
 & & \multicolumn{2}{c}{standard} & \multicolumn{2}{c}{best of 8 timesteps} & \multicolumn{2}{c}{best of 9 timesteps} \\
\cmidrule(lr){3-4}\cmidrule(lr){5-6}\cmidrule(l){7-8}
$p$ & unit & $Z$ & $X$ & $Z$ & $X$ & $Z$ & $X$ \\
\midrule
$8\times10^{-3}$ & $10^{-1}$ & 8.39 (1,665) & 8.48 (1,682) & 8.38 (1,662) & 8.34 (1,654) & 9.58 (1,901) & 9.65 (1,914) \\
$7\times10^{-3}$ & $10^{-1}$ & 6.07 (1,205) & 6.01 (1,192) & 6.00 (1,190) & 5.93 (1,177) & 8.32 (1,651) & 8.42 (1,671) \\
$6\times10^{-3}$ & $10^{-1}$ & 3.24 (643) & 3.05 (605) & 3.06 (607) & 3.09 (613) & 5.66 (1,123) & 5.75 (1,140) \\
$5\times10^{-3}$ & $10^{-1}$ & 1.12 (222) & 1.17 (233) & 1.16 (230) & 1.02 (203) & 2.49 (494) & 2.54 (503) \\
$4\times10^{-3}$ & $10^{-2}$ & 2.44 (139) & 2.60 (118) & 2.38 (116) & 2.67 (107) & 5.85 (116) & 5.65 (112) \\
$3\times10^{-3}$ & $10^{-3}$ & 3.12 (104) & 2.85 (175) & 3.18 (261) & 2.67 (108) & 7.88 (118) & 6.99 (112) \\
$2.5\times10^{-3}$ & $10^{-4}$ & 6.87 (170) & 8.06 (102) & 8.04 (199) & 8.81 (107) & 18.5 (458) & 18.5 (459) \\
\bottomrule
\end{tabular}

\end{table}

\subsection{Scope}
\label{subsec:ibm-scope}

The instruction is modeled from the description in ref.~\citenum{yoder_2025_tour}: the schedule, the timing of
measurement and reset, the noise model and the noise-free boundaries follow the paper; details that the paper does not
state, such as the idle noise of the data qubits in the timestep in which the check qubits are first prepared, are our
choices.
The certified distances and the lower bounds refer to this instruction alone. In a computation the instruction is
repeated and its boundaries are not noise-free; the memory circuit of the two-gross code with 18 cycles, audited in
the main text, also has distance 16, so the shortfall does not come from the boundaries.
The bounds count only failures from the fault sets named above, so the true logical error rates are at least these values.

\FloatBarrier
\section{Codes of logical measurements}
\label{sec:surgery}

A logical measurement on a qLDPC code joins code blocks to an ancilla system for several rounds.
The merged code must keep the distance, and its logical qubits include those of the blocks and possibly new ones.
We certified the merged codes of two published constructions (\texttt{papers/surgery\_codes.py},
\texttt{papers/ssip\_exact.py}, \texttt{papers/ssip\_gauge\_fix.py}; records in \texttt{results/surgery}).

\subsection{Universal adapters}

Ref.~\nocite{swaroop_2024_adapters}\citenum{swaroop_2024_adapters} measures logical operators of a \qc{98,6,12} bivariate bicycle code and a
\qc{200,20,10} lifted-product code through auxiliary graphs and adapters, with distances from CPLEX and BP-OSD searches.
\Cref{tab:surgery-adapters} gives the certified distances of the check matrices in its data record (Zenodo
10.5281/zenodo.17527545), from the weight-bounded search without symmetry, with every lightest logical operator checked
against the matrices.
All reported distances hold, and each code takes at most 0.2~s.

\begin{table}[t]
\centering
\caption{Certified distances of the codes of ref.~\citenum{swaroop_2024_adapters}: the two original codes and the
deformed codes that measure the logical operators named.}
\label{tab:surgery-adapters}
\small
\begin{tabular}{@{}lrrcc@{}}
\toprule
code & $n$ & $k$ & $d_X$ & $d_Z$ \\
\midrule
bivariate bicycle \qc{98,6,12} & 98 & 6 & 12 & 12 \\
lifted product \qc{200,20,10} & 200 & 20 & 10 & 10 \\
BB, auxiliary graph for $\bar Z_1$ & 121 & 5 & 12 & 12 \\
BB, auxiliary graph for $\bar Z_3$ & 115 & 5 & 12 & 12 \\
BB, joint $\bar Z_1\bar Z_3$, naive & 137 & 5 & 12 & 12 \\
LP, auxiliary graph for $\bar Z_2$ & 220 & 19 & 10 & 10 \\
BB, intracode adapter, $\bar Z_1\bar Z_3$ & 150 & 5 & 12 & 12 \\
BB and LP, intercode adapter, $\bar Z_1\bar Z_2$ & 355 & 25 & 10 & 10 \\
\bottomrule
\end{tabular}

\end{table}

\subsection{Surgery on the gross code}
\label{subsec:surgery-ssip}

SSIP\cite{cowtan_2024_ssip,ssip_code} builds external merges between two gross-code blocks, internal merges within one
block, single-qubit measurements and parallel versions of the external merges and single-qubit measurements, and its benchmark
scripts raise the merge depth $r$ until
QDistRnd reports distance 12 for the merged code.
Logical qubits that a merge introduces become gauge qubits, and its paper defines the distance of the resulting
subsystem code by dressed operators (its Section~2, computed as in its Appendix~A): the least weight of an $X$-type operator that commutes with the $Z$ checks and lies
outside the span of the $X$ checks and the gauge $X$ logicals, and the same with $X$ and $Z$ exchanged.
Its routine \texttt{subsystem\_distance\_GAP} evaluates these with QDistRnd as \texttt{DistRandCSS}$([\vb{P}_Z;\vb{G}_Z],\vb{P}_X)$
and \texttt{DistRandCSS}$([\vb{P}_X;\vb{G}_X],\vb{P}_Z)$, where $\vb{G}_Z$ and $\vb{G}_X$ hold the gauge logicals.
\texttt{DistRandCSS}$(\vb{P},\vb{Q})$ returns an upper bound, found by randomized search, on the least weight of a
vector orthogonal to the rows of $\vb{P}$ and independent of the rows of $\vb{Q}$\cite{qdistrnd}, so these calls bound
the bare distances, of operators that also commute with the
gauge logicals of the other type; the dressed distances need the augmented matrix as the second argument.
Dressed distances are at most the bare ones, with equality when a merge leaves no gauge qubit.

We ran SSIP's own benchmark scripts (\texttt{benchmarking/gross\_code\_*.py}) with that routine replaced by exact
computations of both distances: the weight-bounded search up to weight 12, with every lightest operator checked against
the matrices.
In one pass the scripts choose the merge depth from the bare distance, as SSIP's code does, and in another from the
dressed distance, as its paper defines.
\Cref{tab:surgery-ssip} gives the result.
The first pass reproduces SSIP's recorded depths, so its QDistRnd values are tight for the quantity it computes.
At the published depths, the dressed distance is below 12 in 78 of the 141 measurements, down to
2 for the parallel measurements, where 36 to 75 gauge qubits remain.
The second pass needs deeper merges, with 12,834 instead of 5,664 ancilla qubits summed over all
measurements.
Two cases were also checked with plain linear algebra outside our solver: the external $X$ merge of logical operator 2
of each block at depth 1 (306 qubits, 26 logical qubits, 3 of them gauge) has an $X$ operator of weight 9 that commutes
with every $Z$ check and lies outside the span of the $X$ checks and gauge $X$ logicals, and the parallel $X$ measurement
of logical operators 6 to 11 at depth 1 (252 qubits, 42 logical qubits, 36 of them gauge) has one of weight 2 on two
ancilla qubits.
In the second case the gauge logicals that SSIP returns are independent modulo the checks and pair off, 6 logical
qubits remain, and the weight-2 operator acts on them nontrivially.

Which of the two distances applies depends on what happens to the gauge qubits.
In every measurement the shortfall lies in the measured basis: at the published depths the dressed distance in the
other basis is at least 12.
Fixing the gauge logicals of the other basis turns the merged code into the stabilizer code with these logicals added
as checks, whose distance is the bare distance in the measured basis and the dressed distance in the other, so all
141 merged codes then have distance 12 (last column of \cref{tab:surgery-ssip}).
The preparation of the new qubits fixes them.
For an $X$ measurement the new qubits start in $\ket{0}$, under which every $Z$ check of the merged code has a definite
first outcome, and with this preparation every gauge $Z$ logical has a definite value too: its restriction to the
original qubits lies in the span of the original $Z$ checks.
The same holds for $Z$ measurements with $\ket{+}$ and the gauge $X$ logicals.
We checked this for all 141 merged codes, rebuilt by SSIP's scripts at the published depths
(\texttt{papers/ssip\_gauge\_fix.py}).
Gauge logicals commute with every check, so they keep these values during the merge, and measuring the new qubits out
at the split, with the next round of checks, recovers them.
A decoder that compares the two values detects every light dressed operator, among them the two above, which each flip
a definite gauge logical, so the merged code with its gauge fixed has distance 12; with the gauge qubits left free, as
in SSIP's definition, it has the dressed distance.
Like SSIP's own values, these are distances of the merged codes, not fault distances of the whole measurement, which
also depend on the number of rounds and on noisy preparation and readout.
Later constructions fix the gauge explicitly, through the initial state of the new qubits\cite{williamson_2024_gauging}
or through extra checks\cite{cross_2024_surgery}.

SSIP's merged codes depend on the version of the networkx package, which it does not pin: with networkx 3.6, 13 of the 84
external merges at depth 1 differ from those with networkx 3.7 in their parameters or distances.
The table uses Python 3.12, numpy 1.26, the version SSIP pins, and networkx 3.7; the runs with networkx 3.6 (and numpy 2.5) show the
same picture.
The dressed distance depends on which new logical qubits are taken as gauge qubits; we use the gauge logicals that
SSIP's code returns, with the definition of its paper.

\begin{table}[t]
\centering
\caption{SSIP's measurements on the gross code: the number of measurements and SSIP's published depths; at those
depths, how many have a dressed distance below 12 and the least such distance; how many need a deeper merge when the
depth is chosen by the dressed distance, and the ancilla qubits, summed over the measurements, at the published and at
those depths; and the least distance at the published depths with the gauge logicals of the other basis fixed, as the
preparation of the new qubits does. The bare distance is 12 for all of them at the published depths.}
\label{tab:surgery-ssip}
\footnotesize
\resizebox{\linewidth}{!}{\begin{tabular}{@{}llrcrrrrr@{}}
\toprule
 & & & published & \multicolumn{2}{c}{dressed at published depth} & deeper & ancilla qubits & gauge \\
measurement & basis & number & depths & below 12 & least & with dressed & published $\to$ dressed & fixed \\
\midrule
external merges & $X$ & 42 & 1 & 24 & 9 & 24 & 756 $\to$ 1,476 & 12 \\
external merges & $Z$ & 42 & 1 & 21 & 9 & 21 & 756 $\to$ 1,386 & 12 \\
internal merges & $X$ & 12 & 2--3 & 6 & 6 & 6 & 666 $\to$ 1,026 & 12 \\
internal merges & $Z$ & 15 & 1--3 & 10 & 8 & 10 & 750 $\to$ 1,170 & 12 \\
single-qubit measurements & $X$ & 12 & 1--3 & 6 & 8 & 6 & 576 $\to$ 756 & 12 \\
single-qubit measurements & $Z$ & 12 & 1--3 & 6 & 8 & 6 & 576 $\to$ 756 & 12 \\
parallel external merges & $X$ & 1 & 1 & 1 & 2 & 1 & 216 $\to$ 2,016 & 12 \\
parallel external merges & $Z$ & 1 & 1 & 0 & 12 & 0 & 216 $\to$ 216 & 12 \\
parallel single-qubit measurements & $X$ & 2 & 1--3 & 2 & 2 & 2 & 576 $\to$ 2,016 & 12 \\
parallel single-qubit measurements & $Z$ & 2 & 1--3 & 2 & 2 & 2 & 576 $\to$ 2,016 & 12 \\
\bottomrule
\end{tabular}
}
\end{table}

\FloatBarrier
\section{Chips with dead checks}
\label{sec:chips}

\subsection{Model}

Ref.~\citenum{yoder_2025_tour} estimates that a perfect gross-code module with its logical processing unit fabricates about two times in five at a junction yield of 99.98\%, and notes that encoding around defects typically lowers the distance, without giving numbers.
We study the memory block of the gross code.
A dead check qubit removes its check. So does a dead coupler, because a check missing one data qubit would no longer commute with the others.
The check's gates, preparation, measurement and detectors are dropped in every round, and its data qubits idle in those slots.
Logical operators are those of the perfect code.
Dead data qubits are not modeled, so such a chip counts as rejected.
The circuit is the matched circuit of \cref{subsec:ibm-fit} with lists of live checks (\texttt{results/defect\_study/chip.py}), which with no dead check writes the identical circuit, checked on 7 schedules, both bases and 1 to 3 cycles.
The check matrices have rank 66 of 72, so any one dead check, and 99.3\% of pairs, leave the stabilizer group unchanged.

\begin{proposition}[Reduction to one type of dead check]
\label{prop:chip-reduction}
Let $d_{\mathrm{circ}}^{(Z)}(D_X,D_Z)$ be the circuit distance in memory basis $Z$ with the sets $D_X$ and $D_Z$ of dead $X$ and $Z$ checks. Then $d_{\mathrm{circ}}^{(Z)}(D_X,D_Z)\ge d_{\mathrm{circ}}^{(Z)}(\emptyset,D_Z)$, and the same holds with $X$ and $Z$ exchanged.
\end{proposition}

\begin{proof}
In memory basis $Z$ the distance is decided on the sector problem, which keeps the detectors of the $Z$ checks and the $Z$ logical observables, so a fault enters it through the $X$ errors it leaves on the data qubits, including those that an $X$-check qubit spreads to them, and through faults on the $Z$-check qubits. Removing an $X$ check deletes the faults of its own qubit and gates, hook faults included, which can only remove columns. An $X$ error on a data qubit passes a removed gate unchanged, so no remaining column changes. A fault of the new idle step of a data qubit has the same $X$ part as a fault of the removed gate on that qubit alone, so it adds no column. Every column of the sector problem with $D_X$ is therefore a column of the problem without it, and every undetected logical fault set of the former is one of the latter.
\end{proof}

We therefore run only problems with dead checks of the memory basis's own type, and a chip's value is the minimum of the two bases, a certified lower bound on its distance.
Whole-chip runs on 29 chips, in both bases, agree with the values of the dead checks of the memory basis's own type in 56 of the 58 runs; in the other two a dead check of the other type also removes the hook faults of the lightest set, so the whole-chip value is 10 and not 9, and the chip values, the minimum over the two bases, agree on all 29 chips.
Every run uses one thread and the pipeline (\texttt{dp=2}), followed by a replay of the fault set in Stim; on the last 1,692 runs the median time is 11~s and the longest 405~s.

\subsection{Results}

One dead $Z$ check on the standard schedule gives 10 at one cycle in the $Z$ basis and 9 at 2, 3, 4, 6 and 10 cycles, so one cycle is not enough for chips; a dead $X$ check gives 10 in both bases up to three cycles.
On all pairs at three cycles, up to translation, 70 of 71 pairs of dead $Z$ checks give 9, of 71 pairs of dead $X$ checks 62 give 10 and 8 give 9, and the one pair that loses a stabilizer gives 6 in both.
\Cref{tab:chips} gives the distances at three cycles on the standard schedule and on the distance-11 schedule $p_X=(5,6,1,2,3,4)$, $p_Z=(3,2,4,0,1,5)$ of eight timesteps. Ten cycles equal three on all 78 single-basis problems tested, 77 chip patterns, all on the standard schedule; this does not test the distance-11 schedule at ten cycles.
Losing a stabilizer explains most low values, but not all. On the 200 sampled chips each with $m=3$, 4 and 5, every chip with a bound of 8 or below has lost a stabilizer on both schedules (2, 12 and 24 chips), so a rank test rejects it; the test also rejects a few chips at 9 (1, 2 and 2).
The same-type problems show that this is not a rule. Among the 1,228 sampled problems with four or five dead checks of one type, five keep every stabilizer and still fall to 8, for example dead $X$ checks 1, 24, 52 and 55 on the standard schedule, whose fault set of eight replays in Stim. From six dead checks on, a few whole chips keep their rank and fall to 8 or below (standard schedule, 2 of 200 at $m=6$ and at $m=8$; distance-11 schedule, 1 of 200 at $m=8$ and at $m=10$).

\begin{table}[t]
\centering
\caption{Gross-code chips with $m$ dead checks: percentage of chips at each certified lower bound on the circuit distance, three noisy cycles. For $m\le2$, all cases up to translation; for $m\ge3$, 200 random chips, the same on both schedules, every run exact and replayed.}
\label{tab:chips}
\small
\begin{tabular}{@{}lrrrrrrr@{}}
\toprule
 & \multicolumn{3}{c}{standard schedule} & \multicolumn{4}{c}{distance-11 schedule} \\
dead checks & 10 & 9 & $\le$8 & 11 & 10 & 9 & $\le$8 \\
\midrule
0 & 100.0 &  &  & 100.0 &  &  &  \\
1 & 50.0 & 50.0 &  &  & 100.0 &  &  \\
2 & 21.7 & 77.6 & 0.7 &  & 97.9 & 1.4 & 0.7 \\
3 & 8.5 & 90.5 & 1.0 &  & 91.0 & 8.0 & 1.0 \\
4 & 3.5 & 90.5 & 6.0 &  & 84.0 & 10.0 & 6.0 \\
5 & 0.5 & 87.5 & 12.0 &  & 76.0 & 12.0 & 12.0 \\
6 &  & 83.5 & 16.5 &  & 63.5 & 21.0 & 15.5 \\
8 &  & 61.0 & 39.0 &  & 33.0 & 28.5 & 38.5 \\
10 &  & 27.0 & 73.0 &  & 6.5 & 20.0 & 73.5 \\
\bottomrule
\end{tabular}

\end{table}

For a schedule chosen for the chip, we computed one dead check of each type on all 4,660 schedules whose six-cycle memory circuits have distance at least 10, at three cycles (1,165 orbits, all exact and replayed).
Of the 4,232 schedules of distance 10, 1,580 keep 10 under either dead check; of the 428 of distance 11, 400 drop only to 10 under either, and the others to 9.
The worst value, 8, occurs on 64 schedules.

\subsection{Yield}

We take the upper ends of IBM's junction counts, two per qubit and three per coupler, with 144 data qubits, 144 check qubits and 864 data-check couplers. A check dies if its qubit or one of its six couplers fails, and a chip with a dead data qubit is rejected. The number of dead checks follows from these independent failures; among chips with a given number the distance table above applies, with $m=7$ and 9 the mean of their neighbors and $m>10$ counted as 0 (0.01\% of chips at 99.9\%). \Cref{tab:chips-yield} gives the shares.

\begin{table}[t]
\centering
\caption{Percentage of all chips at each junction yield: perfect chips, chips certified at distance at least 10 and at least 9 on each schedule, and chips that keep the stabilizer rank and have no dead data qubit.}
\label{tab:chips-yield}
\small
\begin{tabular}{@{}lrrrrrr@{}}
\toprule
 & & \multicolumn{2}{c}{distance $\ge$10} & \multicolumn{2}{c}{distance $\ge$9} & \\
junction yield & perfect & standard & distance-11 & standard & distance-11 & rank test accepts \\
\midrule
99.98\% & 53 & 70 & 94 & 94.3 & 94.3 & 94 \\
99.95\% & 21 & 41 & 84 & 85.9 & 85.9 & 86 \\
99.9\% & 4 & 16 & 67 & 72.0 & 72.0 & 72 \\
\bottomrule
\end{tabular}

\end{table}

\subsection{Logical error rates}

A lower distance need not mean a much higher rate. From all fault sets up to weight 10 at 3 and 4 cycles, extrapolated linearly in the number of cycles to 10 (the method of \cref{subsec:ibm-leading}, \texttt{results/defect\_study/rate\_chip.py}), the rate of a minimum-weight decoder to order $p^5$ on the standard schedule, at $p=10^{-4}$ and at $10^{-5}$, is 1.09 times the perfect chip's for one dead $Z$ check and 1.03 times for one dead $X$ check.
On the distance-11 schedule, fault sets up to weight 10 at 3 and 4 cycles exist only in the basis of the dead check: 159 and 224 of weight 10 with one dead $Z$ check, 39 and 62 with one dead $X$ check.
The rate per 10-cycle instruction to order $p^5$ at $p=10^{-4}$ is then $3.79\times10^{-18}$ and $8.26\times10^{-19}$, which is 1.6\% and 0.4\% of the $2.33\times10^{-16}$ of the perfect chip on the standard schedule.
The perfect chip on this schedule starts at order $p^6$, estimated at $5.7\times10^{-17}$ per instruction, ten times the
per-cycle value of \cref{subsec:ibm-leading}, so a dead check adds 7\% and 1.4\% to it, and about 66\% and 14\% at $p=10^{-5}$.
We did not compute how a dead check changes the term of order $p^6$.
Simulations with BP-OSD at higher noise (\cref{tab:chips-sim}; settings of \cref{subsec:ibm-sim}, 10 cycles) find, among the tested chips with up to five dead checks, at most 1.6 times the perfect chip's rate at $p=3\times10^{-3}$ on the standard schedule, in the basis that the dead checks affect, and at most 1.4 times the perfect chip's rate on the distance-11 schedule.

\begin{table}[t]
\centering
\caption{BP-OSD failures per run ($\times10^{-3}$) on chips with dead checks, on the standard schedule and on the distance-11 schedule of eight timesteps, 10 cycles, in each memory basis.}
\label{tab:chips-sim}
\small
\begin{tabular}{@{}lrrrrrr@{}}
\toprule
 & \multicolumn{2}{c}{$p=3\times10^{-3}$} & \multicolumn{2}{c}{$p=4\times10^{-3}$} & \multicolumn{2}{c}{$p=5\times10^{-3}$} \\
chip & $Z$ & $X$ & $Z$ & $X$ & $Z$ & $X$ \\
\midrule
\multicolumn{7}{@{}l}{standard schedule} \\
perfect chip & 2.81 & 2.84 & 23.18 & 24.69 & 111.90 & 117.44 \\
one dead $Z$ check & 3.51 & 2.61 & 29.90 & 23.19 & 134.07 & 120.46 \\
one dead $X$ check & 2.71 & 3.51 & 25.50 & 24.91 & 111.39 & 133.06 \\
dead $X$ and $Z$ checks & 3.14 & 3.33 & 27.65 & 26.37 & 136.59 & 127.02 \\
two dead $Z$ checks & 4.45 & 2.81 & 34.59 & 20.98 & 153.23 & 113.41 \\
three dead checks & 3.91 & 3.04 & 32.37 & 28.47 & 144.66 & 117.94 \\
five dead checks & 4.34 & 3.67 & 35.47 & 29.48 & 138.61 & 132.06 \\
\midrule
\multicolumn{7}{@{}l}{distance-11 schedule} \\
perfect chip & 3.18 & 2.67 & 23.85 & 26.75 & 115.93 & 102.32 \\
one dead $Z$ check & 3.69 & 2.62 & 31.89 & 21.90 & 120.16 & 118.30 \\
one dead $X$ check & 2.93 & 3.58 & 22.84 & 29.01 & 111.61 & 125.74 \\
dead $X$ and $Z$ checks & 3.24 & 3.20 & 27.74 & 27.01 & 123.14 & 125.74 \\
two dead $Z$ checks & 4.29 & 2.70 & 32.91 & 25.09 & 145.83 & 107.89 \\
three dead checks & 4.20 & 3.17 & 33.43 & 26.44 & 142.47 & 123.14 \\
five dead checks & 4.37 & 3.80 & 32.02 & 29.91 & 145.83 & 148.52 \\
\bottomrule
\end{tabular}

\end{table}

\subsection{Scope}

The study covers the memory block of the gross code with the same noise on all live qubits. It does not cover the logical processing unit, dead data qubits, or noise that varies across a chip. Exact runs on the two-gross code would take much longer, because a dead check removes the symmetry.

\FloatBarrier
\section{Software and reproducibility}
\label{sec:software}
\suppressfloats[t] %

\subsection{Implementation}
\label{subsec:software-implementation}

The method is implemented in Python with C kernels parallelized with OpenMP\cite{circuit_distance_code}.
It takes a Stim circuit, from which it builds the detector error model without decomposition and merges equal
mechanisms, or a pair of matrices, and returns a record with the certified interval, the log of every threshold
and the error that attains the upper bound, with its replay in Stim for circuits.
\Cref{tab:implementation} lists how the elements of \cref{sec:tropical-rank-dp,sec:sparse-dp,sec:weight-search}
appear in the implementation.
The benchmark, the instruction distances, the two-gross scan and the chips use one configuration of the pipeline
(\texttt{dp=2}, default settings), described in the Methods of the main text; the audit runs use the search with
its reductions, through the drivers in \texttt{papers/}.

\begin{table}[t]
\centering
\caption{Elements of the method and their implementation.}
\label{tab:implementation}
\small
\begin{tabular}{@{}>{\raggedright\arraybackslash}p{0.31\linewidth}>{\raggedright\arraybackslash}p{0.64\linewidth}@{}}
\toprule
element & implementation \\
\midrule
dense Rank DP (\cref{subsec:dense-rank-dp}) & linear decompositions; the column order with the smallest peak cut rank among the given order, its reverse and, up to 2,000 columns outside circuit sectors, greedy orders; one query per logical row; the work of the plan, estimated from one logical row, is computed before the run and converted to a time with a measured rate \\
truncated Rank DP (\cref{sec:sparse-dp}) & minimal-span trellis of the same order; sparse tables of up to 56 bit positions; meet in the middle with a split schedule from probe sweeps; checkpointed traceback; a covering-bound lookahead and a dense fallback, both left off in the reported runs \\
branching (\cref{subsec:threshold-search}) & an unsatisfied row with the fewest available columns; thresholds in increasing order; the tree is split at a fixed depth into tasks that run in parallel \\
anchor reduction (\cref{subsec:search-mitm}) & orbits of lattice translations, cyclic shifts or tiling automorphisms; every generator is checked against \cref{eq:search-symmetry-condition} \\
node pruning (\cref{subsec:search-mitm}) & detector-count bound; covering bound \cref{eq:covering-bound} with $y_r=1/m_r$ and one greedy ascent; dominance bound \cref{eq:kernel-dominance} for zero-label pairs and triples \\
completion tables (\cref{eq:completion-table}) & all supports of at most $k=2$ or $3$ columns, keyed by linear fingerprints of the syndrome, with logical labels or without them; every positive query, including a fingerprint collision, is followed by reconstruction within the full remaining budget \\
projected completion filter (\cref{subsec:search-reductions}) & table of the time-collapsed problem, queried with the projected residual \\
enumeration (\cref{cor:search-enumeration}) & recording mode, used for the minimizers of relaxed problems \\
lifting (\cref{subsec:search-reductions}) & preimages of all minimizers, one per orbit; lifting through weight $W$ (\cref{prop:lifting-W}) for the largest bivariate bicycle circuits \\
restriction (\cref{subsec:search-reductions}) & faults whose detectors lie in one or two rounds (windows), for witnesses \\
choice of evaluator & on the main problem, the dense DP when its predicted time is at most 10~s, after a 1~s head start of the search, and a race when it fits but is slower; on the collapsed problem, the dense DP only when predicted within 10~s; for circuits, the truncated DP at threshold $d_{\mathrm{up}}-1$ after a head start of the search of up to its predicted time, leaving it 1.5 times that time before the deadline \\
\bottomrule
\end{tabular}
\end{table}

\subsection{Verification}
\label{subsec:software-verification}

Every error reported is checked against the original matrices of its problem.
For circuits it is also translated into physical faults, inserted into the noiseless Stim circuit as
deterministic Pauli errors and simulated: no detector fires and a logical observable flips.
Lower bounds come only from completed thresholds of an evaluator whose completeness is proved in
\cref{sec:tropical-rank-dp,sec:sparse-dp,sec:weight-search}.
The test suite compares every kernel with brute force on small instances: the search with and without tables,
symmetry and pruning, the dense DP, the truncated DP with every split (272,076 random queries with every constant split and 66,814 with random split schedules), and the collision counterexample of
\cref{app:search-fingerprints}.
Further checks compare the method with integer programming and MaxSAT on the same matrices, with an independent
implementation for the \qc{72,12,6} circuit, and across our own evaluators wherever more than one finishes.
In the benchmark, a harness checks every record against the ground truth and against the bounds of all other
records.

Up to 20 October 2026, the search kernel limited reconstruction after a table hit to the cost returned by the
table in its labeled mode, the path discussed in \cref{app:search-fingerprints}; the kernel now reconstructs within
the full remaining budget in both modes.
Benchmark runs use label-free tables and were not affected; the final benchmark pass, and the audit runs after that
date, used the corrected kernel.
For every other value reported, either the stored number of failed reconstructions is zero at every negative
threshold, which certifies the value, or the value was recomputed with the corrected kernel: the
\qc{288,12,18} circuit in the $X$ basis, the lifting arguments for the \qc{360,12,24} circuit, the code distances of
the self-dual bivariate bicycle, GALA, Kasai, routing and mitten codes, and the winners of the schedule designs.
All recomputations reproduced the reported values; the records of the new scripts store the hash of the kernel that
produced them, and the others were produced in the same session with the corrected kernel.
On 30 October 2026 four further paths that could end a threshold early were fixed (a cap on stored solutions, the
length of an undo list, a depth limit of 40 and 8-bit costs); no reported run reaches any of them (largest row
degree 242, largest threshold 33, largest DP value 81), and the 238 enumerations behind lifting certificates were
rerun with the fixed kernel and reproduced every bound.

\subsection{Reproduction}
\label{subsec:software-reproduction}

The repository\cite{circuit_distance_code} contains the library (\texttt{cdist/}), the benchmark harness, adapters
and instances (\texttt{bench/}), the builders of all published and designed circuits (\texttt{papers/}), and every
result record (\texttt{results/}), each with its certified bounds, its threshold log and its error.
The benchmark is run by \texttt{bench/run\_matrix.py}, which pins every run to its own cores with bounded memory and
resumable records, and its later passes by \texttt{bench/tier\_b.py}.
The figures and the other tables are drawn by \texttt{docs/make\_ncs\_figures.py} and the benchmark tables by
\texttt{bench/make\_paper\_tables.py}, both from the records; \texttt{docs/sync\_paper.sh} runs both and copies their
output into a checkout of the paper under the names it reads.
The audit inventory is \texttt{docs/AUDIT.md}, with the values of Fig.~3 of the main text and \cref{tab:audit} in
\texttt{docs/audit\_items.csv}.
The memory instructions of IBM's bicycle architecture, with their distances, bounds, enumerations and simulations
(\cref{sec:ibm}), are built and run by \texttt{papers/ibm\_bicycle*.py}, with the records in \texttt{results/ibm},
\texttt{results/ibm\_matched} and \texttt{results/bb288\_scan}; the chips with dead checks are in
\texttt{results/defect\_study}.
The software requires Python 3.9 or later, Stim 1.15 or later, NumPy (2.0 or later for the truncated DP's
lookahead) and a C compiler; each benchmark run used cores of one socket of an AMD EPYC 9654 processor.
Circuits and data files of the audited papers are not redistributed, except the Stim files of
ref.~\citenum{tesseract_code} (Apache 2.0): manifests list the sources and checksums, and the circuits rebuilt from a
paper's text are built by the scripts in \texttt{papers/}.
The software is released under the MIT license.

\FloatBarrier
\putbib[refs,ncs]
\end{bibunit}

\end{document}